\documentclass[12pt, oneside]{article}   	
\usepackage{jheppub}
\usepackage{bbold}
\usepackage{amsmath,breqn, mathtools}
\usepackage{amssymb}
\usepackage{amsfonts}
\usepackage{amsthm}
\usepackage{amsbsy}
\usepackage{array}
\usepackage{comment}

\usepackage{graphics, color}
\usepackage{graphicx}

\usepackage{caption}
\usepackage{subcaption}

\usepackage{tabularx}

\usepackage{tikz}
\tikzstyle{bag} = [align=center]

\usetikzlibrary{decorations.pathreplacing,decorations.markings,arrows.meta}
\usetikzlibrary{decorations.pathmorphing}

\tikzset{
  on each segment/.style={
    decorate,
    decoration={
      show path construction,
      moveto code={},
      lineto code={
        \path [#1]
        (\tikzinputsegmentfirst) -- (\tikzinputsegmentlast);
      },
      curveto code={
        \path [#1] (\tikzinputsegmentfirst)
        .. controls
        (\tikzinputsegmentsupporta) and (\tikzinputsegmentsupportb)
        ..
        (\tikzinputsegmentlast);
      },
      closepath code={
        \path [#1]
        (\tikzinputsegmentfirst) -- (\tikzinputsegmentlast);
      },
    },
  },
  mid arrow/.style={postaction={decorate,decoration={
        markings,
        mark=at position .5 with {\arrow[#1,scale=1.5]{stealth}}
      }}},
}

\numberwithin{equation}{section}

 \def\bz{{\bar z}}
 
\def\0{{(0)}}
\def\1{{(1)}}
\def\2{{(2)}}

\def\<{\langle }
\def\>{\rangle }
\def\[{\left[}
\def\]{\right]}

\newcommand{\lan}{\langle}
\newcommand{\ran}{\rangle}
\usepackage[export]{adjustbox}

\newcommand{\bea}{\begin{eqnarray}}
\newcommand{\eea}{\end{eqnarray}}
\newcommand{\be}{\begin{equation}}
\newcommand{\ee}{\end{equation}}
\newcommand{\ba}{\begin{align}}
\newcommand{\ea}{\end{align}}

\newcommand{\sech}{\mbox{sech}}

\newcommand{\tr}{\mbox{tr}}

\renewcommand{\epsilon}{\varepsilon}

   \makeatletter
  \let\over=\@@over \let\overwithdelims=\@@overwithdelims
  \let\atop=\@@atop \let\atopwithdelims=\@@atopwithdelims
  \let\above=\@@above \let\abovewithdelims=\@@abovewithdelims
\renewcommand\section{\@startsection {section}{1}{\z@}%
                                   {-3.5ex \@plus -1ex \@minus -.2ex}%
                                   {2.3ex \@plus.2ex}%
                                   {\normalfont\large\bfseries}}

\renewcommand\subsection{\@startsection{subsection}{2}{\z@}%
                                     {-3.25ex\@plus -1ex \@minus -.2ex}%
                                     {1.5ex \@plus .2ex}%
                                     {\normalfont\bfseries}}

\def\tr{{\rm tr}}
\def\be{\bea}
\def\ee{\eea}

\def\bdm{\begin{dmath}}
\def\edm{\end{dmath}}

\def\bea{\begin{eqnarray}}
\def\eea{\end{eqnarray}}

\def\ba{\begin{eqnarray}}
\def\ea{\end{eqnarray}}

\def\nspc {} %

\def\bn{\mbox{\textit{\textbf{n}\nspc \,}}}

\def\tr{{\rm tr}}

\enlargethispage{\baselineskip}

\newpage
\renewcommand\Large{\fontsize{15.5}{16}\selectfont}

\def\mO{{\mathcal{O}}}

\usepackage{pgfplots}
\pgfplotsset{compat=1.15}
\usepackage{mathrsfs}
\usetikzlibrary{arrows}

\makeatletter
\def\@fpheader{\ }
\makeatother

\title{Quantum gravity of the Heisenberg algebra}
\author{Ahmed Almheiri${}^{[a,}$,\quad Akash Goel${}^{a^\dagger]}$, \quad Xu-Yao Hu${}^{=1}$  }
\affiliation{${}^{[a,}$New York University Abu Dhabi, P.O. Box 129188, Abu Dhabi, United Arab Emirates}
\affiliation{${}^{[a,a^\dagger]=1}$Center for Cosmology and Particle Physics, New York University, New York, NY 10003, USA}
\affiliation{${}^{=1}$
Institute for Advanced Study, Tsinghua University, Beijing 100084, China}

\abstract{We consider a simplified model of double scaled SYK (DSSYK) in which the Hamiltonian is the position operator of the Harmonic oscillator. This model captures the high temperature limit of DSSYK but could also be defined as a quantum theory in its own right. We study properties of the emergent geometry including its dynamics in response to inserting matter particles. In particular, we find that the model displays de Sitter-like properties such as that infalling matter reduces the rate of growth of geodesic slices between the two boundaries. The simplicity of the model allows us to compute the full generating functional for correlation functions of the length mode or any number of matter operators. We provide evidence that the effective action of the geodesic length between boundary points is non-local. Furthermore, we use the on-shell solution for the geodesic lengths between any two boundary points to reconstruct an effective bulk metric and reverse engineer the dilaton gravity theory that generates this metric as a solution.}
\date{}

\begin{document}

\maketitle

\section{Introduction}

Any complete theory of quantum spacetime ought to provide a way for smooth classical spacetime to emerge from some primitive structure. The prime example of this is gauge/gravity duality in which the bulk spacetime is encoded in strongly-coupled dynamics of the boundary field theory. 

A simple case is the emergence of AdS$_2$ physics from (a subsector of) the SYK model \cite{Sachdev:1992fk,kitaevTalks}, a quantum mechanical model of $N$ Majorana fermions $\psi_i$ with dynamics governed by
\begin{align}
    H = i^{p/2} \sum_{1\leq i_1<\cdots<i_p\leq N} 
    J_{i_1...i_p} \psi_{i_1} ... \psi_{i_p}, \ \lan J_{i_1...i_p}^2  \ran = {{\cal J}^2 \over \lambda \binom{N}{p}}, \  \lambda \equiv {2 p^2 \over N}. 
\end{align}
This emergence can be studied by analyzing the thermal partition function at different temperatures $T \equiv 1/\beta$. In the low temperature limit and $\lambda \ll 1$, the dual geometry of the SYK model is simply the hyperbolic disk \cite{Maldacena:2016hyu,kitaevTalks}. The other extreme of infinite temperature or $\beta = 0$ (and any $\lambda$), the thermal circle vanishes and the Euclidean theory is trivial without any bulk Euclidean geometry.\footnote{This doesn't address the nature of the Lorentzian geometry.} It is natural to expect that intermediate values of $\beta$  probe the gradual emergence of the bulk. 

The origin of this (would be) gradual emergence can be related to a feature of the thermal partition function under a Trotter decomposition, given by
\begin{align}
    Z_\beta = \tr \left[ e^{-\beta H} \right]  = \lim_{k \rightarrow \infty}  \tr \left[ \left( 1 - {\beta \over k} H \right)^{k} \right].
\end{align}
The partition function is expressed as a sum over moments of the Hamiltonian inserted at different locations on the thermal circle. Tuning the value of $\beta$ controls the dominant moment in the expansion of the thermal partition function; larger moments at larger values of $\beta$ and smaller moments for smaller values of $\beta$, suggesting that laying down Hamiltonians gradually builds up the spacetime.\footnote{The exponential $e^x$ is dominated by the term $x^n \over n!$ where $n \sim x$.}

In fact, this expectation is fulfilled in the so-called double scaling limit of the SYK model (DSSYK) where $p,N \rightarrow \infty$ with $\lambda$ fixed \cite{Cotler:2016fpe,Berkooz:2018jqr, Berkooz:2018qkz}. Other works on DSSYK and aspects of the emergent bulk have been studied in \cite{Blommaert:2023opb, Blommaert:2023wad,Okuyama:2022szh,Okuyama:2023iwu, Okuyama:2023byh,Okuyama:2023kdo,Okuyama:2023aup, Okuyama:2023yat, Okuyama:2024yya,Boruch:2023bte,Berkooz:2022mfk,Berkooz:2020xne, Xu:2024hoc, Goel:2023svz}. Upon disorder averaging, each moment of the Hamiltonian is computed by a sum over chord diagrams where each chord represents  Wick contractions between $J_{i_1 ... i_p}$'s in the Hamiltonian insertions, including a penalty factor of $q\equiv e^{-\lambda}$ for every chord intersection. Since the number of chords scales with the number of Hamiltonians, it proliferates in the low temperature limit of DSSYK. The upshot is these chords are the sought-out primitive version of spacetime out of which smooth spacetime emerges; the hyperbolic disk emerges by taking the ``triple scaling" limit $\beta {\cal J} \sim \lambda^{-1}$ as $\lambda \rightarrow 0$ \cite{Berkooz:2018jqr, Berkooz:2018qkz}.\footnote{$\lambda$ behaves as the ratio of the Planck length to the AdS length.}

By slicing the path integral between two boundary points, it can be expressed as a sum of overlaps of states with a definite number of chords. These can be used to define a bulk Hilbert space spanned by orthogonal states $|n\ran$, where $n$ is the number of chords which represents the length of the wormhole between two copies of DSSYK.\footnote{The random couplings of the two copies are drawn from the same Gaussian distribution.} In this Hilbert space interpretation, the boundary partition function can be expressed as the transition amplitude of starting and ending with the zero length wormhole with some amount of Euclidean bulk evolution in between \cite{Berkooz:2018jqr,Berkooz:2018qkz}:
\begin{align}
   \lan  \tr \left[ e^{-\beta H} \right] \ran = \lan 0| e^{-\beta T} | 0 \ran.
\end{align}
where $\lan \cdot \ran$ refers to the disorder averaging. The evolution operator in the bulk is the transfer matrix of DSSYK, given by
\begin{align}
    T = \frac{1}{\sqrt{\lambda}} \left(\alpha^{\dagger} + \alpha  \frac{1-q^{\hat n}}{1-q}\right),
\end{align}
where $\hat{n}$ is the number operator and  $\alpha, \alpha^\dagger$ are the shift operators mapping $|n\ran$ to $|n-1\ran,|n+1\ran$ respectively, without any additional factors. The overall normalization was chosen following \cite{Lin_2022}. The transfer matrix can be mapped to a standard Hermitian bulk Hamiltonian by working in an orthonormal basis of $|n\ran$, resulting in
\begin{align}
   H_B = - \frac{1}{\sqrt{\lambda}} \left( \alpha \sqrt{\frac{1-q^{\hat n}}{1-q}} + \sqrt{\frac{1-q^{\hat n}}{1-q}} \alpha^{\dagger} \right).
\end{align}
This Hamiltonian takes the form of the position operator of the $q$-deformed harmonic oscillator,
\begin{align}
    X_q = a_q + a_q^\dagger, \ [a_q, a_q^\dagger]_q = a_q a_q^\dagger - q a_q^\dagger a_q = 1.
\end{align}
Note that it's not the commutator of the $q$-deformed raising and lowering that's equal to one but their \emph{q-commutator}.

This Hamiltonian is simple enough to diagonalize. It has a continuous spectrum and can be parameterized as
\begin{align}
    \varepsilon(\theta) = - \frac{2 \cos \theta}{\sqrt{\lambda (1-q)}} 
\end{align}
where $\theta\in \[0, \pi\]$. Its eigenfunctions in the number basis $\psi_n(\theta) =  \lan \theta | n \ran$ satisfy the recursion relation
\be 
\sqrt{1 - q^{{n+1}}\over 1 - q}  \psi_{n+1}(\theta) + \sqrt{1 - q^{{n}}\over 1 - q} \, \psi_{n-1}(\theta) = \varepsilon(\theta)\, \psi_{n}(\theta)
\ee 
which is solved using $q-$Hermite polynomials:
\begin{align}
    \psi_n(\theta)  = \frac{H_n(\cos\theta| q)} {\sqrt{(q;q)_{n}}}
\end{align}
Further details on this can be found in \cite{Berkooz:2018jqr, Berkooz:2018qkz, Lin_2022}.

\subsection{High temperature approximation and the Heisenberg algebra}

The goal of this paper is to analyze a toy model of the high temperature limit of DSSYK. See also \cite{Okuyama:2023iwu} where the high temperature limit of DSSYK is discussed. The regime in the high temperature limit that we are interested in is where we take $\beta = \tilde{\beta} \sqrt{\lambda}$ and take $\lambda$ to zero while keeping $\tilde{\beta}$ finite.

This limit favors the center of the spectrum where the density of states becomes Gaussian, $\rho(\theta) \sim e^{-\lambda \epsilon^2(\theta)/2}$; see \cite{Berkooz:2018jqr,Berkooz:2018qkz} for more details. Physics of this regime is well approximated at small $\lambda$ by restricting to $\theta = {\pi \over 2} + \sqrt{\lambda} {\cal E}$ where $\epsilon(\theta) = -{2 \over \sqrt{\lambda}} {\cal E}$. In this regime, states with short wormholes dominate. To see this, consider the average length of the wormhole in the large $p$ theory given by
\begin{align}
    n \lambda = -2 \ln \cos {\pi v \over 2}.
\end{align}
The parameter $v$ is related to the inverse temperature through $\beta \cos (\pi v/2) = \pi v$. Using $\cos{\pi v \over 2} \approx  1 - {\pi^2 v^2 \over 8} \approx 1 - {\lambda \tilde{\beta}^2\over 8}$ at small $\beta$, we get that $n \lambda  \sim \lambda \tilde{\beta}^2$. 

The DSSYK system drastically simplifies in the small $n \lambda$ regime. In particular, by working to leading order in $n \lambda$, up to an overall sign, the Hamiltonian simplifies to
\begin{align}
    -H_B\sqrt{\lambda} &\approx \alpha \sqrt{\hat{n}} + \sqrt{\hat{n}}\alpha^{\dagger} = a + a^{\dagger} ~,
\end{align}
where $[a,a^{\dagger}] = 1$ is the usual Heisenberg algebra. 

This paper will focus on analyzing this model. We will think of it as a model of quantum gravity independent of its DSSYK origin. In section 2, we will analyze the physics of the length mode, the only dynamical mode in this model. Section 3 will be about the properties of matter coupled to the Heisenberg model. We will compute arbitrary matter correlation functions and study the effect of matter perturbations on the dynamics of the length mode. In section 4, we will find the effective bulk metric that gives rise to the classical value of the length. We will then find the dilaton gravity model with the right potential such that this effective metric is a solution.

{\flushleft \bf Note:} As we were completing this work, the papers \cite{berkooz:2024ofm, Berkooz:2024evs} appeared on the arXiv with minor overlap with this one, although their analysis of the path integral of the length mode and its non-local properties is more developed than ours.

\section{Analysis of the Heisenberg model} \label{hmodel}

We consider a chord theory where the dynamics are captured by the simple Hamiltonian
\begin{align}
    \sqrt{2}H = a + a^\dagger, \label{ham}
\end{align}
namely the position operator of the usual Harmonic oscillator. \footnote{This model has a $H \rightarrow -H$ symmetry and we pick the positive sign from here on out.} We can choose to normalize this Hamiltonian with an overall factor of $J$ with units of energy, but for now we imagine setting $J=1$. The spectrum of this Hamiltonian is unbounded both from above and below. It is Hermitian and with eigenvalues $E \in \mathbb{R}$. In keeping with chord picture of DSSYK, this Hamiltonian should be thought of as coming from the transfer matrix $T \sim \alpha \hat{n} + \alpha^\dagger$ after normalizing by the inner product matrix \cite{Berkooz:2018jqr,Berkooz:2018qkz,Lin_2022}.  

The usual Fock space ${\cal H} = \{ |n \ran|\ n = {0,1,2,...} \}$ corresponds to wormholes of different discrete length. It will be convenient to work with unnormalized states satisfying $\lan m| n \ran = n! \delta_{mn}$, where $n!$ is the combinatoric factor counting the different ways of contracting the chords in the bra and ket. The only difference between these chord states and those in \cite{Berkooz:2018jqr,Berkooz:2018qkz,Lin_2022} is that there is no penalty from chord intersections.\footnote{This means that all the terms in the Hamiltonian commute with one another in the double scaling limit. We thank Ohad Mamroud for the discussion on this point. See the recent papers \cite{berkooz:2024ofm, Berkooz:2024evs}. An SYK model with all commuting terms in the Hamiltonian has been studied recently in \cite{Gao:2023gta}, and the Heisenberg model can be understood as the double scaling limit of this.}

Working with these unnormalized states, delta function normalizable energy eigenstates can be expressed in the number basis as
\begin{align}
    | E \ran = \sum_n |n \ran {\psi_n(E) \over n!},
    \label{eq:E-state}
\end{align}
where the wavefunctions satisfy the recursion relation
\begin{align}
    \sqrt{2 } E \psi_n(E) = n \psi_{n-1}(E) +  \psi_{n+1}(E)
\end{align}
which is solved using Hermite polynomials
\begin{align}
    \psi_n(E) = \left(2^n  \sqrt{\pi}\right)^{-1/2} e^{-{E^2 \over 2}} H_n(E).
    \label{eq:psi_n-E}
\end{align}
Note that the number eigenstates in the energy basis are written without the $n!$ normalization factor, 
\begin{align}
    |n \ran = \int dE \, \psi^*_n(E) | E \ran.
\end{align}
Note also that the wavefunctions are real, and so $\psi^* = \psi$.

\subsection{Thermodynamics} \label{thermo}
We use the Hamiltonian \eqref{ham} to compute the partition function of the putative boundary dual of this model. The boundary partition function is given by bulk vacuum expectation value
\begin{align}
    \lan 0 | e^{-\beta H} |0 \ran = {1 \over \sqrt{\pi} }\int dE e^{-\beta E - {E^2 }} =  e^{\beta^2 \over 4}.
\end{align}
The scaling of the free energy as $\beta^2$ suggests that the Planck scale is set by $1/\beta$; one factor of $\beta$ comes from $U(1)$ symmetry while the other is the coupling, namely the Planck length. We will see further evidence in the next section.

We extract the thermal energy in the usual way using $\lan H \ran = -\partial_\beta \ln Z = -\beta/2$. The variance of the energy is $\lan H^2 \ran - \lan H \ran^2 = 1/2$, and hence the fluctuations in the energy are suppressed at large $\beta$. 

The thermal entropy is another interesting quantity to compute. Applying $(1-\beta \partial_\beta)\ln Z$ we get $S = -\beta^2/4$. While unusual, this negative entropy should be understood as the deviation from the maximally mixed state of infinite entropy. Large $\beta$ truncates a large part of the spectrum at higher energies, and hence it makes sense for the entropy to decrease.

\subsection{Wormhole length}\label{wormholelength}

The states $|n \ran$ label wormholes of length $n/J$, where $J$ is the normalization of the Hamiltonian. We will continue measuring everything in terms of this scale and hence set  $J = 1$. It is interesting to study the dynamics of this length by computing its moments in the thermal ensemble. This can be done by considering the generating function for the length given by
\begin{align}
     \lan 0 |e^{-\tau_2 H }r^{\hat{n}} e^{-\tau_1 H } | 0\ran 
\end{align}
and then differentiating its logarithm with respect to $r$ and then taking $r \rightarrow 1$ and setting $\tau_1 + \tau_2 = \beta$. It's important to note our convention of defining $\tau_1, \tau_2$ as going in the same direction (say counterclockwise) around the boundary thermal circle. This generating function is simple to compute by inserting a complete set of states in the energy and number basis to get
\begin{align}
    \lan 0 |e^{-\tau_2 H }r^{\hat{n}} e^{-\tau_1 H } | 0\ran  &= \sum_{n} {1 \over \sqrt{\pi}}\int dE_1 dE_2\, e^{-\sum_i \left(  \tau_i E_i + \frac{E_i^2 }{2} \right)} \, \psi_n(E_1) \psi^*_n(E_2) r^{n} {\lan n| n \ran \over (n!)^2}, \\
    &=\sum_{n} {1 \over \sqrt{\pi}}\int dE_1 dE_2\, e^{-\sum_i \left(  \tau_i E_i + \frac{E_i^2}{2} \right)}  \, \psi_n(E_1) \psi^*_n(E_2) {r^n \over n!}~,
\end{align}
where we used the normalization $\lan n| n\ran = n!$ and the expression for the ground state wavefunction $\psi_0(E) = \pi^{-1/4} e^{-E^2/2}$. Prior to doing the sum, it is simpler to first evaluate the energy integral using the Hermite polynomial formula
\begin{align}
    \int dE e^{{-\tau E } - E^2}H_n(E) = \left({ - \tau} \right)^n e^{\tau^2 \over 4} \sqrt{\pi}
\end{align}
to get
\begin{align}
    \lan 0 |e^{-\tau_2 H }r^{\hat{n}} e^{-\tau_1 H } | 0\ran  &=  e^{\sum_i {\tau_i^2 \over 4} }\sum_{n} {1 \over n!}{\left(  \tau_1 \tau_2 r \over 2 \right)^n}, \\
    &=   {\mathrm{Exp}} \left[ {\sum_i {\tau_i^2 \over 4} } +  { \tau_1 \tau_2 r \over 2 } \right],
\end{align}
which reproduces the partition function $r \rightarrow 1$ as required.

Now we can use this to compute the moments of the length by taking derivatives and then setting $r \rightarrow 1$. Since $r$ appears linearly in the exponent, all the  connected moments (cumulants) are equal
\begin{align}
    Z^{-1}{\lan \tau_2| \hat{n}^k | \tau_1 \ran_c } &= \lim_{r\rightarrow 1} \, (r \partial_r)^k \ln  \lan 0 |e^{-\tau_2 H }r^{\hat{n}} e^{-\tau_1 H } | 0\ran  = {\tau_1 \tau_2 \over 2}.\label{lengthmoments}
\end{align}
This implies that the length of the wormhole becomes a classical variable at large $\beta$ since the fluctuations are suppressed compared to the mean (for instance, the ratio between the variance and the mean $\sim \beta^{-1}$). Hence, we can think of $\beta (\times J)$ as controlling the ratio of the bulk size or IR length scale to the Planck or discreteness scale.

The length of the wormhole in Euclidean and Lorentzian space using \eqref{lengthmoments}. In Euclidean, we set $\tau_1 = \beta - \tau_2 = \tau$ to get
\begin{align}
     \lan n(\tau) \ran = {\tau (\beta - \tau) \over 2} = \includegraphics[width=3cm, valign=c]{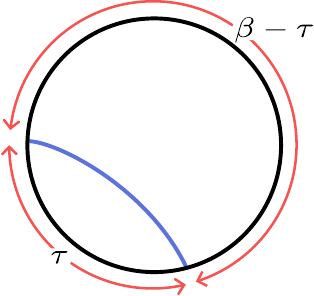},
     \label{eq:<n>_wormhole}
\end{align}
where $0\le \tau \le \beta$. For Lorentzian time, we take the times to be $\tau_1 = \beta/2 + i t$ and $\tau_2 = \beta/2 - i t$ and we find
\begin{align}
    \lan n(t) \ran = {{\beta^2 \over 8 }+ {t^2\over 2}} = \includegraphics[width=3cm, valign=c]{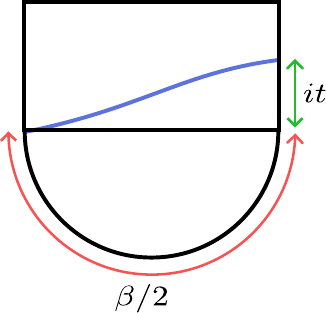}.\label{lorn}
\end{align}
The quadratic growth agrees with the growth of AdS wormholes for short times around $t = 0$.

\subsection{Generating functional for lengths} \label{secgenl}

The simplicity of the model allows us to compute arbitrary correlation functions of the length operator at all times. We quote the result here and leave its derivation to appendix \ref{alllengthgen}. The arbitrary generating functional we find is
\begin{align}
        \lan r_1^{\hat{n}(\tau_1)} ...  \, r_a^{\hat{n}(\tau_a)} \ran_\beta &= e^{- {\beta^2 \over 4}}\mathrm{Exp}\left[ {\sum_{i=1}^{a+1}{(\tau_i - \tau_{i-1})^2 \over 4} + {1\over 2}\sum_{i<j = 1}^{a+1}(\tau_i - \tau_{i-1})(\tau_j - \tau_{j-1}) \prod_{k = i}^{j-1}r_k} \right] \\
        &= \mathrm{Exp}\left[  {1\over 2}\sum_{i<j = 1}^{a+1}(\tau_i - \tau_{i-1})(\tau_j - \tau_{j-1}) \left( \prod_{k = i}^{j-1}r_k - 1 \right) \right]. \label{generatingfunctional}
\end{align}
In the continuum limit, this generating functional can be written as
\begin{align}
    Z[J] \equiv \mathrm{Exp}\left[ {1\over 2} \int_0^\beta \! \! \! d\tau_1 \int_{\tau_1}^\beta \!\!\! d\tau_2  \left( \mathrm{Exp}\left[{\int_{\tau_1}^{\tau_2} \!\!\! d\tau_3 \, J(\tau_3)}\right] - 1\right)\right] \label{contgen}
\end{align}
where $r_k = \mathrm{Exp}[J(\tau_k)]$.

It is interesting to consider the generating functional for perturbations about the classical value of the length. If we write $n(\tau) = {1\over 2} \tau (\beta - \tau) + \eta (\tau)$, the generating functional becomes
\begin{align}
        Z[J] &=  \mathrm{Exp} \left[ {1\over 2} \int_0^\beta \! \! \! d\tau_1 \int_{\tau_1}^\beta \!\!\! d\tau_2  \left( \mathrm{Exp}\left[{\int_{\tau_1}^{\tau_2} \!\!\! d\tau_3 \, J(\tau_3)}\right] - \left( 1 + \int_{\tau_1}^{\tau_2} \! \! \! d\tau \,J(\tau) \right) \right) \right], \\
        &= \mathrm{Exp} \left[ \sum_{k=2}^\infty {1 \over k!}\int_0^\beta \! \!  \prod_{m=1}^k d\tau_m \, G_k(\{ \tau_m \}) \, J(\tau_1)...J(\tau_k)\right],
\end{align}
where 
\begin{align}
    G_k(\{\tau_m \}) \equiv \sum_{m_1 \neq ... \neq m_k = 1}^k{ (\beta - \tau_{m_k})\tau_{m_1} \over 2} \Theta(\tau_{m_k},...,\tau_{m_1})
\end{align}
and $\Theta(\tau_{m_k},...,\tau_{m_1})=1$ enforces $\tau_{m_k}>...>\tau_{m_1}$. The function $G_k$ is the exact connected correlation function of $\eta$.

Recall from the previous section that fluctuations are suppressed for large $\beta$. This can be made manifest by rescaling the fields by a factor of $\beta$ taking $\eta \rightarrow \eta \beta/2 \pi$ (we rescale to an angular coordinate). The scaling of $J$ is fixed by requiring the source term $\int \! d\tau J \eta$ to be $\beta$ independent, and hence we need to take $J \rightarrow  (2 \pi)^2 J/\beta^2$; we require this so that differentiating with respect to $J$ generates the correlators at arbitrary $\beta$. The path integral then becomes
\begin{align}
    Z[J] &= \mathrm{Exp} \left[ \sum_{k=2}^\infty {(\beta/2 \pi)^{2-k} \over k!}\int_0^{2 \pi} \! \!  \prod_{m=1}^k d\theta_m \, \tilde{G}_k(\{ \theta_m \}) \, J(\theta_1)...J(\theta_k)\right], \label{etagen}
\end{align}
where now 
\begin{align}
    \tilde{G}_k(\{\theta_m \}) \equiv \sum_{m_1 \neq ... \neq m_k = 1}^k\!\!\!\! { (2 \pi - \theta_{m_k})\theta_{m_1} \over 2} \Theta(\theta_{m_k},...,\theta_{m_1})
\end{align}
This makes manifest that higher point vertices are suppressed by inverse powers of $\beta$. Note that the quadratic term is $\beta$ independent, implying that the kinetic term of $\eta$ will also be $\beta$ independent, as required.

We can obtain the exact two point function of the fluctuation $\eta$. By differentiating twice we get
\begin{align}
    \lan \eta(\tau_1) \eta (\tau_2) \ran = {(\beta - \tau_1) \tau_2 \over 2} \Theta(\tau_1 - \tau_2) + \tau_1 \leftrightarrow \tau_2, \label{etaprop}
\end{align}
where we've rescaled our field back to the original units. As required, this reproduces the connected second moment of the length in \eqref{lengthmoments}.  The Lorentzian time correlation function grows with time. Setting $\tau_1 = (\beta/2)_+ + i t_1$ and $\tau_2 = (\beta/2)_- + i t_2$ where $\pm$ indicate the order of the insertions, we get
\begin{align}
    \lan \eta(\tau_1) \eta (\tau_2) \ran = {1\over 2} \left({\beta \over 2} - i t_1 \right)\left({\beta \over 2} + i t_2 \right).
\end{align}

\subsection{Path integral}

Here we rederive some results of the previous section from the perspective of the path integral of the length mode $n(\tau)$. First, recall that the length basis states are normalized as $\lan n| n \ran = n!$. Then, we can use the Hamiltonian \eqref{ham} to find the infinitesimal propagator in the length basis
\begin{align}
    \frac{\lan n_2| e^{- \delta \tau H} | n_1 \ran}{\sqrt{\lan n_2 | n_2 \ran} \sqrt{\lan n_1 | n_1 \ran} } 
    &= \frac{\lan n_2 | e^{-\gamma (a+a^{\dagger})}| n_1 \ran}{\sqrt{n_2!} \sqrt{n_1 !}}  \\
    &= \int_0^{2\pi} \frac{dk}{2\pi} e^{i k (n_2 -n_1) } \left(1 - \gamma \sqrt{n_1} e^{i k} - \gamma \sqrt{n_2} e^{-i k} \right) \\
    &\approx \int \frac{dk}{2\pi} e^{\delta \tau \left(i k \dot n_1 - \sqrt{2n_1} \cos k \right)}
\end{align}
with $\gamma \equiv \frac{\delta \tau}{\sqrt{2}}$ and $\dot n_1 \delta \tau \equiv n_2 - n_1$. Taking the small $\delta \tau$ limit and multiplying the infinitesimal propagators in the usual way, we get the following path integral
\begin{align}
    \lan 0 | e^{-\beta H} | 0 \ran &= \int \left(\prod_{i=1}^{M} dn_i \frac{dk_i}{2\pi}\right) e^{\delta \tau \sum_{j = 1}^M\left( i k_j\, \dot{n}_j 
     - \sqrt{2 n_j} \cos k_j \right)}\delta_{n_M,0} \delta_{n_1,0}  \\
     &= \int \! \! Dn(\tau) Dk(\tau) e^{ \int_0^\beta d\tau \left( i k(\tau)\, \dot{n}(\tau) - \sqrt{2 n(\tau)} \cos k(\tau) \right)} \\
     &\equiv \int \! \! Dn(\tau) Dk(\tau)\ e^{-S[n(\tau), k(\tau)]}
\end{align}
with boundary conditions $n(0) = n(\beta) = 0$.

Next, we compute the on-shell solution and compare to the exact answer \eqref{eq:<n>_wormhole} computed in the previous section. The Euler-Lagrange equations are
\begin{align}
    \delta_{n(\tau)} S= 0 &\implies
    i \dot k(\tau) + \frac{\cos[k(\tau)]}{\sqrt{2n(\tau)}} = 0 ~, \\
    \delta_{k(\tau)} S= 0 &\implies
    i \dot n(\tau) + \sqrt{2n(\tau)} \sin[k(\tau)] = 0~.
\end{align}
The solution satisfying the boundary conditions $n(0) = n(\beta) = 0$ is
\begin{align}
    n(\tau) = \frac{\tau (\beta - \tau)}{2}~, \quad 
    k(\tau) = -i \ \text{arctanh}\left(\frac{\beta - 2\tau}{|\beta|}\right)
\end{align}
exactly matching (\ref{eq:<n>_wormhole}).

\subsection{Lagrangian at Large $\beta$} \label{lag}

A direct way of finding the Lagrangian is to integrate out the conjugate momentum $k$ in the previous section. We don't see a simple way of doing and so we leave it for future work. Instead, we present another method using the generating functional \eqref{etagen} to find the Lagrangian for $\eta$. See \cite{berkooz:2024ofm} for a more complete analysis of this.

We work perturbatively in  $1/\beta$. The propagator was found in the previous section in \eqref{etaprop}. It solves the wave equation $\partial_{\theta_1}^2 G(\theta_1, \theta_2) = -\pi \delta(\theta_1 - \theta_2)$  with boundary conditions $G = 0$ when either $\theta_{1,2} = 0, 2\pi$. Hence, the leading term in the Lagrangian is
\begin{align}
    L_0(\eta) = -{1 \over 4} \eta(\theta) \ddot{\eta}(\theta). \label{freelagrangian}
\end{align}

The first subleading correction comes from the three point vertex. We obtain the form of this vertex as follows. We first assume that the interaction term in the action  has the form
\begin{align}
    S_I^{(3)}(\eta) = \beta^{-1} \int_0^{2\pi} d\theta_1 d\theta_2 d\theta_3 F(\theta_1, \theta_2, \theta_3) \eta(\theta_1) \eta (\theta_2) \eta(\theta_3), \label{threepntaction}
\end{align}
for some vertex function $F(\theta_1, \theta_2, \theta_3)$ to be determined. We do so by computing the three point function in two ways, one from the generating functional and another directly from the path integral assuming \eqref{freelagrangian}, \eqref{threepntaction}. This gives  the equation
\begin{align}
    \tilde{G}_3(\theta_1, \theta_2, \theta_3) = 3!\int_0^{2\pi} \! \! \! d\theta_1'd\theta_2'd\theta_3' \, \tilde{G}_{2}(\theta_1,\theta_1')\tilde{G}_{2}(\theta_2,\theta_2')\tilde{G}_{2}(\theta_3,\theta_3') \, F_3(\theta_1', \theta_2', \theta_3').
\end{align}
This can be simplified using the wave equation and differentiating twice with respect to $\theta_{1,2,3}$ to get
\begin{align}
    \partial_1^2\partial_2^2\partial_3^2\tilde{G}_3(\theta_1, \theta_2, \theta_3) = -3!8\,   F_3(\theta_1, \theta_2, \theta_3).
\end{align}
The vertex function $F_3$ comes out to be a linear combination of delta functions and their derivatives. The resulting action is
\begin{align}
    S_I^{(3)} \sim \beta^{-1} \int \!\!\! d \theta \left( (2 \pi -  \theta)\eta \eta' \eta'' + {1 \over 2} \eta^2 \eta'' + {(2 \pi - \theta)\theta\over 2}{\eta'}^2 \eta'' \right).
\end{align}

Next, we consider the next to subleading order that goes as $\beta^{-2}$  which corresponds to the four point vertex. Following the same steps as above, the connected component of the four-point function satisfies
\begin{align}
        \tilde{G}_4(\theta_1, \theta_2, \theta_3,\theta_4) &= \binom{6}{4} {1\over 2}\int_0^{2\pi} \! \! \! \prod_{i = 1}^3d\theta_i'd\theta_i'' \, \prod_{j=1}^2 \tilde{G}_{2}(\theta_j,\theta_j')\tilde{G}_{2}(\theta_{j+2},\theta_j'') \, F_3(\theta_1', \theta_2', \theta_3')F_3(\theta_1'', \theta_2'', \theta_3'') \tilde{G}_{2}(\theta_3',\theta_3'')  \nonumber \\
        &\ \ \ \ \ \ + 4! \int_0^{2\pi} \! \! \! \prod_{i=1}^4 d\theta_i' \tilde{G}_{2}(\theta_i,\theta_i') F_4(\theta_1',\theta_2',\theta_3',\theta_4').
\end{align}
We already have $F_3$ and $\tilde{G}_4$ can be read off from the generating functional, which means we can find $F_4$. As before, we differentiate both sides with respect to all angles twice, and  get the equation
\begin{align}
    \partial_1\partial_2\partial_3\partial_4 \tilde{G}_4 +  \binom{6}{4} {8 }\int_{0}^{2\pi} d\theta_3' d\theta_3''  F_3(\theta_1, \theta_2, \theta_3')F_3(\theta_3, \theta_4, \theta_3'') \tilde{G}_{2}(\theta_3',\theta_3'')  = - 4!16\, F_4
\end{align}
In addition to local terms, the solution for the vertex function $F_4$ contains a non-local part. We focus on this contribution since the full expression is somewhat involved. The non-local piece contributes to the as
\begin{align}    
S_{\mathrm{NL}}^{(4)} &\sim {\beta^{-2}}\int_0^{2 \pi} \! \! \! d\theta_1 d\theta_2 \, \Theta(\theta_2- \theta_1) (\eta(\theta_1)')^2 (\eta(\theta_2)')^2 {(2 \pi - \theta_2)\theta_1 }, \\
&\sim {\beta^{-2}}\int_0^{2 \pi} \! \! \! d\theta_1 d\theta_2 \,  (\eta'(\theta_1))^2 \tilde{G}_2(\theta_1, \theta_2)(\eta'(\theta_2))^2.
\end{align}
So we have two independent integrals over the circle of ${\eta'}^2$ connected by the $\eta$ propagator $\tilde{G}_2$. This form is suggestive of having integrated out a field that couples linearly to ${\eta'}^2$ with propagator $\tilde{G}_2$.

Finally, we note that there's an additional reason to suspect the theory is non-local independent of the explicit term found in the four point vertex. This is the observation that all interaction terms involve derivatives of the order of the degree of the vertices, and hence the Lagrangian has an infinite number of derivatives. This is often a feature of non-local theories. We emphasize that despite this non-locality, the theory makes sense since it is defined through a Hamiltonian formulation.

\section{The Heisenberg model with particles}

Next we consider including particles to the Heisenberg model. This means we include matter cords in addition to Hamiltonian chords that build up the spacetime.\footnote{We use L. Susskind's convention of using c{\bf h}ords exclusively for {\bf H}amiltonian insertions.} Matter cords connect boundary insertions of boundary matter operators, which in DSSYK correspond to (random) products of fermions. The only rule we follow is to assign a penalty factor $e^{-\Delta}$ for every cord-chord intersection, and $e^{-\Delta_i \Delta_j}$ for cord-cord intersection. The $\Delta$'s are the dimensions of the matter particles. Note that from the DSSYK perspective, these matter fields have dimensions that scale as $\lambda^{-1}$; this is needed to have a non-trivial penalty factor in the $\lambda \rightarrow 0$ limit since the factor goes as $e^{-\lambda \Delta_{\mathrm{DSSYK}}}$.

\subsection{Correlation functions}\label{correlationfunctions}

Here we compute correlation functions of matter insertions using the rules outlined above. This model is so simple that we can write down the general $n$-point correlation function with any ordering. Suppose we have $n/2$ pairs of insertions of different operators. In the case of duplicate pairs, one considers all wick contractions. We denote $\tau_1,...,\tau_n$ as the time separations between the operators as we go around the circle, and hence they sum up to $\beta$. The general (normalized) correlation function is
\begin{align}
    \mathrm{Exp}\left[ -{\beta^2 \over 4} + \sum_{i=1}^{4} {\tau^2_i \over 4} + \sum_{i<j=1}^n {e^{-\sum_k \Delta_k} \over 2}\tau_i \tau_j   \right]e^{\sum \prod_i \Delta_i}, \label{generaln}
\end{align}
where $\sum_k \Delta_k$ runs over all matter cords encountered by a Hamiltonian chord connecting the boundary regions $\tau_i$ and $\tau_j$. The factor on the far right accounts for matter cord intersections.

Let's look at a few examples. The simplest case is the two point function. There are two time intervals $\tau_1, \tau_2$ and any chord between them encounters the single matter line. We get (see Appendix \ref{appendix:2-pt} for details)
\begin{align}
    {1 \over Z}\tr[e^{-\tau_2 H} V e^{-\tau_1 H}V] &=  \frac{\lan 0 |e^{-\tau_2 H} e^{-\Delta_V \hat n} e^{-\tau_1 H} | 0 \ran }{\lan 0 | e^{-\beta H} | 0 \ran}
    \label{eq:bdy-bulk-2pt} \\
    &= 
    \mathrm{Exp} \left[ -{\beta^2 \over 4} + \sum_{i=1}^2 {\tau_i^2 \over 4}  + {e^{-\Delta_V} \over 2} \tau_1 \tau_2 \right],
    \label{eq:2-pt}
\end{align}
where $\tau_1 + \tau_2 = \beta$. In this expression and all that follows, the left-hand side should be understood from the boundary perspective, where the trace is over the boundary Hilbert space and the $H$ is the putative boundary microscopic Hamiltonian. Furthermore, these are disorder averaged boundary quantities, since only then can they be computed via the chord techniques in the bulk. The Lorentzian continuation of this is the 2-sided correlation function obtained by setting $\tau_1 = {\beta \over 2} -  i t, \tau_2 = {\beta \over 2} +  i t$. This gives
\begin{align}
    \frac{1}{Z}\tr[e^{-\tau_2 H} V e^{-\tau_1 H}V] = \mathrm{Exp}\left[-{1 - e^{-\Delta_V} \over 8}\left( \beta^2 + 4 t^2\right) \right].
\end{align}
This is the correlation function across the wormhole whose length appears in the exponent. It agrees with what we found earlier in \eqref{lorn}. This expression also agrees with that found in \cite{Gao:2023gta} after taking the dimension of our operator as $\sim 1/q$ and setting our $J^2/q$ to ${\cal J}^2$ of that paper and taking large $q$.

Next up is the four point function. We consider pairs of different operators to avoid summing over Wick contractions. The time-ordered correlator (TOC) is (see Appendix \ref{appendix:TOC} for details)
\begin{align}
&{1 \over Z}\tr[e^{-\tau_4 H}W e^{-\tau_3 H} W e^{-\tau_2 H} V e^{-\tau_1 H}V] = 
\includegraphics[width=3.5cm, valign=c]{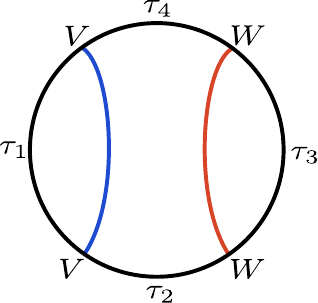} \nonumber \\
&= \mathrm{Exp}\left[  -{\beta^2 \over 4} + \sum_{i=1}^{4} {\tau^2_i \over 4} + {e^{-\Delta_V} \over 2} (\tau_1 \tau_2 + \tau_1 \tau_4) + {e^{-\Delta_W} \over 2} (\tau_2 \tau_3 + \tau_3 \tau_4)  +  {e^{-\Delta_V-\Delta_W} \over 2} \tau_1 \tau_3 + {1 \over 2} \tau_2 \tau_4     \right],
\label{eq:TOC} 
\end{align}
while the out-of-time-order correlator (OTOC) is (see Appendix \ref{appendix:OTOC} for details)
\begin{align}
&{1 \over Z}\tr[e^{-\tau_4 H}W e^{-\tau_3 H} V e^{-\tau_2 H} W e^{-\tau_1 H}V] = 
\includegraphics[width=3.5cm, valign=c]{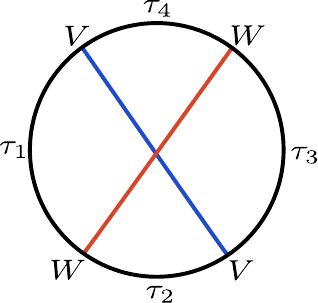} \nonumber \\
&= \mathrm{Exp}\left[  -{\beta^2 \over 4} + \sum_{i=1}^{4} {\tau^2_i \over 4} + {e^{-\Delta_V} \over 2} (\tau_1 \tau_4 + \tau_2 \tau_3) + {e^{-\Delta_W} \over 2} (\tau_1 \tau_2 + \tau_3 \tau_4)  +  {e^{-\Delta_V - \Delta_W} \over 2} (\tau_1 \tau_3 + \tau_2 \tau_4)     \right]e^{-\Delta_V \Delta_W}.
\label{eq:OTOC} 
\end{align}
Notice how in the OTOC all pairs of boundary intervals are separated by at least one matter chord. The last factor of the OTOC accounts for the intersection between matter cords; it's an additional contribution to the length measured by one operator in units of the dimension of the other. As a consistency check, one can check that the OTOC (\ref{eq:OTOC}) reduces to the two-point function (\ref{eq:2-pt}) when setting 
$\tau_2 = \tau_4 = 0$
and $\Delta_W = 0$ followed by relabeling $\tau_3 \to \tau_2$, and the TOC (\ref{eq:TOC}) reduces to the two-point function (\ref{eq:2-pt}) upon setting $\tau_3 = 0$ and relabeling $\tau_2 + \tau_4 \to \tau_2$. We note that \eqref{eq:TOC} and \eqref{eq:OTOC} look similar to those in \cite{Gao:2023gta} but we expect agreement in the double scaling limit.

The TOC and OTOC can be used to compute a variety of interesting quantities. For instance, we can compute the OTO commutator of two operators on the same side of the wormhole. We find
\begin{align}
{1 \over Z} \tr\left[e^{-\beta H} [W(0),V(t)]^2\right] &= -2  + 2  \cos \left[ {(1 - e^{-\Delta_V })(1 - e^{-\Delta_W })J^2 \beta t / 2 }\right] \,  e^{-\Delta_V \Delta_W -(1 - e^{-\Delta_V })(1 - e^{-\Delta_W }) J^2t^2}, \\
 &\approx -2  + 2  \cos \left[ {\Delta_V \Delta_WJ^2 \beta t / 2 }\right] \,  e^{-\Delta_V \Delta_W (  1 + J^2t^2) }.
\end{align}
where we restored the normalization of the Hamiltonian $J$. The second equation is the small dimension limit of both operators. In extracting the scrambling time, we ignore the oscillating factor and define it as the time when the exponential factor becomes small. This happens when
\begin{align}
    t_{\mathrm{scr}} = {1 \over J}\,  {1\over \sqrt{ (1 - e^{-\Delta_V })(1 - e^{-\Delta_W })}} \sim {1 \over J}\,  {1\over \sqrt{ \Delta_V \Delta_W }}. \label{tscr}
\end{align}
Note the dependence of the scrambling time on the dimensions. This dependence drops out in the large dimension limit, a curious result. On the far right we have the small dimension limit; the scrambling time goes off to infinity for small perturbations.

\subsection{Dynamics of the length}

We can use these correlation functions to study the dynamics of the length under perturbations. This can be read off from the OTOC where one pair of operators creates a shockwave on one side and one pair on the two boundaries to measure the length of the wormhole. The length will be extracted by differentiating with respect to the dimension of the two-sided length-measuring/probe operator $\Delta_W$ before sending it to zero, that is
\begin{align}
    n(t_L,t_R) = \lim_{\Delta_W \to 0}\left(-\frac{d}{d\Delta_W} \text{OTOC}(\tau_1,\tau_2,\tau_3,\tau_4)\right)
\end{align}
with the dimension of the perturbation operator $\Delta_V = \Delta$. We consider the following continuation to Lorentzian time
\begin{align}
    \tau_1 = \frac{\beta}{2} + i(t_L-t_S)~,
    \quad 
    \tau_2 = \frac{\beta}{2} - i(t_L-t_S) ~,\quad 
    \tau_3 = i(t_R - t_S)~, \quad 
    \tau_4 = -i (t_R - t_S)~.
\end{align}

Working in Killing time where $t_L$ goes down on the left and $t_R$ goes up on the right, the exact result for throwing a perturbation of dimension $\Delta$ on the right at time $t_S$ is
\begin{align}
    n(t_L, t_R) &= {1 \over 2} \left( t_L- t_R\right)^2 + {\beta^2 \over 8} + \left( 1 - e^{-\Delta}\right)(t_L - t_S)(t_R - t_S) + \Delta, \\
    &= \includegraphics[width=3.5cm, valign=c]{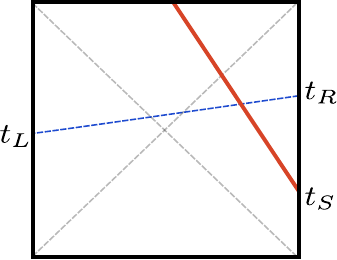} \nonumber 
\end{align}
The last two terms are introduced by the perturbation. The last term $\Delta$ is the contribution to the length from the ``spatial size'' of the operator; recall from \eqref{generaln} that matter cords contribute a length equal to their dimension. We study this formula in various regimes. Going forward, we will use the boost symmetry to set, $t_S = 0$.

In the first regime we consider, we anchor one endpoint of the geodesic at the origin. Surprisingly, the result doesn't depend on which of the two endpoints is fixed. The length is
\begin{align}
   n(t) = {t^2 \over 2} + {\beta^2 \over 8} + \Delta \Theta(t)= \includegraphics[width=3.5cm, valign=c]{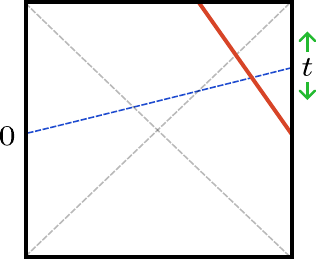}. 
\end{align}
Here (and only here) we are considering the time-dependent problem of inserting the shockwave at a particular time and continuing to evolve to the future. The presence of the shock does not change the rate of growth of the wormhole, and only gives a fixed outward shift. This possibly is a reflection of the absence of scrambling. 

The next regime is where we set the two Killing times to be equal. This corresponds to boost evolution. This gives
\begin{align}
    n(t) =  {\beta^2 \over 8} + \left( 1 - e^{-\Delta}\right)t^2 + \Delta= \includegraphics[width=3.5cm, valign=c]{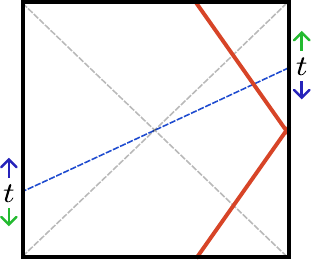}, 
\end{align}
while the unperturbed case is constant and equal to $\beta^2/8$.

The last, and most confusing, regime is where both times are evolved upwards, $-t_L = t_R = t$. The length is
\begin{align}
    n(t) = \left(1 + e^{-\Delta} \right)t^2 + {\beta^2 \over 8} + \Delta +  \left(1 - e^{-\Delta} \right)t_S^2 = \includegraphics[width=3.5cm, valign=c]{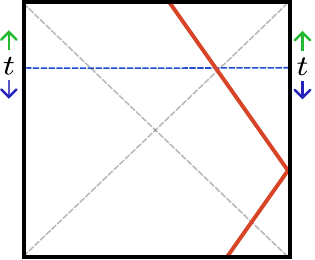} 
\end{align}
where we restored the location of the perturbation. This result has a feature that makes sense, another that doesn't, and one that is unusual. What makes sense is that it grows. What doesn't make sense is that it depends only on the square of perturbation insertion time $t_S$. This means the length of the wormhole at any time is independent of whether the shock was inserted at either $\pm t_S$. However, we will see in the next section that there is a distinction between the two signs when analyzing the lengths to the left/right of the shockwave.

The unusual property is that the growth rate of the perturbed wormhole is \emph{slower} than that of the empty wormhole. Slowing the growth of the wormhole is usually associated with traversability \cite{Gao:2016bin,Maldacena:2017axo}, and therefore it seems that the wormholes in this model can be made traversable through single-sided perturbations. This would be problematic if this model is to be considered the bulk dual of two decoupled boundary systems. However, we checked that the two-sided correlator in the perturbed state is real, and hence the commutator of operators on the two sides vanishes.

Finally, we note that the phenomenon of slowed growth under a perturbation is reminiscent of what happens in de Sitter space. As discussed in \cite{Aalsma:2020aib,Aalsma:2019rpt}, throwing in a shockwave decreases the rate of expansion of the space and induces a time advance that allows signals to travel from one static patch to the other.

\subsection{Individual lengths and folded space}\label{secindl}

So far we have discussed the total length of the wormhole. In this section, we will provide a prescription for measuring lengths to, from, and between matter cords. We will set this up for two colliding shockwaves inside the wormhole before going back to reanalyze the case of a single shockwave. We point out some issues with the prescription when analyzing the simpler single shockwave case.

\subsubsection{Colliding shockwaves inside the wormhole}
This model exhibits the phenomenon of negative lengths discussed in \cite{Lin:2023trc,Haehl:2021tft} between two shockwaves that collide behind the horizon of a black hole. The total length can be extracted from the six-point function composed of four operators for preparing the shocks and a pair for measuring the length. For simplicity, we will take the shocks to have the same dimension $\Delta$. Using our rule above, this is given by
\begin{align}
    &\mathrm{Exp}\Bigg[  -{\beta^2 \over 4} + \sum_{i=1}^{6} {\tau^2_i \over 4} + {e^{-\Delta} \over 2} (\tau_1 \tau_2 + \tau_2 \tau_3 + \tau_4 \tau_5 + \tau_5 \tau_6)    \nonumber \\
     &+  {e^{-2 \Delta} \over 2} (\tau_1 \tau_3 + \tau_4 \tau_6) + {r \over 2} \tau_1 \tau_6 +{r \over 2} \tau_2 \tau_5 + {r \over 2} \tau_3 \tau_4   + {e^{-2 \Delta}\over 2} (r \, \tau_3 \tau_6 + r \, \tau_1 \tau_4) \label{six} \\
     &+ {e^{-\Delta}\over 2}(r\, \tau_1 \tau_5 + r\, \tau_3 \tau_5 + r\, \tau_2 \tau_6 + r \,\tau_2 \tau_4)\Bigg], \nonumber  \\ & \qquad \qquad\qquad \qquad \qquad \qquad= \includegraphics[width=3.5cm, valign=c]{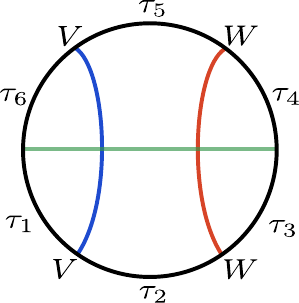}, 
\end{align}
where the green line represents the matter cord used to measure length and represents the variable $r$ in the expression above.

This is a six-point function where the dimension of the operator threading the wormhole is $-\ln r$. To set up the colliding shockwave configuration, we set the times to be
\begin{align}
    \tau_1 &= \epsilon + i t - i t_S, \ \tau_2 = {\beta \over 2} -2\epsilon + 2 i t_S, \ \tau_3 = \epsilon + i t - i t_S, \label{sixkin1} \\
    \tau_4 &= \epsilon - i t + i t_S, \tau_5 = {\beta \over 2} -2 \epsilon - 2i t_S, \tau_6 = \epsilon - i t + i t_S, \label{sixkin2}
\end{align}
where $t_S$ is the right particle insertion time, and $\epsilon$ determines the order and is eventually set to 0. The total length of the wormhole is computed by taking $\partial_r$ and then $r \rightarrow 1$. We get
\begin{align}
     \lan n_{\mathrm{Total}} \ran &= (1 + e^{-2 \Delta})\bigg[ t - (1 - \sech[\Delta]) \,  t_S \bigg]^2 
     + 2 \tanh[\Delta]t_S^2 + {\beta^2 \over 8}, \\
     &= \includegraphics[width=3.5cm, valign=c]{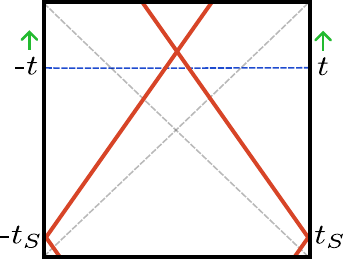}
\end{align}
In the small $\Delta$ limit this becomes
\begin{align}
    \lan n_{\mathrm{Total}} \ran \approx  {\beta^2 \over 8} + 2 (1 - \Delta) t^2 + 2 \Delta t_S^2 + ...
\end{align}
If we compare this to the unperturbed case of $\Delta \rightarrow 0$, the two expressions agree at $t = t_S$, namely when we measure the length between the insertion points. As we noted earlier, an unusual property of these expressions is that the coefficient of the $t^2$ terms is \emph{smaller} for the case with matter than without; the perturbed wormhole grows more slowly than the empty wormhole. While the length in the perturbed case gets an extra kick in the first derivative of the length, the unperturbed wormhole length grows faster and eventually overtakes the perturbed wormhole.

So far we have discussed the total length of the wormhole. Next, we will give a prescription for measuring the lengths in the wormhole to the left and right of the pair of matter cords and also the length between them. The prescription is to slice the six-point function \eqref{six} along the matter cord we used to measure the total length which splits the overlap into $\tau_1, \tau_2, \tau_3$ on one side and $\tau_4, \tau_5, \tau_6$ on the other. We will think of the six-point function as the overlap
\begin{align}
       \sum_{\substack{n_L, n_M, n_R \\ n'_L, n'_M, n'_R}} F(\tau_i) \lan n'_L, n'_M, n'_R | n_L, n_M, n_R \ran = \sum_{\substack{n_L, n_M, n_R \\ n'_L, n'_M, n'_R}} \includegraphics[width=3.5cm, valign=c]{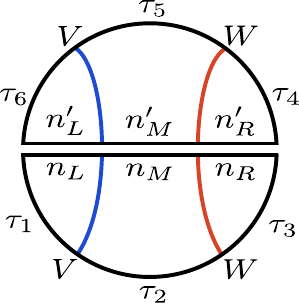} .\label{sixoverlap}
\end{align}
Thus, each half of the path integral prepares a superposition we label as $|n_L,n_M, n_R \ran$ following the conventions of \cite{Lin_2022}. The integers $n_L, n_M, n_R$ represent the number of chords that begin from regions $\tau_1, \tau_2, \tau_3$, respectively,  and  connect to any of $\tau_4, \tau_5, \tau_6$. Similarly, the integers $n'_L, n'_M, n'_R$ are those that begin from regions $\tau_6, \tau_5, \tau_4$, respectively,  and  connect to any of $\tau_1, \tau_2, \tau_3$.

We want to extract the typical values of the numbers that run in the correlation function. However, a subtlety is that the states $|n_L,n_M, n_R \ran$ are not orthogonal \cite{Lin_2022}. This introduces an ambiguity of whether we want to measure the lengths using the bra or ket in the overlap \eqref{sixoverlap}. The choice we adopt is to take the average between the bra and ket, which we implement by considering 
\begin{align}
       \sum_{n'_L, n'_M, n'_R} \sum_{n_L, n_M, n_R} F(\tau_i) \lan n'_L, n'_M, n'_R |r_L^{n_L + n'_L\over 2} r_M^{n_M + n'_M \over 2} r_R^{n_R + n'_R \over 2} | n_L, n_M, n_R \ran. \label{sixoverlap}
\end{align}
Note that if we take $r_L = r_M = r_R = r$ then the inserted operator is just $r^{n}$ where $n$ is the total length given by $n = n_L + n_M + n_R = n_L' + n_M' + n_R'$ and there is no ambiguity. In terms of the expression for the six-point function \eqref{six}, inserting this operator amounts to replacing the factors of $r$ as follows
\begin{align}
    &\mathrm{Exp}\Bigg[  -{\beta^2 \over 4} + \sum_{i=1}^{6} {\tau^2_i \over 4} + {e^{-\Delta} \over 2} (\tau_1 \tau_2 + \tau_2 \tau_3 + \tau_4 \tau_5 + \tau_5 \tau_6)  +  {e^{-2 \Delta} \over 2} \Big(\tau_1 \tau_3 + \tau_4 \tau_6 + r_L^{1\over 2} r_R^{1\over 2} (   \tau_3 \tau_6 + \tau_1 \tau_4) \Big) \nonumber \\   
     &+ {r_L \over 2} \tau_1 \tau_6 +{r_M \over 2} \tau_2 \tau_5 + {r_R \over 2} \tau_3 \tau_4  + {e^{-\Delta}\over 2}\Big(r_L^{1\over 2}r_M^{1\over 2} \, (\tau_1 \tau_5+ \tau_2 \tau_6)  + r_M^{1\over 2}r_R^{1\over 2} \, (\tau_3 \tau_5 + \tau_2 \tau_4) \Big)  \Bigg].
\end{align}
The choices of the prefactors $r_L,...$ can be understood as follows. Note first that the $r$'s only multiply terms pairing regions $1,2,3$ to $4,5,6$. Which $r$'s multiply which term is determined by the overlap \eqref{sixoverlap}. Take for instance the term $r_L^{1\over 2} r_M^{1\over 2} \, \tau_1 \tau_5$. The prefactor $r_L^{1\over 2}$ is included  because $\tau_1$ is associated to $n_L$ while $r^{1 \over 2}_M$ is included because $\tau_5$ is associated with $n'_M$. The same reasoning applies to the rest of the terms.

Now we just have to differentiate with respect to the $r$ factors to compute the lengths. Working in the same kinematics as in \eqref{sixkin1}, \eqref{sixkin2} we find
\begin{align}
    &n_L = n_R = {1 + e^{-2 \Delta}\over 2} \left( t - {t_S \over 2}\left(2 - \cosh[\Delta]^{-1} \right) \right)^2 - {1 \over 4}  t_S^2 (1 - \tanh[\Delta]), \\
    &n_M = 2 e^{-\Delta} t_S \, t - 2(1 - e^{-\Delta})t_S^2 + {\beta^2 \over 8}.
\end{align}
There are several notable aspects of this result. As before, the distances $n_L$ and $n_R$ that measure the length between the boundaries and the particles grow slower than the unperturbed case. Furthermore, these distances can be negative and lower bounded by
\begin{align}
    n_{\mathrm{min}}=  - {t_S^2 \over 4}  \left(1-\tanh[\Delta]\right),
\end{align}
when
\begin{align}
    t = -{t_S \over 2}\left(2 - \cosh[\Delta]^{-1} \right).
\end{align}

\begin{align}
     \includegraphics[width=10cm, valign=c]{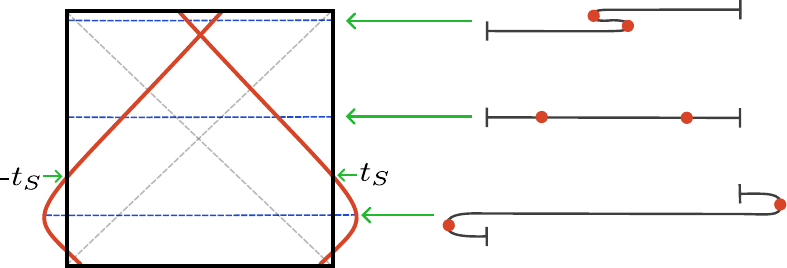}
\end{align}

Another interesting feature is the behavior of $n_M$ which measures the length between the particles. It either grows or shrinks depending on the sign of $t_S$
(the time when the particles are inserted). This suggests that the geometry in between the particles grows to the future, and the sign of 
$t_S$
determines whether the particles can compete with this growth or not. This feature is also somewhat de Sitter/growing cosmology-like. It is also noteworthy to point out that this length grows indefinitely as we move the boundary times to the future. This is very different from the behavior in standard AdS examples where the geodesic between the two boundaries piles up on a limiting surface \cite{Hartman:2013qma} where the distance between the particles saturates.

\subsubsection{Single shockwave lengths}

Following the prescription of the previous section, we can use the OTOC \eqref{eq:OTOC} to find the generating function for the individual lengths. We will use this to show some peculiar features of the lengths.

The generating function is 
\begin{align}
    \mathrm{Exp}\left[  -{\beta^2 \over 4} + \sum_{i=1}^{4} {\tau^2_i \over 4} + {1 \over 2} (r_L \tau_1 \tau_4 + r_R \tau_2 \tau_3) + {e^{-\Delta_W} \over 2} (\tau_1 \tau_2 + \tau_3 \tau_4)  +  {e^{ - \Delta_W}r_L^{1/2}r_R^{1/2} \over 2} (\tau_1 \tau_3 + \tau_2 \tau_4)     \right]e^{-\Delta_V \Delta_W}.
\end{align}
Using the convention where time runs upwards on the right and down on the left, the continuation to Lorentzian time we use is
\begin{align}
    \tau_1 &= {\beta \over 2} + i t_L - i t_S, \ \tau_2 =  i t_S - i t_R, \ \tau_3 = i t_R - i t_S, \ \tau_4 = {\beta \over 2} + i t_S - i t_L, \label{fourkin2}
\end{align}
and  find
\begin{align}
    n_R &= {t_R^2 \over 2} -{1 \over 2}e^{-\Delta}t_L t_R - {t_S\over 2} \left( (2 - e^{-\Delta})t_R - e^{-\Delta}t_L \right)+{t_S^2 \over 2}(1- e^{-\Delta}),\\
    n_L &={t_L^2 \over 2} -{1 \over 2}e^{-\Delta}t_L t_R - {t_S\over 2} \left( (2 - e^{-\Delta})t_L - e^{-\Delta}t_R \right)+{t_S^2 \over 2}(1- e^{-\Delta}) + {\beta^2 \over 8}.
\end{align}

Let's consider the dynamics of the lengths by evolving the boundary times in different ways. The first case we consider is evolving the boundary times upwards in a symmetric fashion $t_R = - t_L = t$, then we get
\begin{align}
    n_R &= {t^2 \over 2}(1 + e^{-\Delta}) -t\, t_S +{t_S^2 \over 2}(1- e^{-\Delta}),\\
    n_L &={t^2 \over 2}(1 + e^{-\Delta}) +t\, t_S +{t_S^2 \over 2}(1- e^{-\Delta})+ {\beta^2 \over 8}.
\end{align}
Just like the collision case in the previous section, there's a time when the right length becomes negative where we have
\begin{align}
    n_R\! \left[ t_S/(1 + e^{-\Delta})\right] = - {e^{\Delta}t_S^2 \over 2(1 + e^{\Delta})},
\end{align}
and $n_L$ remains positive. A picture of this looks like
\begin{align}
     \includegraphics[width=10cm, valign=c]{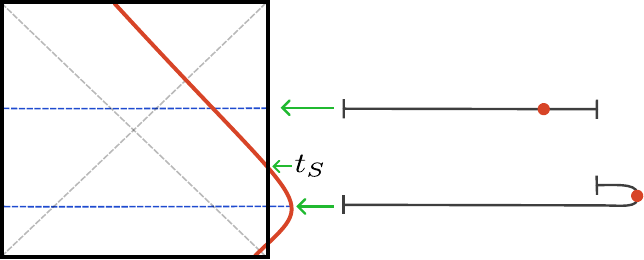}.
\end{align}

So far, this has been consistent with our previous analysis. We find strange behavior if we set the right time $t_R=0$ and scan over different values of $t_L = -t$. We get
\begin{align}
    n_R &=  -{e^{-2 \Delta} \over 2}t\, t_S +{t_S^2 \over 2}(1- e^{-\Delta}),\\
    n_L &={t^2 \over 2}  + {t_S t\over 2}  (2 - e^{-\Delta}) +{t_S^2 \over 2}(1- e^{-\Delta}) + {\beta^2 \over 8}.
\end{align}
This result is odd for two reasons: one is that the right length can become arbitrarily negative, and two is that this happens as we move the \emph{left} boundary time. Furthermore, the right length vanishes at a different time from the shockwave insertion time
\begin{align}
    n_R\left[( e^{\Delta}-1)t_S \right] = 0.
\end{align}
We should note, however, that some of these phenomena are present even without the shockwave.

\subsection{Overlap and bulk correlations}

These matter states are not an orthonormal basis for the Hilbert space of single particle states. For instance, consider the overlap of single matter cord states with any number of spacetime chords as was considered for DSSYK in \cite{Lin_2022}. Following this work, we label these states as $|n_L, n_R \ran$ where the comma represents the insertion of a single matter cord. While such states are orthogonal for different total lengths, they are not orthogonal for different locations of the matter cord. The overlap we find is
\begin{align}
    \lan n_L',n_R'| n_L,n_R \ran &= {\Gamma\left[ 1 + {n\over 2} + x\right]\Gamma\left[ 1 + {n\over 2} + x'\right] e^{-\Delta (n + x + x')} \over \Gamma\left[ 1 - x - x'\right]} \nonumber\\
    &\times {}_2F_1\left[ -{n\over 2} - x, -{n\over 2} - x', 1 - (x + x'), e^{2\Delta} \right] \label{overlap}
\end{align}
where $2x = n_L - n_R, \ 2x' = n'_L - n'_R,\  n = n_L + n_R = n'_L + n'_R $. The distance between the two matter cords is $x + x'$.

The non-orthogonality of the overlap is not surprising because it, in a sense, computes the (unnormalized) bulk correlation function of the two matter cords separated by some distance. The normalized correlation function is
\begin{align}
     {\lan n_L',n_R'| n_L,n_R \ran \over \sqrt{\lan n_L',n_R'| n_L',n_R' \ran\lan n_L,n_R| n_L,n_R \ran}} \equiv {\lan \mO(x') \mO(x) \ran_{\mathrm{bulk} }\over \sqrt{\lan \mO(x') \mO(x') \ran_{\mathrm{bulk} } \lan \mO(x) \mO(x) \ran_{\mathrm{bulk} }}}.
\end{align}
Note that this is an equal-time two-point function.

The overlap \eqref{overlap} can be approximated in the large total length limit. Here we follow ideas in \cite{Lin:2023trc}. Recall from section \ref{wormholelength} that the length wavefunction is highly peaked in the semi-classical limit where $\beta \rightarrow \infty$, which coincides with the large length limit due to the relation $\lan n \ran = \beta^2/8$. This property suggests the relation
\begin{align}
    \lan n_L,n_R| n_L,n_R \ran_{\mathrm{bulk}} \approx  \lan  \mO(\beta - \tau)  \mO(\tau)\ran_{\mathrm{boundary}}
\end{align}
where $\beta, \tau$ are functions of $n_L,n_R$ and where we used $\mO^\dagger(\tau)=  \mO(\beta - \tau)$. The relation between the boundary time and the bulk length on the $\mathbb{Z}_2$ symmetric slice can be extracted from the four-point function. We find
\begin{align}
     n &= n_L + n_R  = {\beta^2 \over 8}, \\
     x &= {n_L - n_R \over 2} = {\beta (\tau - \beta)\over 16}.
\end{align}
While we stated the correspondence between the bulk and boundary at the level of expectation values, we can establish a relation between the bulk and boundary Hilbert spaces by slicing both path integrals across the $\mathbb{Z}_2$, giving
\begin{align}
    | n_L,n_R \ran_{\mathrm{bulk}} \approx   \mO(\tau)| \beta \ran_{\mathrm{boundary}}.
\end{align}
where $|\beta \ran_{\mathrm{boundary}}$ is the thermofield double state.
Finally, using our results for the correlation functions in \eqref{correlationfunctions} to find
\begin{align}
     {\lan n_L',n_R'| n_L,n_R \ran \over \sqrt{\lan n_L',n_R'| n_L',n_R' \ran\lan n_L,n_R| n_L,n_R \ran}} &\approx {\lan  \mO(\beta - \tau)  \mO(\tau)\ran_{\mathrm{boundary}} \over \sqrt{\lan  \mO(\beta - \tau)  \mO(\tau)\ran_{\mathrm{boundary}} \lan  \mO(\beta - \tau)  \mO(\tau)\ran_{\mathrm{boundary}}}}, \\
     &\approx e^{-8 (x-x')^2 (1 - e^{-\Delta}) \over \beta^2}.
\end{align}
It would be interesting to understand the bulk matter system that gives rise to this propagator. We leave this for future work.

\section{Dilaton-gravity model}
In this section, we analyze whether aspects of the Heisenberg model can be reproduced from a dilaton-gravity theory. We do this by interpreting the lengths in \eqref{eq:<n>_wormhole} as geodesic lengths on some background, and we use these lengths to reverse engineer a metric. We use techniques discussed in \cite{Hashimoto:2020mrx}.\footnote{We thank Alexey Milekhin for the discussion on the point.} Then we can find the dilaton-gravity theory that reproduces this metric, namely by finding the appropriate dilaton potential. We end by analyzing the thermodynamics of this dilaton-gravity model.
\subsection{Metric from lengths}
\begin{figure}[!h]
\centering 
    \begin{tikzpicture}[scale=0.7, baseline={([yshift=0cm]current bounding box.center)}]
    \draw[very thick] (0,0) circle (3cm); %
    \draw[very thick, color=red] (2.,{sqrt(3^2 - 2^2)}) .. controls (0.,1.5) and (0.,1.5) .. (-2.,{sqrt(3^2 - 2^2)});
    \draw[fill=black] (0,0) circle (0.05);
    \draw[dashed,thick] (0,0) -- (2.,{sqrt(3^2 - 2^2)});
    \draw[dashed,thick] (0,0) -- (-2.,{sqrt(3^2 - 2^2)});
    \pgfmathsetmacro{\arcStart}{atan(sqrt(3^2 - 2^2)/2)}
    \pgfmathsetmacro{\arcEnd}{-atan(sqrt(3^2 - 2^2)/2)}
    \draw[thick] (1/3,{sqrt(5)/6}) arc (\arcStart:180+\arcEnd:0.5);
    \node at (0.,0.85) {$\Theta$};
    \node[text=red] at (0.,2) {$l$};
    \draw[->, >=latex, thick] (3.5,0) arc (0:30:3.5);
    \node at (4.5,1) {$\theta$ or $\tau$};
    \draw[->, >=latex, thick] (0,0) -- (3,0);
    \node at (2,0.3) {$r$ or $\rho$};
    \end{tikzpicture}   
\caption{A geodesic of length $l$ spans a central angle $\Theta$. $r$ ($\rho$) and $\theta$ ($\tau$) are the radial and angular coordinates, respectively, in the rotationally invariant metric (\ref{eq:rotationally invariant metric ansatz}) and the black hole metric (\ref{rhometric}).}
\label{fig:geodesic-l-theta}
\end{figure}
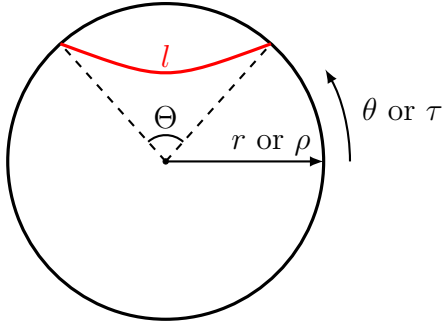

The starting point for finding the metric is the length\footnote{For this section only we change our notation for the length from $n$ to $l$.} formula, obtained by performing a rescaling of $\tau \to \frac{\Theta}{2\pi} \beta$ from \eqref{eq:<n>_wormhole},
\begin{align}
    l(\Theta) = \frac{\beta^2}{8\pi^2} \Theta(2\pi-\Theta)
    \label{ltheta}
\end{align}
where $\Theta$ is the angle along the unit circle with $\Theta \sim \Theta + 2 \pi$. This formula characterizes the length of a geodesic as a function of the central angle it spans (as shown in Figure \ref{fig:geodesic-l-theta}). The maximum geodesic length, or the diameter of the disk, is  $l(\pi) = \beta^2/8$.

We are looking for a rotationally invariant space with the topology of a disk. The metric will be determined by the geodesic lengths \eqref{ltheta}. We assume a metric ansatz of the form
\begin{align}
    ds^2 = f^2(r)dr^2 +  r^2 d\theta^2.
    \label{eq:rotationally invariant metric ansatz}
\end{align}
The radial coordinate $r$ vanishes at the origin and goes to a finite value at the boundary. The value of $r$ at the boundary is fixed by length element along $\theta$ determined by \eqref{ltheta} and must satisfy
\begin{align}
     r^2_\partial = \left( {dl \over d\Theta}\right)^2 \Bigg|_{\Theta \rightarrow 0} = \frac{\beta^4}{16\pi^2}. \label{boundaryradius}
\end{align}
We pick the positive branch of this equation. Regularity of the metric near the origin requires 
$f(r) \sim 1$ plus possibly higher order terms.

Along any geodesic, a conserved quantity associated with the Killing vector $\partial_{\theta}$ is given by $M = r^2 \dot{\theta}$, where the dot denotes the derivative with respect to an affine parameter of the geodesic.
From the geodesic equation $f(r)^2 \dot r^2 + r^2 \dot \theta^2 = f^2 \frac{M^2}{r^4} \left(\frac{dr}{d \theta}\right)^2 + \frac{M^2}{r^2} = 1$, we get the following integral equation
\begin{align}
    \Theta(l) = 2 M \int_{r_{\mathrm{min}}}^{r_\partial} {f(r) dr \over r \sqrt{r^2 - M^2}}~, \label{thetaint}
\end{align}
where $r_{\mathrm{min}}$ is the minimum radial coordinate along the geodesic and satisfies $r_{\mathrm{min}} = M$. The function $\Theta(l)$ is the angular separation of a geodesic of length $l$, and is obtained by inverting \eqref{ltheta}. It is given by
\begin{align}
    \Theta(l) = \pi \left( 1 - \sqrt{1-\frac{8l}{\beta^2}}\right)~.
\end{align}
The equation \eqref{thetaint} is an example of a generalized Abel equation, and it can be inverted to obtain $f(r)$ as an integral of $\Theta(l)$; the details of this are presented in Appendix \ref{appendix:f(r)}. The solution we find\footnote{This solution was first found by Alexey Milekhin (unpublished).} is
\begin{align}
    f(r) = \sqrt{1 - \frac{r^2}{r^2_{\partial}}}
    = \sqrt{1-\frac{16 \pi^2 r^2}{\beta^4}}
    ~,
    \label{f}
\end{align}
and the full metric is given by
\begin{align}
    ds^2 = \left(1 - \frac{16\pi^2 r^2}{\beta^4}\right) dr^2 + r^2 d\theta^2 ~.
    \label{rmetric}
\end{align}
where $0\le \theta \le 2 \pi$ and $0\le r \le \beta^2/4 \pi$. 
For later use, we also express this metric in a ``black hole gauge'' with an emblackening factor, as in
\begin{align}
    ds^2 = g(\rho) d\tau^2 + \frac{d\rho^2}{g(\rho)}~, \qquad 
    g(\rho) = \frac{\beta^2}{4} \left(
    1 - \left( 1 - \frac{24\pi \rho }{\beta^3} \right)^{\frac{2}{3}}
    \right)~,
    \label{rhometric}
\end{align}
where $0\le \tau \le \beta$.
The boundary in this gauge is at $\rho = \frac{\beta^3}{24\pi}$ while the ``horizon'' is at the origin $\rho =0$.\footnote{The map between the two radial coordinates $\rho$ and $r$ is explicitly given by 
$\rho(r) = \frac{\beta^3}{24\pi}\left[1 - \left(1 - \frac{16\pi^2 r^2}{\beta^4}\right)^{3/2}\right]$.}
The space described by this metric is not a hyperbolic space. The Ricci scalar is
\begin{align}
    R = - \frac{2r_{\partial}^2}{(r_{\partial}^2 - r^2)^2} = -\frac{32\pi^2 \beta^4}{(\beta^4 - 16\pi^2 r^2)^2} 
    =-\frac{32\pi^2}{\beta^4 \left(1-\frac{24\pi \rho}{\beta^3}\right)^{4/3}} ~.
    \label{eq:Ricci-R}
\end{align}
The space is still negatively curved everywhere. At large $\beta$, there's a large proper volume region near the origin where the metric is approximately AdS$_2$.\footnote{To see this, we can write the metric \eqref{rmetric} near the origin in terms of a radial proper distance $\sigma$ as
\begin{align}
    ds^2 \approx d\sigma^2 + \left(\sigma + {8 \pi^2 \sigma^3 \over 3 \beta^4} \right)^2 d\theta^2,
\end{align}
which indeed agrees with the near origin expansion of the hyperbolic metric $ds^2 = d\sigma^2 + (\sinh \sigma)^2 d\theta^2$ after a rescaling of the coordinates.} The blow-up of the curvature near the boundary descends from the microscopic description of the model where the two-point function of the Majorana fermions goes to one at coincident points. This property means that the bulk length between boundary points decreases to zero as they approach one another, something that cannot hold for a constant negative curvature space.

\subsection{Dilaton potential from the metric}

Next, we determine the dilaton potential that would give rise to the metric derived above. We assume the general dilaton-gravity action
\begin{align}
    S = \int_{M} \sqrt{g} \left[ \phi R +  U(\phi)\right] + 2 \int_{\partial M} \sqrt{\gamma} \phi_b K.
\end{align}
The aim is to find $U(\phi)$ using the metric obtained in the previous section. The equations of motion of this action imply
\begin{align}
    U(\phi) = \nabla^2 \phi,\quad  \ 
    R = - \partial_\phi U(\phi). \label{ee}
\end{align}
We will solve for $\phi$ and $U(\phi)$ by assuming the rotationally invariant metric above. Working in the gauge in \eqref{rmetric} and writing $\partial_r \phi = \partial_r U/\partial_\phi U$, we find that $U(r)$ satisfies the following differential equation
\begin{align}
    {r f \over 2 f'}U'' + \left[ {f\over f'} + r - {r f f'' \over 2 f^{'2}}\right] U' + U = 0,
\end{align}
where primes denote $r$ derivatives, and $f$ is the metric function \eqref{f}. Regularity at the origin determines the solution to be
\begin{align}
    U = \frac{U_0}{\sqrt{1-\frac{16\pi^2 r^2}{\beta^4}}} = \frac{U_0}{\left(1- \frac{24\pi}{\beta^3} \rho\right)^{1/3}} ~,
    \label{potentialr}
\end{align}
where $U_0$ is the value of the dilaton potential at the origin; it is a free parameter that we leave unfixed in this subsection.  We've also expressed the potential in the black hole gauge \eqref{rhometric}. Next, we can obtain the dilaton by integrating $\partial_r U / \partial_\phi U$. We find
\begin{align}
    \phi &= \phi_h + \frac{\beta^4 U_0}{96 \pi^2}\left[1 - \left(1-\frac{16\pi^2 r^2}{\beta^4}\right)^{3/2}\right] \\ 
    &= \phi_h + \frac{\beta U_0}{4\pi} \rho ~,
\end{align}
where $\phi_h$ is the dilaton value at the horizon.
Finally, the dilaton potential can be obtained by expressing either radial coordinate in terms of the dilaton, and then plugging the answer into \eqref{potentialr}. We find
\begin{align}
    U(\phi) = U_0 \left(1 - \frac{96 \pi^2 (\phi-\phi_h)}{\beta^4 U_0}\right)^{-\frac{1}{3}}
\end{align}
We observed earlier that by keeping everything fixed and taking the large $\beta$ limit we retrieve AdS space near the origin, and we can see that again here since the potential becomes linear in $\phi$.

Notice that the dilaton potential depends on the inverse temperature $\beta$. Hence this bulk model is not governed by a local Lagrangian integrated along the time direction; the Lagrangian parameters are sensitive to the overall length of the time circle. This is perhaps connected to the non-local aspects discussed in section \ref{lag}.

\subsection{Thermodynamics and boundary conditions}

Here we compute the thermodynamics of the dilaton gravity model constructed in the previous section. We found a solution for the dilaton with boundary condition $\phi = \phi_h$ at the horizon. It is more natural to express the solution in terms of a boundary value for the dilaton, $\phi_b$. The horizon value is then
\begin{align}
    \phi_h = \phi_b - \frac{\beta^4 U_0}{96\pi^2}
    \label{phih}
\end{align}
and the dilaton profile is
\begin{align}
    \phi = \phi_b - \frac{\beta^4 U_0}{96\pi^2} \left(1- \frac{16\pi^2 r^2}{\beta^4}\right)^{3/2}
\end{align}
The value of the dilaton at the horizon \eqref{phih} computes the thermal entropy of this dilaton-gravity model. We have to pick $\phi_b$ and $U_0$ before comparing this to the thermal entropy of the Heisenberg model in section \ref{thermo} where $S = - \beta^2/4$. However, we don't have a method of fixing these constants other than matching to the entropy. This matching sets
\begin{align}
    \phi_b = \frac{\beta^4 U_0}{96 \pi^2} - \frac{\beta^2}{4}    ~.
\end{align}

It is perhaps not too surprising that this dilaton gravity theory with a fixed $\beta$ independent $\phi_b$ and $U_0$ disagrees with the Heisenberg model. Such a dilaton-gravity model will produce maximal scrambling with Lyapunov exponent equal to $2\pi/\beta$, while in \ref{correlationfunctions} we found that the Heisenberg model doesn't really scramble because the scrambling time grows without bound as the perturbation becomes light \eqref{tscr}. Furthermore, we found in section \ref{lag} that the effective action for the length is non-local, and so it seems suspicious that it can be described by a simple model of dilaton gravity.

\section{Summary}

This paper featured the analysis of a simple quantum mechanical model that captures the high temperature limit of DSSYK and describes the physics of a two dimensional theory of quantum gravity. The Hamiltonian of the model \eqref{ham} governs the dynamics of the chords that make up the spacetime. The only gravitational degree of freedom is the length mode which measures the length of a one dimensional wormhole connecting two boundary times. We computed the full generating functional for arbitrary insertions of the length mode. Furthermore,  we provided evidence that the effective boundary theory of the length is not local, in contrast to the Schwarzian description of the low temperature limit of SYK.

We also studied the interaction between the length mode and (non-interacting) matter particles. The general matter correlation function is presented in section \ref{correlationfunctions}. In the analysis of the four-point function, we found that the model displays unusual scrambling behavior, and we identified what can be called a Lyapunov exponent that: (a) doesn't depend on the temperature, (b) scales as $1/\Delta$ where $\Delta$ is the dimension of the perturbation, and (c) scales as $1/J$ where $J$ is the coefficient of the Hamiltonian.

Using the matter correlation functions, we computed the dynamics of the length mode in response to particle insertions, and we investigated situations where lengths become negative \cite{Lin:2023trc,Haehl:2021tft}. These include colliding shockwaves behind the horizon and particles folding beyond the asymptotic boundary; this latter phenomenon is what was called exploring the `fake region' in \cite{Lin:2023trc}.

There are many recent works suggesting a connection between DSSYK and de Sitter space \cite{Milekhin:2023bjv, Susskind:2022dfz, Susskind:2022bia, Susskind:2023hnj, Rahman:2023pgt, Lin:2022nss, Rahman:2024vyg, Rahman:2022jsf, Narovlansky:2023lfz, Verlinde:2024znh, Verlinde:2024zrh, Aguilar-Gutierrez:2024nau}. While we do not find a smoking gun pointing at de Sitter, we nevertheless find evidence suggesting that the bulk looks like a growing cosmology. The first piece is that the perturbed wormhole grows more slowly than the vacuum wormhole, similar to the perturbed dynamics of a growing cosmology such as de Sitter \cite{Aalsma:2019rpt,Aalsma:2020aib}. Another is from the behavior of the distance between two particles inserted symmetrically from the two boundaries in the thermofield double (TFD). This distance depends linearly on the boundary time with a coefficient linear in the particle insertion time. This leads to several observations: (a) for positive insertion times, the distance between the particles grows without bound and the particles never collide, (b) for negative insertion times, the distance between the particles eventually becomes negative, signaling a collision, and then decreases without bound, (c)  the distance between the two particles does not saturate for large boundary times, suggesting the absence of the limiting surface phenomenon discussed in \cite{Hartman:2013qma} in the AdS case.

We showed how to extract an effective bulk metric from the boundary two-point function of probe matter fields. If we assume that the bulk theory is a dilaton gravity theory, then we showed how to find the dilaton potential that gives the effective metric. We find a single-parameter family of Lagrangians. We can use this model to reproduce the thermodynamics of the Heisenberg model only after fine tuning the dilaton boundary condition, although we don't have a principle that fixes this choice. An odd feature of this Lagrangian is that it depends on the inverse temperature, and hence the action shouldn't be thought of as the integral of a local density over the thermal circle.

For future work, it would be interesting to study the traversable wormhole protocol  \cite{Gao:2016bin} and the eternal traversable wormhole \cite{Maldacena:2018lmt} in the Heisenberg model. Weakly coupling the two sides by simple operators amounts to deforming the Hamiltonian by the length $\delta H = e^{-\mu a^\dagger a} \approx 1 -\mu a^\dagger a$. Note that since the boundary microscopic Hamiltonian is a sum of commuting terms \cite{berkooz:2024ofm, Berkooz:2024evs}, this model might be relevant for the recent work on experimentally simulating a traversable wormhole in the lab \cite{Jafferis:2022crx,Kobrin:2023rzr,Jafferis:2023moh}.

\section*{Acknowledgements}

We thank Micha Berkooz, Yiyang Jia, Henry Lin, Simon Lin, Adam Levine, Juan Maldacena, Ohad Mamroud, Alexey Milekhin, Fedor Popov, Xiaoliang Qi, Adel Rahman, Douglas Stanford, Jiuci Xu and Ying Zhao for insightful discussion and correspondence. This research was supported in part by grant NSF PHY-2309135 to the Kavli Institute for Theoretical Physics (KITP). The work of XYH is supported in part by NSF grant PHY-1820814 and by the Shuimu Tsinghua Scholar Program of Tsinghua University.

\appendix 

\section{All-point length generating functional} \label{alllengthgen}
Consider the following generating functional
\begin{align}
    \lan r_1^{\hat{n}(t_1)} ...  \, r_a^{\hat{n}(t_a)} \ran_\beta \label{genf}
\end{align}
for some constants $r_1,..., r_a$, and where the correlators generated by differentiating with respect to $r_i$ followed by taking all $r_i \rightarrow 1$.

The method for evaluating \eqref{genf} is to divide the boundary circle into $a+1$ intervals. We can think of each interval as preparing a bulk state of the form
\begin{align}
    e^{-H(t_i - t_{i-1})} |0\ran &= \sum_{m_i=0}^{\infty}\frac{1}{m_i!} | m_{i} \ran \lan m_i | e^{-H(t_{i} - t_{i-1})} |0\ran 
    \\
    &=\sum_{m_i=0}^{\infty}\frac{1}{m_i!} \int dE_i\ e^{-E_i(t_i-t_{i-1})}\psi_{m_i}(E_i) \psi_0^*(E_i) | m_{i} \ran
    \\
    &= \sum_{m_i=0}^{\infty} \frac{1}{m_i!} \int {dE_i \over \sqrt{\pi}} e^{-E_i^2-E_i(t_{i} - t_{i-1})} {H_{m_i}(E_i) \over 2^{m_i \over 2}} |m_i \ran \\
    &=e^{\frac{(t_i - t_{i-1})^2}{4}}\sum_{m_i=0}^{\infty} \frac{1}{m_i!}\left( - \frac{t_{i} - t_{i-1}}{\sqrt{2}}\right)^{m_i} | m_{i} \ran
    \label{segment}
\end{align}
where $i$ goes from $1$ to $a$, and $\psi_n(E)$ is given by (\ref{eq:psi_n-E}) in the main text. This replaces each boundary segment with a superposition of definite chord number states.  What remains is to enumerate all the possible contractions between them while weighing each with the appropriate power of $\{r_i \}$. The systematic way of doing this is to write  $m_i=\sum_{j\neq i}m_{ij}$ where $m_{ij}$ is the number of chords shared between segments $i$ and $j$ ($m_{ij}$ is symmetric in $i,j$). 
\begin{align}
    \includegraphics[width=5cm, valign=c]{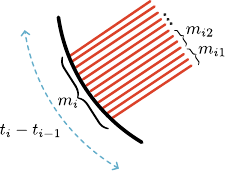}
\end{align}
This division comes with a combinatoric factor of $\binom{m_i}{\{ m_{ij}\}} = \frac{m_i!}{\prod_{j\neq i} m_{ij}!}$
for each $i$ coming from all the different ways of dividing $m_i$ into $\{ m_{ij} \}$. Note that each $m_{ij}$ appears exactly twice in the denominator of this expression. However, there's an additional factor of $m_{ij}!$ coming from the number of ways of joining the chords together; these chord states are not normalized. The final combinatoric factor is
\begin{align}
   {\prod_{i=1}^{a+1} m_i! 
    \over \prod_{j \neq i} m_{ij}!}.
\end{align}
Note that the numerator and denominator products are independent of each other. The numerators eventually cancel against a normalizing factor in the original chord states in \eqref{segment}. Furthermore, from the constraint that $m_i = \sum_{j\neq i}m_{ij}$ for all $i$, the $m_i$'s are determined in terms of the $m_{ij}$, all the sums can be expressed as sums over the $m_{ij}$'s. So far we have
\begin{align}
    \sum_{m_{i<j} = 1}^{a+1}  {1 \over \prod_{j \neq i} m_{ij}!} \times \left( (t_{j} - t_{j-1})(t_{i} - t_{i-1}) \over 2\right)^{m_{ij}} \times \ r_i \mathrm{\ factors.}
\end{align}
The last thing to be determined are the $r_i$ factors. This is determined by noting that $m_{ij}$ intersects $j-i$ slices corresponding to $r_i \times...\times r_{j-1}$.
\begin{align}
    \includegraphics[width=5cm, valign=c]{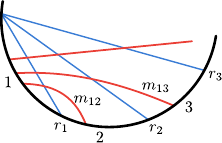}
\end{align}
Hence we get the factor
\begin{align}
    r_1^{m_{12}} (r_1 r_2)^{m_{13}} ... = \prod_{i<j} \left( \prod_{k = i}^{j-1} r_k \right)^{m_{ij}}
\end{align}
Putting everything together and taking the sum, we obtain
\begin{align}
        \lan r_1^{\hat{n}(t_1)} ...  \, r_a^{\hat{n}(t_a)} \ran_\beta &= e^{- {\beta^2 \over 4}}\mathrm{Exp}\left[ {\sum_{i=1}^{a+1}{(t_i - t_{i-1})^2 \over 4} + {1\over 2}\sum_{i<j = 1}^{a+1}(t_i - t_{i-1})(t_j - t_{j-1}) \prod_{k = i}^{j-1}r_k} \right] \\
        &= \mathrm{Exp}\left[  {1\over 2}\sum_{i<j = 1}^{a+1}(t_i - t_{i-1})(t_j - t_{j-1}) \left( \prod_{k = i}^{j-1}r_k - 1 \right) \right].
\end{align}

\section{Correlation functions}
In this appendix, we summarize the details involved in calculating various types of matter correlation functions discussed in the main text.
\subsection{Two-point function}
\label{appendix:2-pt}
The boundary two-point function of two matter operators $V$ can be written as a bulk transition amplitude from an `initial' state $|n(\tau_1)\ran \equiv e^{-\tau_1 H} | 0 \ran$ to a `final' state $|n(\tau_2)\ran \equiv e^{-\tau_2 H} |0\ran$. Insertion of two operators $V$ on the boundary induces a single matter line insertion in the bulk. This means that the transition amplitude involves a bulk operator $e^{-\Delta_V \hat n}$, where $\Delta_V $ is the dimension of particle $V$.
\begin{align}
    &\quad \tr[e^{-\tau_2 H} V e^{-\tau_1 H} V]
    =\!\!\!\!\!\!\!\!\!\!\!\!
    \begin{tikzpicture}[scale=0.35, baseline={([yshift=0cm]current bounding box.center)}]
    \definecolor{AGHblue}{rgb}{0.118, 0.294, 0.812}
    \draw[very thick] (0,0) circle (3cm);
    \draw[thick, color=AGHblue] (2.5,{-sqrt(2.75)}) .. controls (1.5,0) and (-1.5,0) .. (-2.5,{-sqrt(2.75)});
    \draw (0,-3.5) node[text width=2 cm,align=center] { \small $\tau_1$};
    \draw (0,3.5) node[text width=2 cm,align=center] { \small $\tau_2$};
    \draw (-3,-2.3) node[text width=2 cm,align=center,AGHblue] { \small $V$};
    \draw (3,-2.3) node[text width=2 cm,align=center,AGHblue] { \small $V$};
    \draw[fill=black, color=AGHblue] (-2.5,{-sqrt(2.75)}) circle (0.15);
    \draw[fill=black, color=AGHblue] (2.5,{-sqrt(2.75)}) circle (0.15);
    \end{tikzpicture}
    \!\!\!\!\!\!\!\!\!\!\!\!
    =
    \lan 0 |e^{-\tau_2 H} e^{-\Delta_V \hat n} e^{-\tau_1 H} | 0 \ran
    \notag \\
    &=
    \!\!\!\!\!\!\!\!\!\!\!\!
    \begin{tikzpicture}[scale=0.35, baseline={([yshift=0cm]current bounding box.center)}]
    \definecolor{AGHblue}{rgb}{0.118, 0.294, 0.812}
    \pgfmathsetmacro{\startAngle}{asin(-sqrt(2.75)/3)}
    \pgfmathsetmacro{\endAngle}{180-asin(-sqrt(2.75)/3)}
    \draw[very thick] (2.5,1.65831+0.2) arc (\startAngle:\endAngle:3cm);
    \pgfmathsetmacro{\xStartupper}{3*cos(\startAngle)}
    \pgfmathsetmacro{\yStartupper}{3*sin(\startAngle)+2*1.65831+0.2}
    \pgfmathsetmacro{\xEndupper}{3*cos(\endAngle)}
    \pgfmathsetmacro{\yEndupper}{3*sin(\endAngle)+2*1.65831+0.2}
    \pgfmathsetmacro{\xStartlower}{3*cos(\startAngle)}
    \pgfmathsetmacro{\yStartlower}{3*sin(\startAngle)+2*1.65831-0.2}
    \pgfmathsetmacro{\xEndlower}{3*cos(\endAngle)}
    \pgfmathsetmacro{\yEndlower}{3*sin(\endAngle)+2*1.65831-0.2}
    \pgfmathsetmacro{\xStart}{3*cos(\startAngle)}
    \pgfmathsetmacro{\yStart}{3*sin(\startAngle)+2*1.65831}
    \pgfmathsetmacro{\xEnd}{3*cos(\endAngle)}
    \pgfmathsetmacro{\yEnd}{3*sin(\endAngle)+2*1.65831}
    \draw[gray] (\xStartupper,\yStartupper) .. controls (0,2.7) and (0,2.7) .. (\xEndupper,\yEndupper);
    \draw[AGHblue, thick] (\xStart,\yStart) .. controls (0,2.5) and (0,2.5) .. (\xEnd,\yEnd);
    \draw[very thick] (-2.5,1.65831-0.2) arc (\endAngle:360+\startAngle:3cm);
    \draw[gray] (\xStartlower,\yStartlower) .. controls (0,2.3) and (0,2.3) .. (\xEndlower,\yEndlower);
    \draw (\xStart + 0.8,\yStart-0.2) node[text width=2 cm,align=center,AGHblue] { \small $V$};
    \draw (\xEnd - 0.8,\yEnd - 0.2) node[text width=2 cm,align=center,AGHblue] { \small $V$};
    \draw[fill=black, color=AGHblue] (\xStart,\yStart) circle (0.15);
    \draw[fill=black, color=AGHblue] (\xEnd,\yEnd) circle (0.15);
    \draw (0,-0.5) node[text width=2 cm,align=center] { \small $\tau_1$};
    \draw (0,7) node[text width=2 cm,align=center] { \small $\tau_2$};
    \draw (0,1.4) node[text width=2 cm,align=center] { \footnotesize $n(\tau_1)$};
    \draw (0,3.1) node[text width=2 cm,align=center] { \footnotesize $n(\tau_2)$};
    \end{tikzpicture}
    \!\!\!\!\!\!\!\!\!\!\!\!
    = \sum_{n_1,n_2=0}^{\infty} \frac{1}{n_1! n_2!} \lan n(\tau_2)|n_2\ran \lan n_2|e^{-\Delta_V \hat n} |n_1 \ran \lan n_1|n(\tau_1)\ran
    \notag\\
    &=
    \left(\prod_{i = 1}^2 \sum_{n_i} \frac{1}{n_i!} \int {dE_i \over \sqrt{\pi}} e^{-E_i^2-E_i \tau_i} {H_{n_i}(E_i) \over 2^{n_i \over 2}}\right) \lan n_2|e^{-\Delta_V \hat n} |n_1 \ran 
    \notag
    \\
    &=\left(\prod_{i = 1}^2 \sum_{n_i} \frac{1}{n_i!} \int {dE_i \over \sqrt{\pi}} e^{-E_i^2-E_i \tau_i} {H_{n_i}(E_i) \over 2^{n_i \over 2}}\right)
    \times e^{-\Delta_V n_1} \times  \delta_{n_1,n_2}  n_1!
    \notag
    \\
    &= \exp\left[\sum_{i=1}^{2} \frac{\tau_i^2}{4} + \frac{e^{-\Delta_V}}{2} \tau_1 \tau_2\right]
\end{align}
where we used the fact that the segment $\tau_i$ of the circle prepares the bulk state
\begin{align}
    |n(\tau_i)\ran \equiv e^{-\tau_i H} |0\ran 
    =
    \sum_{n_i=0}^{\infty} \frac{1}{n_i!} \int {dE_i \over \sqrt{\pi}} e^{-E_i^2-E_i \tau_i} {H_{n_i}(E_i) \over 2^{n_i \over 2}} |n_i \ran ~.
    \label{eq:n(tau_i)}
\end{align}
Including the normalization factor $Z^{-1} = e^{-\frac{\beta^2}{4}}$ for $\tau_1 + \tau_2 = \beta$, we obtain (\ref{eq:2-pt}).

\subsection{Out-of-time-order correlation function (OTOC)}
\label{appendix:OTOC}
The (unnormalized) OTOC of the boundary theory can be expressed in the bulk theory as
\begin{align}
    \tr[e^{-\tau_4 H}W e^{-\tau_3 H} V e^{-\tau_2 H} W e^{-\tau_1 H} V]
    =\!\!\!\!\!\!\!\!\!\!
    \!\!\!\!\!\!\!
    \begin{tikzpicture}[scale=0.25, baseline={([yshift=0cm]current bounding box.center)}]
    \definecolor{AGHblue}{rgb}{0.118, 0.294, 0.812}
    \definecolor{AGHred}{rgb}{0.843, 0.267, 0.153}
    \draw[very thick] (0,0) circle (5cm);
    \draw[thick, color = AGHred] (-3.5,-3.5) -- (3.5, 3.5);
    \draw[thick, color = AGHblue] (-3.5, 3.5) -- (3.5, -3.5);
    \draw (0,-5.8)  node[text width=3.1 cm,align=center] { \small $\tau_2$};
    \draw (0,5.8)  node[text width=3.1 cm,align=center] { \small $\tau_4$};
    \draw (-5.8,0)  node[text width=3.1 cm,align=center] { \small $\tau_1$};
    \draw (5.8,0)  node[text width=3.1 cm,align=center] { \small $\tau_3$};
    \draw[fill=black, color=AGHred] (-3.5, -3.5) circle (0.2);
    \draw (-4.5,-4.) node[text width=2 cm,align=center,AGHred] { \small $W$};
    \draw[fill=black, color=AGHred] (3.5, 3.5) circle (0.2);
    \draw (4.5,4.) node[text width=2 cm,align=center,AGHred] { \small $W$};
    \draw[fill=black, color=AGHblue] (3.5, -3.5) circle (0.2);
    \draw (-4.5,4.) node[text width=2 cm,align=center,AGHblue] { \small $V$};
    \draw[fill=black, color=AGHblue] (-3.5, 3.5) circle (0.2);
    \draw (4.5,-4.) node[text width=2 cm,align=center,AGHblue] { \small $V$};
    \end{tikzpicture}
    \label{eq:OTOC-C4}
\end{align}
Our goal is to evaluate the graphical representation on the right-hand side of (\ref{eq:OTOC-C4}) using a bulk calculation.

\begin{figure}[!h]
\centering 
\begin{tikzpicture}[scale=0.35, baseline={([yshift=0cm]current bounding box.center)}]
    \definecolor{AGHblue}{rgb}{0.118, 0.294, 0.812}
    \definecolor{AGHred}{rgb}{0.843, 0.267, 0.153}
    \draw[very thick] (0,0) circle (5cm);

   \draw[thick, color = AGHred] (-3.5,-3.5) -- (3.5, 3.5);
   \draw[thick, color = AGHblue] (-3.5, 3.5) -- (3.5, -3.5);

   \draw (0,3)  node[text width=3.1 cm,align=center] { \large IV};
   \draw (0,-3)  node[text width=3.1 cm,align=center] { \large II};
   \draw (3,0)  node[text width=3.1 cm,align=center] { \large III};
   \draw (-3,0)  node[text width=3.1 cm,align=center] { \large I};

   \draw[thick, dotted] (-3.2,3.7) -- (0, 0.5);
   \draw[thick, dotted] (3.2,3.7) -- (0, 0.5);
   \draw[thick, dotted] (-3.2,-3.7) -- (0, -0.5);
   \draw[thick, dotted] (3.2,-3.7) -- (0, -0.5);
   \draw[thick, dotted] (-3.7, -3.2) -- (-0.5, 0);
   \draw[thick, dotted] (-3.7, 3.2) -- (-0.5, 0);
   \draw[thick, dotted] (3.7, -3.2) -- (0.5, 0);
   \draw[thick, dotted] (3.7, 3.2) -- (0.5, 0);

   \draw[fill=black, color=AGHred] (-3.5, -3.5) circle (0.2);
   \draw (-4.5,-4.) node[text width=2 cm,align=center,AGHred] { \small $W$};
   \draw[fill=black, color=AGHred] (3.5, 3.5) circle (0.2);
   \draw (4.5,4.) node[text width=2 cm,align=center,AGHred] { \small $W$};
   \draw[fill=black, color=AGHblue] (3.5, -3.5) circle (0.2);
   \draw (-4.5,4.) node[text width=2 cm,align=center,AGHblue] { \small $V$};
   \draw[fill=black, color=AGHblue] (-3.5, 3.5) circle (0.2);
   \draw (4.5,-4.) node[text width=2 cm,align=center,AGHblue] { \small $V$};

   \draw (-5.8,0)  node[text width=3.1 cm,align=center] { \small $\tau_1$};
   \draw (0,-5.8)  node[text width=3.1 cm,align=center] { \small $\tau_2$};
   \draw (5.8,0)  node[text width=3.1 cm,align=center] { \small $\tau_3$};
   \draw (0,5.8)  node[text width=3.1 cm,align=center] { \small $\tau_4$};
\end{tikzpicture}
\caption{The two matter cords split the diagram into four regions.}
\label{fig:OTOC}
\end{figure}
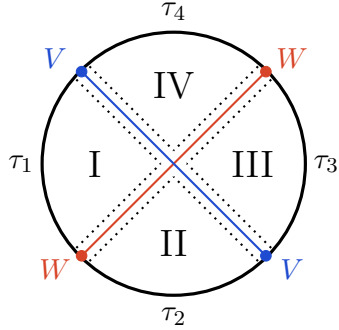

First, we note that 
the two matter cords split the diagram into four regions as shown in Figure \ref{fig:OTOC}. 

Each segment $\tau_i$ prepares on the dotted bulk slice (right before crossing any matter cords) a bulk state
$|n(\tau_i)\ran$ given by (\ref{eq:n(tau_i)}). The split segments prepares the product of four states,
\begin{align}
    \!\!\!\!\!\!\!\!\!\!\!\!\!\!\!\!\!
    \begin{tikzpicture}[scale=0.25, baseline={([yshift=0cm]current bounding box.center)}]
   \draw[very thick] (-3.7,3.2) arc (140:220:5cm); %
   \draw[very thick] (3.7,3.2) arc (40:-40:5cm); %
   \draw[very thick] (-3.2,-3.7) arc (-130:-50:5cm); %
   \draw[very thick] (3.2,3.7) arc (50:130:5cm); %
   \draw (0,3)  node[text width=3.1 cm,align=center] { \small IV};
   \draw (0,-3)  node[text width=3.1 cm,align=center] { \small II};
   \draw (3,0)  node[text width=3.1 cm,align=center] { \small III};
   \draw (-3,0)  node[text width=3.1 cm,align=center] { \small I};
   \draw[thick, dotted] (-3.2,3.7) -- (0, 0.5);
   \draw[thick, dotted] (3.2,3.7) -- (0, 0.5);
   \draw[thick, dotted] (-3.2,-3.7) -- (0, -0.5);
   \draw[thick, dotted] (3.2,-3.7) -- (0, -0.5);
   \draw[thick, dotted] (-3.7, -3.2) -- (-0.5, 0);
   \draw[thick, dotted] (-3.7, 3.2) -- (-0.5, 0);
   \draw[thick, dotted] (3.7, -3.2) -- (0.5, 0);
   \draw[thick, dotted] (3.7, 3.2) -- (0.5, 0);
   \draw (-5.8,0)  node[text width=3.1 cm,align=center] { \small $\tau_1$};
   \draw (0,-5.8)  node[text width=3.1 cm,align=center] { \small $\tau_2$};
   \draw (5.8,0)  node[text width=3.1 cm,align=center] { \small $\tau_3$};
   \draw (0,5.8)  node[text width=3.1 cm,align=center] { \small $\tau_4$};
   \end{tikzpicture}
   \!\!\!\!\!\!\!\!\!\!\!\!\!\!\!\!\!
   &= |n(\tau_1)\ran \otimes
   |n(\tau_2)\ran \otimes
   |n(\tau_3)\ran \otimes
   |n(\tau_4)\ran
   \notag \\
   &=\left(\prod_{i=1}^{4}\sum_{n_i=0}^{\infty} \int {dE_i \over \sqrt{\pi}} e^{-E_i^2-E_i \tau_i} \frac{H_{n_i}(E_i)}{2^{n_i \over 2}n_i!}\right)
    \left(|n_1 \ran \otimes |n_2 \ran \otimes |n_3 \ran \otimes |n_4 \ran \right)
    \label{eq:product-of-4-segment}
\end{align}
Next, we stitch the four quadrants together. This amounts to accounting for the correct combinatorial factor in connecting the Hamiltonian chords.

We use the following gluing rule associated with gluing chords, 
\begin{align}
    |n_1 \ran \otimes | n_2 \ran= 
    \!\!\!\!\!\!\!\!\!\!\!\!\!\!\!\!
    \begin{tikzpicture}[scale=0.4, baseline={([yshift=0cm]current bounding box.center)}] %
        \draw[very thick] (-2,0) arc (270:90:2);
        \draw[fill=black, draw=none] (-2,0) circle (0.2cm);
        \draw[fill=black, draw=none] (-1,0) circle (0.2cm);
        \draw (-4.5,2)  node[text width=3.1 cm,align=center] { \small $\tau_1$};
        \draw (1,-2.5)  node[text width=3.1 cm,align=center] { \small $\tau_2$};
        \draw[color=black] (-2,0) -- (-2,4);
        \draw[color=gray] (-2.2, 0.4) -- (-1.8, 0.4); 
        \draw[color=gray] (-2.2, 0.8) -- (-1.8, 0.8); 
        \draw[color=gray] (-2.2, 1.2) -- (-1.8, 1.2);
        \draw[color=gray] (-2.2, 1.6) -- (-1.8, 1.6); 
        \draw[color=gray] (-2.2, 2.0) -- (-1.8, 2.0); 
        \draw[color=gray] (-2.2, 2.4) -- (-1.8, 2.4); 
        \draw[color=gray] (-2.2, 2.8) -- (-1.8, 2.8); 
        \draw[color=gray] (-2.2, 3.2) -- (-1.8, 3.2); 
        \draw[color=gray] (-2.2, 3.6) -- (-1.8, 3.6); 
        \draw (-1.1,2) node[text width=3.1 cm,align=center] { \large $n_1$};
        \draw[very thick] (-1,0) arc (180:360:2);
        \draw[color=black] (-1,0) -- (3,0);
        \draw[color=gray] (-0.6, -0.2) -- (-0.6, 0.2); 
        \draw[color=gray] (-0.2, -0.2) -- (-0.2, 0.2); 
        \draw[color=gray] (0.2, -0.2) -- (0.2, 0.2); 
        \draw[color=gray] (0.6, -0.2) -- (0.6, 0.2); 
        \draw[color=gray] (1, -0.2) -- (1, 0.2); 
        \draw[color=gray] (1.4, -0.2) -- (1.4, 0.2); 
        \draw[color=gray] (1.8, -0.2) -- (1.8, 0.2); 
        \draw[color=gray] (2.2, -0.2) -- (2.2, 0.2); 
        \draw[color=gray] (2.6, -0.2) -- (2.6, 0.2); 
        \draw (1,0.8) node[text width=3.1 cm,align=center] { \large $n_2$};
    \end{tikzpicture} 
    \!\!\!\!\!\!
    =\quad \sum_{k=0}^{\text{min}(n_1,n_2)} D_{k}^{n_1, n_2}
    \!\!\!\!\!\!\!\!\!\!\!\!\!\!\!\!
    \begin{tikzpicture}[scale=0.4, baseline={([yshift=0cm]current bounding box.center)}] %
        \draw[very thick] (0,-3) arc (270:90:3);
        \draw[fill=black, draw=none] (0,0) circle (0.2cm);
        \draw[fill=black, draw=none] (-3,0) circle (0.2cm);
        \draw (-3,2.5)  node[text width=3.1 cm,align=center] { \small $\tau_1$};
        \draw (-3,-2.5)  node[text width=3.1 cm,align=center] { \small $\tau_2$};
        \draw[color=black] (0,-3) -- (0,3);
        \draw[color=black] (0,0) -- (-3,0);
        \draw (1.6,1) node[text width=1 cm,align=center] {$n_1-k$};
        \draw (1.6,-1) node[text width=1 cm,align=center] {$n_2-k$};
        \draw (-1.3,0.8) node[text width=1 cm,align=center] {$k$};
        \draw[color=gray] (-0.2, -2.6) -- (0.2, -2.6);
        \draw[color=gray] (-0.2, -2.2) -- (0.2, -2.2);
        \draw[color=gray] (-0.2, -1.8) -- (0.2, -1.8);
        \draw[color=gray] (-0.2, -1.4) -- (0.2, -1.4);
        \draw[color=gray] (-0.2, -1) -- (0.2, -1);
        \draw[color=gray] (-0.2, -0.6) -- (0.2, -0.6);
        \draw[color=gray] (-0.2, -0.2) -- (0.2, -0.2);
        \draw[color=gray] (-0.2, 2.6) -- (0.2, 2.6);
        \draw[color=gray] (-0.2, 2.2) -- (0.2, 2.2);
        \draw[color=gray] (-0.2, 1.8) -- (0.2, 1.8);
        \draw[color=gray] (-0.2, 1.4) -- (0.2, 1.4);
        \draw[color=gray] (-0.2, 1) -- (0.2, 1);
        \draw[color=gray] (-0.2, 0.6) -- (0.2, 0.6);
        \draw[color=gray] (-0.2, 0.2) -- (0.2, 0.2);
        \draw[color=gray] (-2.6,-0.2) -- (-2.6,0.2);
        \draw[color=gray] (-2.2,-0.2) -- (-2.2,0.2);
        \draw[color=gray] (-1.8,-0.2) -- (-1.8,0.2);
        \draw[color=gray] (-1.4,-0.2) -- (-1.4,0.2);
        \draw[color=gray] (-1,-0.2) -- (-1,0.2);
        \draw[color=gray] (-0.6,-0.2) -- (-0.6,0.2);
    \end{tikzpicture}
    \label{eq:gluing}
\end{align}
where
$k$ is the number of chords connecting the $\tau_1$ and $\tau_2$ boundary segments. The remaining $n_1-k$ open chords connect $\tau_1$ to segments other than $\tau_2$. Similarly, $n_2-k$ open chords connect $\tau_2$ to segments other than $\tau_1$.
The combinatorial factor
\begin{align}
    D_{k}^{n_1, n_2} \equiv
    \binom{n_1}{k}\binom{n_2}{k}k!
    = \frac{n_1! n_2 !}{(n_1 - k)! (n_2 - k)! k!}
    \label{eq:D-combinatorial}
\end{align}
represents the number of ways $k$ chords from the $\tau_1$ segment can connect to $k$ chords from the $\tau_2$ segment, out of $n_1$ and $n_2$ total chords emerging from $\tau_1$ and $\tau_2$, respectively.
This count further implies an upper bound on the sum, given that $k \leq \min(n_1, n_2)$.

In the presence of matter cords, we assign a penalty factor $q_V=e^{-\Delta_V}$ for every $V$-cord-chord intersection, $q_W = e^{-\Delta_W}$ for every $W$-cord-chord intersection, and $q_{VW}=e^{-\Delta_V\Delta_W}$ for $V$-cord-$W$-cord intersection.

The gluing rule (\ref{eq:gluing}) can be generalized to account for the intersection of the $k$ Hamiltonian chords with matter cords. Gluing the regions I and II can be expressed diagrammatically as 
\begin{align}
    \begin{tikzpicture}[scale=0.4, baseline={([yshift=0cm]current bounding box.center)}] 
    \definecolor{AGHblue}{rgb}{0.118, 0.294, 0.812}
    \definecolor{AGHred}{rgb}{0.843, 0.267, 0.153}
        \draw[very thick] (-2,0) arc (270:90:2);
        \draw[fill=AGHred, draw=none] (-1,0.5) circle (0.2cm);
        \draw[fill=AGHred, draw=none] (2,3.5) circle (0.2cm);
        \draw[color=AGHred,very thick] (-1,0.5) -- (2,3.5);
        \draw[color=AGHred] (3.1,3.5) node[text width=3.1 cm,align=center] {  $W$};
        \draw[color=black] (-2,0) -- (-2,4);
        \draw[color=gray] (-2.2, 0.4) -- (-1.8, 0.4); 
        \draw[color=gray] (-2.2, 0.8) -- (-1.8, 0.8); 
        \draw[color=gray] (-2.2, 1.2) -- (-1.8, 1.2);
        \draw[color=gray] (-2.2, 1.6) -- (-1.8, 1.6); 
        \draw[color=gray] (-2.2, 2.0) -- (-1.8, 2.0); 
        \draw[color=gray] (-2.2, 2.4) -- (-1.8, 2.4); 
        \draw[color=gray] (-2.2, 2.8) -- (-1.8, 2.8); 
        \draw[color=gray] (-2.2, 3.2) -- (-1.8, 3.2); 
        \draw[color=gray] (-2.2, 3.6) -- (-1.8, 3.6); 
        \draw (-1.1,2) node[text width=3.1 cm,align=center] {  $n_1$};
        \draw[very thick] (-1,0) arc (180:360:2);
        \draw[color=black] (-1,0) -- (3,0);
        \draw[color=gray] (-0.6, -0.2) -- (-0.6, 0.2); 
        \draw[color=gray] (-0.2, -0.2) -- (-0.2, 0.2); 
        \draw[color=gray] (0.2, -0.2) -- (0.2, 0.2); 
        \draw[color=gray] (0.6, -0.2) -- (0.6, 0.2); 
        \draw[color=gray] (1, -0.2) -- (1, 0.2); 
        \draw[color=gray] (1.4, -0.2) -- (1.4, 0.2); 
        \draw[color=gray] (1.8, -0.2) -- (1.8, 0.2); 
        \draw[color=gray] (2.2, -0.2) -- (2.2, 0.2); 
        \draw[color=gray] (2.6, -0.2) -- (2.6, 0.2); 
        \draw (1,0.8) node[text width=3.1 cm,align=center] {  $n_2$};
    \end{tikzpicture} 
    \!\!\!\!\!\!\!\!\!\!\!\!\!
    =
    \quad \sum_{k=0}^{\min(n_1,n_2)} D_{k}^{n_1, n_2} q_W^k \times \quad
    \begin{tikzpicture}[scale=0.4, baseline={([yshift=0cm]current bounding box.center)}, rotate=45]
    \definecolor{AGHblue}{rgb}{0.118, 0.294, 0.812}
    \definecolor{AGHred}{rgb}{0.843, 0.267, 0.153}
        \draw[very thick] (0,-3) arc (270:90:3);
        \draw[fill=AGHred, draw=none] (3,0) circle (0.2cm);
        \draw[fill=AGHred, draw=none] (-3,0) circle (0.2cm);
        \draw[color=black] (0,-3) -- (0,3);
        \draw[color=AGHred, very thick] (3,0) -- (-3,0);
        \draw[color=gray] (-0.2, -2.6) -- (0.2, -2.6);
        \draw[color=gray] (-0.2, -2.2) -- (0.2, -2.2);
        \draw[color=gray] (-0.2, -1.8) -- (0.2, -1.8);
        \draw[color=gray] (-0.2, -1.4) -- (0.2, -1.4);
        \draw[color=gray] (-0.2, -1) -- (0.2, -1);
        \draw[color=gray] (-0.2, -0.6) -- (0.2, -0.6);
        \draw[color=gray] (-0.2, -0.2) -- (0.2, -0.2);
        \draw[color=gray] (-0.2, 2.6) -- (0.2, 2.6);
        \draw[color=gray] (-0.2, 2.2) -- (0.2, 2.2);
        \draw[color=gray] (-0.2, 1.8) -- (0.2, 1.8);
        \draw[color=gray] (-0.2, 1.4) -- (0.2, 1.4);
        \draw[color=gray] (-0.2, 1) -- (0.2, 1);
        \draw[color=gray] (-0.2, 0.6) -- (0.2, 0.6);
        \draw[color=gray] (-0.2, 0.2) -- (0.2, 0.2);
        \draw[color=gray] (-2.6,-0.2) -- (-2.6,0.2);
        \draw[color=gray] (-2.2,-0.2) -- (-2.2,0.2);
        \draw[color=gray] (-1.8,-0.2) -- (-1.8,0.2);
        \draw[color=gray] (-1.4,-0.2) -- (-1.4,0.2);
        \draw[color=gray] (-1,-0.2) -- (-1,0.2);
        \draw[color=gray] (-0.6,-0.2) -- (-0.6,0.2);
        \draw (0.8,1.5) node[rotate=-45, text width=1 cm, align=center] {\scriptsize $n_1-k$};
        \draw (0.8,-1.5) node[rotate=-45, text width=1 cm, align=center] {\scriptsize $n_2-k$};
        \draw (-1.3,0.8) node[text width=1 cm, align=center, AGHred] {\scriptsize $k$};
       \draw (-2.7,2.7)  node[text width=.1 cm,align=center] { \small $\tau_1$};
       \draw (-2.5,-2.5)  node[text width=.1 cm,align=center] { \small $\tau_2$};
    \end{tikzpicture}
\end{align}

Following the same rule, joining a portion of the remaining $n_2 - k$ chords of region II with $m$ out of the $n_3$ chords of region III yields
\begin{align}
    \begin{tikzpicture}[scale=0.4, baseline={([yshift=0cm]current bounding box.center)}, rotate=45]
    \definecolor{AGHblue}{rgb}{0.118, 0.294, 0.812}
    \definecolor{AGHred}{rgb}{0.843, 0.267, 0.153}
        \draw[very thick] (-1.026,-2.819) arc (250:110:3);
        \draw[black](-1.026,-2.819) -- (-1.026,2.819);
        \draw[fill=AGHred, draw=none] (3,0) circle (0.2cm);
        \draw[fill=AGHred, draw=none] (-3,0) circle (0.2cm);
        \draw[fill=AGHblue, draw=none] (0,3) circle (0.2cm);
        \draw[fill=AGHblue, draw=none] (0,-3) circle (0.2cm);
        \draw[color=black] (0,-3) -- (0,3);
        \draw[color=AGHred, very thick] (3,0) -- (-3,0);
        \draw[color=AGHblue, very thick] (0,3) -- (0,-3);
        \draw (-0.7,1.5) node[rotate=-45, text width=1 cm, align=center] {\scriptsize $n_1-k$};
        \draw (-0.7,-1.5) node[rotate=-45, text width=1 cm, align=center] {\scriptsize $n_2-k$};
        \draw (-2,0.5) node[text width=1 cm, align=center, AGHred] {\scriptsize $k$};
       \draw (-2.7,2.7)  node[text width=.1 cm,align=center] { \small $\tau_1$};
       \draw (-2.5,-2.5)  node[text width=.1 cm,align=center] { \small $\tau_2$};
       \draw (0.,3.8)  node[text width=.1 cm,align=center,AGHblue] { \small $V$};
       \draw (3.8,0)  node[text width=.1 cm,align=center,AGHred] { \small $W$};
       \draw[very thick] (1.,-4.) arc (-90:0:3);
       \draw[black] (1.,-4.) .. controls (0,-2) and (0,0.5) .. (4.,-1);
       \draw (1.5,-1.5) node[rotate=0, text width=1 cm, align=center] {\scriptsize $n_3$};
    \end{tikzpicture}
    =\sum_{m=0}^{\min(n_3,n_2-k)} D_m^{n_3,n_2-k} q_V^{m}
    \times
    \begin{tikzpicture}[scale=0.35, baseline={([yshift=0cm]current bounding box.center)}, rotate=45]
    \definecolor{AGHblue}{rgb}{0.118, 0.294, 0.812}
    \definecolor{AGHred}{rgb}{0.843, 0.267, 0.153}
         \draw[very thick] (0,-5) arc (270:110:5);
         \draw[fill=AGHred, draw=none] (5,0) circle (0.2cm);
         \draw[fill=AGHred, draw=none] (-5,0) circle (0.2cm);
         \draw[fill=AGHblue, draw=none] (0,5) circle (0.2cm);
         \draw[fill=AGHblue, draw=none] (0,-5) circle (0.2cm);
         \draw[color=black] (0,-3) -- (0,3);
         \draw[color=AGHred, very thick] (5,0) -- (-5,0);
         \draw[color=AGHblue, very thick] (0,5) -- (0,-5);
         \draw (-2.,2.) node[rotate=-45, text width=1 cm, align=center] {\scriptsize $n_1-k$};
        \draw (-4.2,4.2)  node[text width=.1 cm,align=center] { \small $\tau_1$};
        \draw (-4.,-4.)  node[text width=.1 cm,align=center] { \small $\tau_2$};
        \draw (3.,-4.5)  node[text width=.1 cm,align=center] { \small $\tau_3$};
        \draw[very thick] (0,-5) arc (270:340:5);
        \draw[color=black] (0,-1.71) -- (4.70,-1.71);
        \draw (0.5,-3.5) node[text width=1 cm, align=center, AGHblue] {\scriptsize $m$};
        \draw[color=black] (-1.71,0) arc (180:270:1.71);
        \draw (-1.3,-1.6) node[rotate=0, text width=1.2 cm, align=center] {\fontsize{5pt}{6pt}\selectfont $n_2-k-m$};
        \draw (2.5,-2) node[rotate=45, text width=1.2 cm, align=center] {\scriptsize $n_3-m$};
        \draw[color=black] (-1.71,0) -- (-1.71,4.70);
        \draw (-3.5,0.5) node[text width=1 cm, align=center, AGHred] {\scriptsize $k$};
\end{tikzpicture}
\end{align}
Notice the remaining $n_2-k-m$ chords of region II; these will ultimately connect to chords in Region IV. 

Next, we proceed to join regions III and IV, resulting in
\begin{align}
    \begin{tikzpicture}[scale=0.35, baseline={([yshift=0cm]current bounding box.center)}, rotate=45]
    \definecolor{AGHblue}{rgb}{0.118, 0.294, 0.812}
    \definecolor{AGHred}{rgb}{0.843, 0.267, 0.153}
     \draw[very thick] (0,-5) arc (270:110:5);
     \draw[fill=AGHred, draw=none] (5,0) circle (0.2cm);
     \draw[fill=AGHred, draw=none] (-5,0) circle (0.2cm);
     \draw[fill=AGHblue, draw=none] (0,5) circle (0.2cm);
     \draw[fill=AGHblue, draw=none] (0,-5) circle (0.2cm);
     \draw[color=black] (0,-3) -- (0,3);
     \draw[color=AGHred, very thick] (5,0) -- (-5,0);
     \draw[color=AGHblue, very thick] (0,5) -- (0,-5);
     \draw (-2.,2.) node[rotate=-45, text width=1 cm, align=center] {\scriptsize $n_1-k$};
    \draw (-4.2,4.2)  node[text width=.1 cm,align=center] { \small $\tau_1$};
    \draw (-4.,-4.)  node[text width=.1 cm,align=center] { \small $\tau_2$};
    \draw (3.,-4.5)  node[text width=.1 cm,align=center] { \small $\tau_3$};
    \draw[very thick] (0,-5) arc (270:340:5);
    \draw[color=black] (0,-1.71) -- (4.70,-1.71);
    \draw (0.5,-3.5) node[text width=1 cm, align=center, AGHblue] {\scriptsize $m$};
    \draw[color=black] (-1.71,0) arc (180:270:1.71);
    \draw (-1.3,-1.6) node[rotate=0, text width=1.2 cm, align=center] {\fontsize{5pt}{6pt}\selectfont $n_2-k-m$};
    \draw (2.5,-2) node[rotate=45, text width=1.2 cm, align=center] {\scriptsize $n_3-m$};
    \draw[color=black] (-1.71,0) -- (-1.71,4.70);
    \draw (-3.5,0.5) node[text width=1 cm, align=center, AGHred] {\scriptsize $k$};
    \draw[very thick] (6.,1.2) arc (0:90:5);
    \draw[black] (6.,1.2) .. controls (0,-1) and (0,5) .. (1,6.2);
    \draw (2,2) node[rotate=0, text width=1 cm, align=center] {\scriptsize $n_4$};
    \end{tikzpicture}
    =
    \sum_{s=0}^{\min(n_4,n_3-m)}D_s^{n_4,n_3-m} q_W^s \times
    \quad 
    \begin{tikzpicture}[scale=0.35, baseline={([yshift=0cm]current bounding box.center)}, rotate=45]
    \definecolor{AGHblue}{rgb}{0.118, 0.294, 0.812}
    \definecolor{AGHred}{rgb}{0.843, 0.267, 0.153}
        \draw[very thick] (0,-5) arc (270:110:5);
        \draw[fill=AGHred, draw=none] (5,0) circle (0.2cm);
        \draw[fill=AGHred, draw=none] (-5,0) circle (0.2cm);
        \draw[fill=AGHblue, draw=none] (0,5) circle (0.2cm);
        \draw[fill=AGHblue, draw=none] (0,-5) circle (0.2cm);
        \draw[color=black] (0,-3) -- (0,3);
        \draw[color=AGHred, very thick] (5,0) -- (-5,0);
        \draw[color=AGHblue, very thick] (0,5) -- (0,-5);
       \draw (-4.2,4.2)  node[text width=.1 cm,align=center] { \small $\tau_1$};
       \draw (-4.,-4.)  node[text width=.1 cm,align=center] { \small $\tau_2$};
       \draw (3.,-4.5)  node[text width=.1 cm,align=center] { \small $\tau_3$};
       \draw (4.5,3)  node[text width=.1 cm,align=center] { \small $\tau_4$};
       \draw[very thick] (0,-5) arc (270:360:5);
       \draw[color=black] (0,-1.71) arc (270:360:1.71);
       \draw (2.3,-2.3) node[rotate=0, text width=1.05 cm, align=center] {\fontsize{5pt}{6pt}\selectfont $n_3-m-s$};
       \draw (0.5,-3.5) node[text width=1 cm, align=center, AGHblue] {\scriptsize $m$};
       \draw[color=black] (-1.71,0) arc (180:270:1.71);
       \draw (-1.3,-1.6) node[rotate=0, text width=1.2 cm, align=center] {\fontsize{5pt}{6pt}\selectfont $n_2-k-m$};
      \draw[very thick] (5.,0) arc (0:70:5);
      \draw[color=black] (1.71,0) -- (1.71,4.7);
      \draw (3.5,-0.5) node[text width=1 cm, align=center, AGHred] {\scriptsize $s$};
      \draw[color=black] (-1.71,0) -- (-1.71,4.7);
      \draw (-2.1,2.5) node[rotate=-45, text width=1 cm, align=center] {\scriptsize $n_1-k$};
      \draw[color=black] (-1.71,0) -- (-1.71,4.7);
      \draw (2.1,2.5) node[rotate=-45, text width=1 cm, align=center] {\scriptsize $n_4-s$};
      \draw (-3.5,0.5) node[text width=1 cm, align=center, AGHred] {\scriptsize $k$};
    \end{tikzpicture}
\end{align}
There are $s$ chords connecting the $\tau_3$ and $\tau_4$ segments, while the remaining $n_3 - m - s$ chords originating from $\tau_3$ will eventually connect to chords in region I that are generated from the $\tau_1$ segment.

Finally, we connect I and IV and get
\begin{align}
    \begin{tikzpicture}[scale=0.35, baseline={([yshift=0cm]current bounding box.center)}, rotate=45]
    \definecolor{AGHblue}{rgb}{0.118, 0.294, 0.812}
    \definecolor{AGHred}{rgb}{0.843, 0.267, 0.153}
        \draw[very thick] (0,-5) arc (270:110:5);
        \draw[fill=AGHred, draw=none] (5,0) circle (0.2cm);
        \draw[fill=AGHred, draw=none] (-5,0) circle (0.2cm);
        \draw[fill=AGHblue, draw=none] (0,5) circle (0.2cm);
        \draw[fill=AGHblue, draw=none] (0,-5) circle (0.2cm);
        \draw[color=black] (0,-3) -- (0,3);
        \draw[color=AGHred, very thick] (5,0) -- (-5,0);
        \draw[color=AGHblue, very thick] (0,5) -- (0,-5);
       \draw (-4.2,4.2)  node[text width=.1 cm,align=center] { \small $\tau_1$};
       \draw (-4.,-4.)  node[text width=.1 cm,align=center] { \small $\tau_2$};
       \draw (3.,-4.5)  node[text width=.1 cm,align=center] { \small $\tau_3$};
       \draw (4.5,3)  node[text width=.1 cm,align=center] { \small $\tau_4$};
       \draw[very thick] (0,-5) arc (270:360:5);
       \draw[color=black] (0,-1.71) arc (270:360:1.71);
       \draw (2.3,-2.3) node[rotate=0, text width=1.05 cm, align=center] {\fontsize{5pt}{6pt}\selectfont $n_3-m-s$};
       \draw (0.5,-3.5) node[text width=1 cm, align=center, AGHblue] {\scriptsize $m$};
       \draw[color=black] (-1.71,0) arc (180:270:1.71);
       \draw (-1.3,-1.6) node[rotate=0, text width=1.2 cm, align=center] {\fontsize{5pt}{6pt}\selectfont $n_2-k-m$};
      \draw[very thick] (5.,0) arc (0:70:5);
      \draw[color=black] (1.71,0) -- (1.71,4.7);
      \draw (3.5,-0.5) node[text width=1 cm, align=center, AGHred] {\scriptsize $s$};
      \draw[color=black] (-1.71,0) -- (-1.71,4.7);
      \draw (-2.1,2.5) node[rotate=-45, text width=1 cm, align=center] {\scriptsize $n_1-k$};
      \draw[color=black] (-1.71,0) -- (-1.71,4.7);
      \draw (2.1,2.5) node[rotate=-45, text width=1 cm, align=center] {\scriptsize $n_4-s$};
      \draw (-3.5,0.5) node[text width=1 cm, align=center, AGHred] {\scriptsize $k$};
    \end{tikzpicture}
    =\quad 
    \sum_{t=0}^{\min(n_1-k,n_4-s)} D_{t}^{n_1-k, n_4-s} q_{V}^t  \times
    \quad
    \begin{tikzpicture}[scale=0.35, baseline={([yshift=0cm]current bounding box.center)}, rotate=45]
    \definecolor{AGHblue}{rgb}{0.118, 0.294, 0.812}
    \definecolor{AGHred}{rgb}{0.843, 0.267, 0.153}
        \draw[very thick] (0,-5) arc (270:270+360:5);
        \draw[fill=AGHred, draw=none] (5,0) circle (0.2cm);
        \draw[fill=AGHred, draw=none] (-5,0) circle (0.2cm);
        \draw[fill=AGHblue, draw=none] (0,5) circle (0.2cm);
        \draw[fill=AGHblue, draw=none] (0,-5) circle (0.2cm);
        \draw[color=black] (0,-3) -- (0,3);
        \draw[color=AGHred, very thick] (5,0) -- (-5,0);
        \draw[color=AGHblue, very thick] (0,5) -- (0,-5);
       \draw (-4.2,4.2)  node[text width=.1 cm,align=center] { \small $\tau_1$};
       \draw (-4.,-4.)  node[text width=.1 cm,align=center] { \small $\tau_2$};
       \draw (3.,-4.5)  node[text width=.1 cm,align=center] { \small $\tau_3$};
       \draw (4,4)  node[text width=.1 cm,align=center] { \small $\tau_4$};
       \draw[color=black] (0,-1.71) arc (270:270+360:1.71);
       \draw (2.3,-2.3) node[rotate=0, text width=1.05 cm, align=center] {\fontsize{5pt}{6pt}\selectfont $n_3-m-s$};
       \draw (0.5,-3.5) node[text width=1 cm, align=center, AGHblue] {\scriptsize $m$};
       \draw (-1.3,-1.6) node[rotate=0, text width=1.2 cm, align=center] {\fontsize{5pt}{6pt}\selectfont $n_2-k-m$};
      \draw (3.5,-0.5) node[text width=1 cm, align=center, AGHred] {\scriptsize $s$};
      \draw (0.5,3.5) node[text width=1 cm, align=center, AGHblue] {\scriptsize $t$};
      \draw (1.5,1.5) node[rotate=0, text width=1.05 cm, align=center] {\fontsize{5pt}{6pt}\selectfont $n_4-s-t$};
      \draw (-3.5,0.5) node[text width=1 cm, align=center, AGHred] {\scriptsize $k$};
      \draw (-2.3,2.3) node[rotate=0, text width=1.05 cm, align=center] {\fontsize{5pt}{6pt}\selectfont $n_1-k-t$};
    \end{tikzpicture}
\end{align}
What's left is joining II with IV and I with III. This amounts to multiplying what we derived so far by the following factors:
\begin{itemize}
    \item $\delta_{n_1 - k - t, n_3 - m - s} \delta_{n_4 - s - t, n_2 - k - m}$
    due to the conservation of chord numbers,
    \item $(n_1 - k - t)! (n_2 - k - m)!$ arising from the combinatorics of connecting I-III chords and II-IV chords,
    \item $(q_V q_W)^{(n_1 - k - t)+(n_2 - k - m)}$ for the crossings of matter cords and Hamiltonian chords,
    \item $q_{VW}$ for the intersection between $V$-chord and $W$-chord.
\end{itemize}
Putting everything together, we get
\begin{align}
    &\!\!\!\!\!\!\!\!\!\!\!\!\!\!\!\!\!\!\!
    \begin{tikzpicture}[scale=0.25, baseline={([yshift=0cm]current bounding box.center)}]
    \definecolor{AGHblue}{rgb}{0.118, 0.294, 0.812}
    \definecolor{AGHred}{rgb}{0.843, 0.267, 0.153}
    \draw[very thick] (0,0) circle (5cm);
    \draw[thick, color = AGHred] (-3.5,-3.5) -- (3.5, 3.5);
    \draw[thick, color = AGHblue] (-3.5, 3.5) -- (3.5, -3.5);
    \draw (0,-5.8)  node[text width=3.1 cm,align=center] { \small $\tau_2$};
    \draw (0,5.8)  node[text width=3.1 cm,align=center] { \small $\tau_4$};
    \draw (-5.8,0)  node[text width=3.1 cm,align=center] { \small $\tau_1$};
    \draw (5.8,0)  node[text width=3.1 cm,align=center] { \small $\tau_3$};
    \draw[fill=black, color=AGHred] (-3.5, -3.5) circle (0.2);
    \draw (-4.5,-4.) node[text width=2 cm,align=center,AGHred] { \small $W$};
    \draw[fill=black, color=AGHred] (3.5, 3.5) circle (0.2);
    \draw (4.5,4.) node[text width=2 cm,align=center,AGHred] { \small $W$};
    \draw[fill=black, color=AGHblue] (3.5, -3.5) circle (0.2);
    \draw (-4.5,4.) node[text width=2 cm,align=center,AGHblue] { \small $V$};
    \draw[fill=black, color=AGHblue] (-3.5, 3.5) circle (0.2);
    \draw (4.5,-4.) node[text width=2 cm,align=center,AGHblue] { \small $V$};
    \end{tikzpicture}
    \!\!\!\!\!\!\!\!\!\!\!\!\!\!\!\!\!\!\!
    = \left(\prod_{i=1}^{4}\sum_{n_i=0}^{\infty} \int {dE_i \over \sqrt{\pi}} e^{-E_i^2-E_i \tau_i} \frac{H_{n_i}(E_i)}{2^{n_i \over 2}n_i!}\right)\times \delta_{n_1 - k - t, n_3 - m - s} \delta_{n_4 - s - t, n_2 - k - m} \notag \\
    &\times\sideset{}{'}\sum_{k,m,s,t}
    \frac{n_1! n_2! n_3 ! n_4 !}{k!m! s!t! (n_3 -m -s)!(n_4-s-t)!}\times q_V^{n_1+n_2-2k}q_W^{n_1+n_4-2t} q_{VW}
    \notag \\
    =&e^{\sum_{i=1}^4 \frac{\tau_i^2}{4}} \sum_{n_1,n_2,n_3,n_4=0}^{\infty}\left(-\frac{\tau_1}{\sqrt{2}}\right)^{n_1} \left(-\frac{\tau_2}{\sqrt{2}}\right)^{n_2}
    \left(-\frac{\tau_3}{\sqrt{2}}\right)^{n_3}
    \left(-\frac{\tau_4}{\sqrt{2}}\right)^{n_4}
    \delta_{n_1 - k - t, n_3 - m - s} \delta_{n_4 - s - t, n_2 - k - m}
    \notag \\
    &\times \sideset{}{'}\sum_{k,m,s,t} \frac{1}{k! m! s! t! (n_3 -m -s)! (n_4 - t - s)!} 
    \times q_V^{n_1+n_2-2k}q_W^{n_1+n_4-2t} q_{VW}
    \label{eq:OTOC-C11}
\end{align}
where $\sideset{}{'}\sum_{k,m,s,t}$ is a shorthand for $\sum_{k=0}^{\min(n_1,n_2)}
\sum_{m=0}^{\min(n_3,n_2-k)}
\sum_{s=0}^{\min(n_4,n_3-m)}
\sum_{t=0}^{\min(n_1-k,n_4-s)}$.
In the second step, we have performed the $E_i$ integrals using the fact
\begin{align}
\int \! \!  {dE\over \sqrt{\pi}} e^{- {E^2} - \tau E} H_{n}(E)= e^{\tau^2 \over 4}(- \tau)^n~,
\label{eq:integral_C12}
\end{align}
The two Kronecker deltas represent constraints that allow us to express $n_1$ and $n_2$ in terms of $n_3$ and $n_4$. This is useful because it enables us to make all sums free, namely:
\begin{align}
    \sum_{n_1,n_2,n_3,n_4}\sideset{}{'}\sum_{k,m,s,t}
    \to \sum_{k,m,s,t=0}^{\infty} \sum_{n_3 = m+s}^{\infty} \sum_{n_4 = s+t}^{\infty}
    \label{eq:sum-transform}
\end{align}
It follows that 
\begin{align}
    \text{eq.(\ref{eq:OTOC-C11})} 
    =& 
    q_{VW}\times e^{\sum_{i=1}^4 \frac{\tau_i^2}{4}} \sum_{k,m,s,t,\tilde n_3,\tilde n_4 = 0}^{\infty}
    \left[\frac{1}{k!}\left(\frac{\tau_1 \tau_2}{2}q_W\right)^k\right]
    \left[\frac{1}{m!}\left(\frac{\tau_2 \tau_3}{2}q_V\right)^m\right]
    \left[\frac{1}{s!}\left(\frac{\tau_3 \tau_4}{2}q_W\right)^s\right]
    \notag \\
    &\times \left[\frac{1}{t!}\left(\frac{\tau_1 \tau_4}{2} q_V\right)^t\right]
    \left[\frac{1}{\tilde n_3}\left(\frac{\tau_1 \tau_3}{2}q_V q_W\right)^{\tilde n_3}\right]
    \left[\frac{1}{\tilde n_4!} \left(\frac{\tau_2 \tau_4}{2}q_V q_W\right)^{\tilde n_4}\right]
\end{align}
where 
$\tilde n_3 \equiv n_3 - m -s$, $\tilde n_4 \equiv n_4 - s -t$.
All the sums can be written as exponentials. We then arrive at
\begin{align}
    \!\!\!\!\!\!\!\!\!\!\!\!\!\!\!\!\!\!\!
    \begin{tikzpicture}[scale=0.25, baseline={([yshift=0cm]current bounding box.center)}]
    \definecolor{AGHblue}{rgb}{0.118, 0.294, 0.812}
    \definecolor{AGHred}{rgb}{0.843, 0.267, 0.153}
    \draw[very thick] (0,0) circle (5cm);
    \draw[thick, color = AGHred] (-3.5,-3.5) -- (3.5, 3.5);
    \draw[thick, color = AGHblue] (-3.5, 3.5) -- (3.5, -3.5);
    \draw (0,-5.8)  node[text width=3.1 cm,align=center] { \small $\tau_2$};
    \draw (0,5.8)  node[text width=3.1 cm,align=center] { \small $\tau_4$};
    \draw (-5.8,0)  node[text width=3.1 cm,align=center] { \small $\tau_1$};
    \draw (5.8,0)  node[text width=3.1 cm,align=center] { \small $\tau_3$};
    \draw[fill=black, color=AGHred] (-3.5, -3.5) circle (0.2);
    \draw (-4.5,-4.) node[text width=2 cm,align=center,AGHred] { \small $W$};
    \draw[fill=black, color=AGHred] (3.5, 3.5) circle (0.2);
    \draw (4.5,4.) node[text width=2 cm,align=center,AGHred] { \small $W$};
    \draw[fill=black, color=AGHblue] (3.5, -3.5) circle (0.2);
    \draw (-4.5,4.) node[text width=2 cm,align=center,AGHblue] { \small $V$};
    \draw[fill=black, color=AGHblue] (-3.5, 3.5) circle (0.2);
    \draw (4.5,-4.) node[text width=2 cm,align=center,AGHblue] { \small $V$};
    \end{tikzpicture}
    \!\!\!\!\!\!\!\!\!\!\!\!\!\!\!\!\!\!\!
    &= e^{-\Delta_V \Delta_W} \exp \Bigg[\sum_{i=1}^{4} \frac{\tau_i^2}{4} + \frac{e^{-\Delta_V}}{2}(\tau_1 \tau_4 +\tau_2 \tau_3) 
    + \frac{e^{-\Delta_W}}{2}(\tau_1 \tau_2+\tau_3 \tau_4) \notag \\
    &\quad \quad + \frac{e^{-\Delta_V - \Delta_W}}{2}(\tau_1 \tau_3 + \tau_2 \tau_4)\Bigg]
\end{align}
Including the normalization factor $Z^{-1} = e^{-\frac{\beta^2}{4}}$ for $\sum_{i=1}^4 \tau_i = \beta$, we get (\ref{eq:OTOC}).

\subsection{Time-ordered four-point correlation function (TOC)}
\label{appendix:TOC}
The (unnormalized) time-ordered four-point correlation function of the boundary theory can be represented in the bulk theory as
\begin{align}
    \tr[e^{-\tau_4 H}W e^{-\tau_3 H} W e^{-\tau_2 H} V e^{-\tau_1 H} V]
    =
    \begin{tikzpicture}[scale=0.25, baseline={([yshift=0cm]current bounding box.center)}]
    \definecolor{AGHblue}{rgb}{0.118, 0.294, 0.812}
    \definecolor{AGHred}{rgb}{0.843, 0.267, 0.153}
       \draw[very thick] (0,0) circle (5cm);
       \draw (-6.5,0)  node[text width=.1 cm,align=center] { \small $\tau_1$};
       \draw (0,-5.8)  node[text width=.1 cm,align=center] { \small $\tau_2$};
       \draw (5.8,0)  node[text width=.1 cm,align=center] { \small $\tau_3$};
       \draw (0,5.8)  node[text width=.1 cm,align=center] { \small $\tau_4$};
       \draw[fill=black, color=AGHblue] (-3.5, -3.5) circle (0.2);
       \draw (-4.5,-4.) node[text width=1 cm,align=center,AGHblue] { \small $V$};
       \draw[fill=black, color=AGHred] (3.5, 3.5) circle (0.2);
       \draw (4.5,4.) node[text width=1 cm,align=center,AGHred] { \small $W$};
       \draw[fill=black, color=AGHred] (3.5, -3.5) circle (0.2);
       \draw (-4.5,4.) node[text width=1 cm,align=center,AGHblue] { \small $V$};
       \draw[fill=black, color=AGHblue] (-3.5, 3.5) circle (0.2);
       \draw (4.5,-4.) node[text width=1 cm,align=center,AGHred] { \small $W$};
       \draw[thick,AGHblue] (-3.5, -3.5) .. controls (-2,-0.5) and (-2,0.5) .. (-3.5, 3.5);
       \draw[thick,AGHred] (3.5, -3.5) .. controls (2,-0.5) and (2,0.5) .. (3.5, 3.5);
    \end{tikzpicture}
    \label{eq:TOC-C16}
\end{align}
We now evaluate the graphical representation in (\ref{eq:TOC-C16}). Similar to the OTOC calculation, we observe that the insertions of four operators divide the boundary into four segments: $\tau_1$, $\tau_2$, $\tau_3$, and $\tau_4$.

First, we will calculate the state prepared by each segment, then combine them, taking into account the crossings of matter cords and Hamiltonian chords, to construct the complete computation of Eq. (\ref{eq:TOC-C16}).

As shown in (\ref{eq:product-of-4-segment}), the split boundary segments prepare the product of four states: $|n(\tau_1)\rangle \otimes |n(\tau_2)\rangle \otimes |n(\tau_3)\rangle \otimes |n(\tau_4)\rangle$.

Following the same gluing rules that are used in Appendix \ref{appendix:OTOC}, we first connect $\tau_1$ and $\tau_2$ by
\begin{align}
    \begin{tikzpicture}[scale=0.35, baseline={([yshift=0cm]current bounding box.center)}]
    \definecolor{AGHblue}{rgb}{0.118, 0.294, 0.812}
    \definecolor{AGHred}{rgb}{0.843, 0.267, 0.153}
       \draw[very thick] (5,0) arc (0:35:5);
       \draw[very thick] (5,0) arc (0:-35:5);
       \draw[very thick] (0,5) arc (90:125:5);
       \draw[very thick] (0,5) arc (90:55:5);
       \draw[very thick] (-5,0) arc (180:145:5);
       \draw[very thick] (-5,0) arc (180:215:5);
       \draw[very thick] (0,-5) arc (270:235:5);
       \draw[very thick] (0,-5) arc (270:305:5);
       \draw (-6.5,0)  node[text width=.1 cm,align=center] { \small $\tau_1$};
       \draw (0,-5.8)  node[text width=.1 cm,align=center] { \small $\tau_2$};
       \draw (5.8,0)  node[text width=.1 cm,align=center] { \small $\tau_3$};
       \draw (0,5.8)  node[text width=.1 cm,align=center] { \small $\tau_4$};
       \draw[fill=black, color=AGHblue] (-3.5, -3.5) circle (0.2);
       \draw (-4.5,-4.) node[text width=1 cm,align=center,AGHblue] { \small $V$};
       \draw[fill=black, color=AGHred] (3.5, 3.5) circle (0.2);
       \draw (4.5,4.) node[text width=1 cm,align=center,AGHred] { \small $W$};
       \draw[fill=black, color=AGHred] (3.5, -3.5) circle (0.2);
       \draw (-4.5,4.) node[text width=1 cm,align=center,AGHblue] { \small $V$};
       \draw[fill=black, color=AGHblue] (-3.5, 3.5) circle (0.2);
       \draw (4.5,-4.) node[text width=1 cm,align=center,AGHred] { \small $W$};
        \draw[thick,AGHblue] (-3.5, -3.5) .. controls (-0.5,-1) and (-0.5,1) .. (-3.5, 3.5);
       \draw[thick,AGHred] (3.5, -3.5) .. controls (0.5,-0.5) and (0.5,0.5) .. (3.5, 3.5);
       \draw[black] (-2.87, -4.10)
       .. controls (-1,-1) and (1,-1) .. (2.87, -4.10);
       \draw[black] (2.87, 4.10)
       .. controls (1,1) and (-1,1) .. (-2.87, 4.10);
       \draw[black] (-4.10, 2.87) .. controls (-1,1) and (-1,-1) .. (-4.10, -2.87);
       \draw[black] (4.10, -2.87) .. controls (1,-1) and (1,1) .. (4.10, 2.87);
       \draw (-2.5,0) node[text width=1 cm,align=center] { \small $n_1$};
       \draw (0,-2.5) node[text width=1 cm,align=center] { \small $n_2$};
       \draw (2.5,0) node[text width=1 cm,align=center] { \small $n_3$};
       \draw (0,2.5) node[text width=1 cm,align=center] { \small $n_4$};
    \end{tikzpicture}
    \quad
    =
    \sum_{k=0}^{\min(n_1,n_2)} D_{k}^{n_1,n_2} q_V^k \times
    \quad
    \begin{tikzpicture}[scale=0.35, baseline={([yshift=0cm]current bounding box.center)}]
    \definecolor{AGHblue}{rgb}{0.118, 0.294, 0.812}
    \definecolor{AGHred}{rgb}{0.843, 0.267, 0.153}
       \draw[very thick] (5,0) arc (0:35:5);
       \draw[very thick] (5,0) arc (0:-35:5);
       \draw[very thick] (0,5) arc (90:125:5);
       \draw[very thick] (0,5) arc (90:55:5);
       \draw[very thick] (-5,0) arc (180:145:5);
       \draw[very thick] (-5,0) arc (180:225:5);
       \draw[very thick] (0,-5) arc (270:225:5);
       \draw[very thick] (0,-5) arc (270:305:5);
       \draw (-6.5,0)  node[text width=.1 cm,align=center] { \small $\tau_1$};
       \draw (0,-5.8)  node[text width=.1 cm,align=center] { \small $\tau_2$};
       \draw (5.8,0)  node[text width=.1 cm,align=center] { \small $\tau_3$};
       \draw (0,5.8)  node[text width=.1 cm,align=center] { \small $\tau_4$};
       \draw[fill=black, color=AGHblue] (-3.5, -3.5) circle (0.2);
       \draw (-4.5,-4.) node[text width=1 cm,align=center,AGHblue] { \small $V$};
       \draw[fill=black, color=AGHred] (3.5, 3.5) circle (0.2);
       \draw (4.5,4.) node[text width=1 cm,align=center,AGHred] { \small $W$};
       \draw[fill=black, color=AGHred] (3.5, -3.5) circle (0.2);
       \draw (-4.5,4.) node[text width=1 cm,align=center,AGHblue] { \small $V$};
       \draw[fill=black, color=AGHblue] (-3.5, 3.5) circle (0.2);
       \draw (4.5,-4.) node[text width=1 cm,align=center,AGHred] { \small $W$};
        \draw[thick,AGHblue] (-3.5, -3.5) .. controls (-0.5,-1) and (-0.5,1) .. (-3.5, 3.5);
       \draw[thick,AGHred] (3.5, -3.5) .. controls (0.5,-0.5) and (0.5,0.5) .. (3.5, 3.5);
       \draw[black] (-4.10,2.87)
       -- (2.87, -4.10);
       \draw[black] (2.87, 4.10)
       .. controls (1,1) and (-1,1) .. (-2.87, 4.10);
       \draw[black] (4.10, -2.87) .. controls (1,-1) and (1,1) .. (4.10, 2.87);
       \draw (-3,1) node[text width=1 cm,align=center,rotate = -45] { \scriptsize $n_1-k$};
       \draw (1,-3) node[text width=1 cm,align=center,rotate=-45] { \scriptsize $n_2-k$};
       \draw (-2.5,-2) node[text width=1 cm,align=center,rotate = 0,AGHblue] { \scriptsize $k$};
       \draw (2.5,0) node[text width=1 cm,align=center] { \scriptsize $n_3$};
       \draw (0,2.5) node[text width=1 cm,align=center] { \scriptsize $n_4$};
    \end{tikzpicture}
\end{align}
where $D_{k}^{n_1,n_2}$ is the combinatorial factor given by (\ref{eq:D-combinatorial}).
There are $k$ chords connecting the boundary segments $\tau_1$ and $\tau_2$, while crossing the matter $V$ chord. The remaining $n_1-k$ open chords originating from $\tau_1$ and $n_2-k$ open chords from $\tau_2$ will connect to $\tau_3$ and $\tau_4$, as will be illustrated in the subsequent gluing process.

For the process of gluing the $\tau_1$-$\tau_2$ segment to $\tau_3$, we have
\begin{align}
    \!\!\!\!\!\!
    \begin{tikzpicture}[scale=0.35, baseline={([yshift=0cm]current bounding box.center)}]
    \definecolor{AGHblue}{rgb}{0.118, 0.294, 0.812}
    \definecolor{AGHred}{rgb}{0.843, 0.267, 0.153}
       \draw[very thick] (5,0) arc (0:35:5);
       \draw[very thick] (5,0) arc (0:-35:5);
       \draw[very thick] (0,5) arc (90:125:5);
       \draw[very thick] (0,5) arc (90:55:5);
       \draw[very thick] (-5,0) arc (180:145:5);
       \draw[very thick] (-5,0) arc (180:225:5);
       \draw[very thick] (0,-5) arc (270:225:5);
       \draw[very thick] (0,-5) arc (270:305:5);
       \draw (-6.5,0)  node[text width=.1 cm,align=center] { \small $\tau_1$};
       \draw (0,-5.8)  node[text width=.1 cm,align=center] { \small $\tau_2$};
       \draw (5.8,0)  node[text width=.1 cm,align=center] { \small $\tau_3$};
       \draw (0,5.8)  node[text width=.1 cm,align=center] { \small $\tau_4$};
       \draw[fill=black, color=AGHblue] (-3.5, -3.5) circle (0.2);
       \draw (-4.5,-4.) node[text width=1 cm,align=center,AGHblue] { \small $V$};
       \draw[fill=black, color=AGHred] (3.5, 3.5) circle (0.2);
       \draw (4.5,4.) node[text width=1 cm,align=center,AGHred] { \small $W$};
       \draw[fill=black, color=AGHred] (3.5, -3.5) circle (0.2);
       \draw (-4.5,4.) node[text width=1 cm,align=center,AGHblue] { \small $V$};
       \draw[fill=black, color=AGHblue] (-3.5, 3.5) circle (0.2);
       \draw (4.5,-4.) node[text width=1 cm,align=center,AGHred] { \small $W$};
        \draw[thick,AGHblue] (-3.5, -3.5) .. controls (-0.5,-1) and (-0.5,1) .. (-3.5, 3.5);
       \draw[thick,AGHred] (3.5, -3.5) .. controls (0.5,-0.5) and (0.5,0.5) .. (3.5, 3.5);
       \draw[black] (-4.10,2.87)
       -- (2.87, -4.10);
       \draw[black] (2.87, 4.10)
       .. controls (1,1) and (-1,1) .. (-2.87, 4.10);
       \draw[black] (4.10, -2.87) .. controls (1,-1) and (1,1) .. (4.10, 2.87);
       \draw (-3,1) node[text width=1 cm,align=center,rotate = -45] { \scriptsize $n_1-k$};
       \draw (1,-3) node[text width=1 cm,align=center,rotate=-45] { \scriptsize $n_2-k$};
       \draw (-2.5,-2) node[text width=1 cm,align=center,rotate = 0,AGHblue] { \scriptsize $k$};
       \draw (2.5,0) node[text width=1 cm,align=center] { \scriptsize $n_3$};
       \draw (0,2.5) node[text width=1 cm,align=center] { \scriptsize $n_4$};
    \end{tikzpicture}
    \quad
    =
    \sum_{m=0}^{\min(n_2-k,n_3)} D_m^{n_2-k,n_3} q_W^m \times \quad
    \begin{tikzpicture}[scale=0.35, baseline={([yshift=0cm]current bounding box.center)}]
    \definecolor{AGHblue}{rgb}{0.118, 0.294, 0.812}
    \definecolor{AGHred}{rgb}{0.843, 0.267, 0.153}
       \draw[very thick] (5,0) arc (0:35:5);
       \draw[very thick] (5,0) arc (0:-45:5);
       \draw[very thick] (0,5) arc (90:125:5);
       \draw[very thick] (0,5) arc (90:55:5);
       \draw[very thick] (-5,0) arc (180:145:5);
       \draw[very thick] (-5,0) arc (180:225:5);
       \draw[very thick] (0,-5) arc (270:225:5);
       \draw[very thick] (0,-5) arc (270:315:5);
       \draw (-6.5,0)  node[text width=.1 cm,align=center] { \small $\tau_1$};
       \draw (0,-5.8)  node[text width=.1 cm,align=center] { \small $\tau_2$};
       \draw (5.8,0)  node[text width=.1 cm,align=center] { \small $\tau_3$};
       \draw (0,5.8)  node[text width=.1 cm,align=center] { \small $\tau_4$};
       \draw[fill=black, color=AGHblue] (-3.5, -3.5) circle (0.2);
       \draw (-4.5,-4.) node[text width=1 cm,align=center,AGHblue] { \small $V$};
       \draw[fill=black, color=AGHred] (3.5, 3.5) circle (0.2);
       \draw (4.5,4.) node[text width=1 cm,align=center,AGHred] { \small $W$};
       \draw[fill=black, color=AGHred] (3.5, -3.5) circle (0.2);
       \draw (-4.5,4.) node[text width=1 cm,align=center,AGHblue] { \small $V$};
       \draw[fill=black, color=AGHblue] (-3.5, 3.5) circle (0.2);
       \draw (4.5,-4.) node[text width=1 cm,align=center,AGHred] { \small $W$};
        \draw[thick,AGHblue] (-3.5, -3.5) .. controls (-0.5,-1) and (-0.5,1) .. (-3.5, 3.5);
        \draw[thick,AGHred] (3.5, -3.5) .. controls (0.5,-0.5) and (0.5,0.5) .. (3.5, 3.5);
       \draw[black] (2.87, 4.10)
       .. controls (1,1) and (-1,1) .. (-2.87, 4.10);
        \draw[black] (-4.10,2.87)
       .. controls (1,-2) and (-1,-2) .. (4.10,2.87);
       \draw (-3,1) node[text width=1 cm,align=center,rotate = -45] { \scriptsize $n_1-k$};
       \draw (3,1) node[text width=1 cm,align=center,rotate = 45] { \scriptsize $n_3-m$};
       \draw (0.2,-1.9) node[text width=1.2 cm,align=center,rotate=0] { \fontsize{5pt}{6pt}\selectfont $n_2-k-m$};
       \draw (-2.8,-2.) node[text width=1 cm,align=center,rotate = 0,AGHblue] { \scriptsize $k$};
       \draw (2.8,-2.) node[text width=1 cm,align=center,rotate = 0,AGHred] { \scriptsize $m$};
       \draw (0,2.5) node[text width=1 cm,align=center] { \scriptsize $n_4$};
    \end{tikzpicture}
\end{align}
For the $n_2-k$ chords emerging from the $\tau_2$ boundary segment, $m$ of them connect to the chords generated by $\tau_3$, while the remaining $n_2-k-m$ connect to those generated by $\tau_4$, as will be seen.

Further gluing the $\tau_1$-$\tau_2$-$\tau_3$ segment to $\tau_4$ gives
\begin{align}
        \begin{tikzpicture}[scale=0.35, baseline={([yshift=0cm]current bounding box.center)}]
        \definecolor{AGHblue}{rgb}{0.118, 0.294, 0.812}
        \definecolor{AGHred}{rgb}{0.843, 0.267, 0.153}
       \draw[very thick] (5,0) arc (0:35:5);
       \draw[very thick] (5,0) arc (0:-45:5);
       \draw[very thick] (0,5) arc (90:125:5);
       \draw[very thick] (0,5) arc (90:55:5);
       \draw[very thick] (-5,0) arc (180:145:5);
       \draw[very thick] (-5,0) arc (180:225:5);
       \draw[very thick] (0,-5) arc (270:225:5);
       \draw[very thick] (0,-5) arc (270:315:5);
       \draw (-6.5,0)  node[text width=.1 cm,align=center] { \small $\tau_1$};
       \draw (0,-5.8)  node[text width=.1 cm,align=center] { \small $\tau_2$};
       \draw (5.8,0)  node[text width=.1 cm,align=center] { \small $\tau_3$};
       \draw (0,5.8)  node[text width=.1 cm,align=center] { \small $\tau_4$};
       \draw[fill=black, color=AGHblue] (-3.5, -3.5) circle (0.2);
       \draw (-4.5,-4.) node[text width=1 cm,align=center,AGHblue] { \small $V$};
       \draw[fill=black, color=AGHred] (3.5, 3.5) circle (0.2);
       \draw (4.5,4.) node[text width=1 cm,align=center,AGHred] { \small $W$};
       \draw[fill=black, color=AGHred] (3.5, -3.5) circle (0.2);
       \draw (-4.5,4.) node[text width=1 cm,align=center,AGHblue] { \small $V$};
       \draw[fill=black, color=AGHblue] (-3.5, 3.5) circle (0.2);
       \draw (4.5,-4.) node[text width=1 cm,align=center,AGHred] { \small $W$};
        \draw[thick,AGHblue] (-3.5, -3.5) .. controls (-0.5,-1) and (-0.5,1) .. (-3.5, 3.5);
       \draw[thick,AGHred] (3.5, -3.5) .. controls (0.5,-0.5) and (0.5,0.5) .. (3.5, 3.5);
       \draw[black] (2.87, 4.10)
       .. controls (1,1) and (-1,1) .. (-2.87, 4.10);
        \draw[black] (-4.10,2.87)
       .. controls (1,-2) and (-1,-2) .. (4.10,2.87);
       \draw (-3,1) node[text width=1 cm,align=center,rotate = -45] { \scriptsize $n_1-k$};
       \draw (3,1) node[text width=1 cm,align=center,rotate = 45] { \scriptsize $n_3-m$};
       \draw (0.2,-1.9) node[text width=1.2 cm,align=center,rotate=0] { \fontsize{5pt}{6pt}\selectfont $n_2-k-m$};
       \draw (-2.8,-2.) node[text width=1 cm,align=center,rotate = 0,AGHblue] { \scriptsize $k$};
       \draw (2.8,-2.) node[text width=1 cm,align=center,rotate = 0,AGHred] { \scriptsize $m$};
       \draw (0,2.5) node[text width=1 cm,align=center] { \scriptsize $n_4$};
    \end{tikzpicture}
    \quad
    =
    \sum_{s=0}^{\min(n_4,n_3-m)}
    D_{s}^{n_4,n_3-m} q_W^s
    \times \quad
    \begin{tikzpicture}[scale=0.35, baseline={([yshift=0cm]current bounding box.center)}]
    \definecolor{AGHblue}{rgb}{0.118, 0.294, 0.812}
    \definecolor{AGHred}{rgb}{0.843, 0.267, 0.153}
       \draw[very thick] (5,0) arc (0:45:5);
       \draw[very thick] (5,0) arc (0:-45:5);
       \draw[very thick] (0,5) arc (90:125:5);
       \draw[very thick] (0,5) arc (90:45:5);
       \draw[very thick] (-5,0) arc (180:145:5);
       \draw[very thick] (-5,0) arc (180:225:5);
       \draw[very thick] (0,-5) arc (270:225:5);
       \draw[very thick] (0,-5) arc (270:315:5);
       \draw (-6.5,0)  node[text width=.1 cm,align=center] { \small $\tau_1$};
       \draw (0,-5.8)  node[text width=.1 cm,align=center] { \small $\tau_2$};
       \draw (5.8,0)  node[text width=.1 cm,align=center] { \small $\tau_3$};
       \draw (0,5.8)  node[text width=.1 cm,align=center] { \small $\tau_4$};
       \draw[fill=black, color=AGHblue] (-3.5, -3.5) circle (0.2);
       \draw (-4.5,-4.) node[text width=1 cm,align=center,AGHblue] { \small $V$};
       \draw[fill=black, color=AGHred] (3.5, 3.5) circle (0.2);
       \draw (4.5,4.) node[text width=1 cm,align=center,AGHred] { \small $W$};
       \draw[fill=black, color=AGHred] (3.5, -3.5) circle (0.2);
       \draw (-4.5,4.) node[text width=1 cm,align=center,AGHblue] { \small $V$};
       \draw[fill=black, color=AGHblue] (-3.5, 3.5) circle (0.2);
       \draw (4.5,-4.) node[text width=1 cm,align=center,AGHred] { \small $W$};
        \draw[thick,AGHblue] (-3.5, -3.5) .. controls (0.8,-1) and (0.8,1) .. (-3.5, 3.5);
       \draw[thick,AGHred] (3.5, -3.5) .. controls (0.1,-0.5) and (0.1,0.5) .. (3.5, 3.5);
        \draw[black] (-2.87,4.10)
       .. controls (6,-3) and (6,-3) .. (-4.10,2.87);
       \draw (-2.5,1.2) node[text width=1 cm,align=center,rotate = -30] { \scriptsize $n_1-k$};
       \draw (0,2.5) node[text width=1 cm,align=center,rotate = -30] { \scriptsize $n_4-s$};
       \draw (3.2,-0.1) node[text width=1.2 cm,align=center,rotate = -30] { \fontsize{5pt}{6pt}\selectfont  $n_3-m-s$};
       \draw (0.2,-2.5) node[text width=1.2 cm,align=center,rotate=0] { \fontsize{5pt}{6pt}\selectfont $n_2-k-m$};
       \draw[->,line width=0.8pt] (0.3,-2) -- (0.3,0.);
       \draw (-2.,-1.5) node[text width=1 cm,align=center,rotate = 0,AGHblue] { \scriptsize $k$};
       \draw (2.8,-2.) node[text width=1 cm,align=center,rotate = 0,AGHred] { \scriptsize $m$};
       \draw (2.8,2.2) node[text width=1 cm,align=center,rotate = 0,AGHred] { \scriptsize $s$};
    \end{tikzpicture}
\end{align}
Connecting a portion of $n_4-s$ chords arising from $\tau_4$ and a portion of $n_1-k$ from $\tau_1$ gives
\begin{align}
        \begin{tikzpicture}[scale=0.35, baseline={([yshift=0cm]current bounding box.center)}]
        \definecolor{AGHblue}{rgb}{0.118, 0.294, 0.812}
        \definecolor{AGHred}{rgb}{0.843, 0.267, 0.153}
        \draw[very thick] (5,0) arc (0:45:5);
       \draw[very thick] (5,0) arc (0:-45:5);
       \draw[very thick] (0,5) arc (90:125:5);
       \draw[very thick] (0,5) arc (90:45:5);
       \draw[very thick] (-5,0) arc (180:145:5);
       \draw[very thick] (-5,0) arc (180:225:5);
       \draw[very thick] (0,-5) arc (270:225:5);
       \draw[very thick] (0,-5) arc (270:315:5);
       \draw (-6.5,0)  node[text width=.1 cm,align=center] { \small $\tau_1$};
       \draw (0,-5.8)  node[text width=.1 cm,align=center] { \small $\tau_2$};
       \draw (5.8,0)  node[text width=.1 cm,align=center] { \small $\tau_3$};
       \draw (0,5.8)  node[text width=.1 cm,align=center] { \small $\tau_4$};
       \draw[fill=black, color=AGHblue] (-3.5, -3.5) circle (0.2);
       \draw (-4.5,-4.) node[text width=1 cm,align=center,AGHblue] { \small $V$};
       \draw[fill=black, color=AGHred] (3.5, 3.5) circle (0.2);
       \draw (4.5,4.) node[text width=1 cm,align=center,AGHred] { \small $W$};
       \draw[fill=black, color=AGHred] (3.5, -3.5) circle (0.2);
       \draw (-4.5,4.) node[text width=1 cm,align=center,AGHblue] { \small $V$};
       \draw[fill=black, color=AGHblue] (-3.5, 3.5) circle (0.2);
       \draw (4.5,-4.) node[text width=1 cm,align=center,AGHred] { \small $W$};
        \draw[thick,AGHblue] (-3.5, -3.5) .. controls (0.8,-1) and (0.8,1) .. (-3.5, 3.5);
       \draw[thick,AGHred] (3.5, -3.5) .. controls (0.1,-0.5) and (0.1,0.5) .. (3.5, 3.5);
        \draw[black] (-2.87,4.10)
       .. controls (6,-3) and (6,-3) .. (-4.10,2.87);
       \draw (-2.5,1.2) node[text width=1 cm,align=center,rotate = -30] { \scriptsize $n_1-k$};
       \draw (0,2.5) node[text width=1 cm,align=center,rotate = -30] { \scriptsize $n_4-s$};
       \draw (3.2,-0.1) node[text width=1.2 cm,align=center,rotate = -30] { \fontsize{5pt}{6pt}\selectfont  $n_3-m-s$};
       \draw (0.2,-2.5) node[text width=1.2 cm,align=center,rotate=0] { \fontsize{5pt}{6pt}\selectfont $n_2-k-m$};
       \draw[->,line width=0.8pt] (0.3,-2) -- (0.3,0.);
       \draw (-2.,-1.5) node[text width=1 cm,align=center,rotate = 0,AGHblue] { \scriptsize $k$};
       \draw (2.8,-2.) node[text width=1 cm,align=center,rotate = 0,AGHred] { \scriptsize $m$};
       \draw (2.8,2.2) node[text width=1 cm,align=center,rotate = 0,AGHred] { \scriptsize $s$};
    \end{tikzpicture}
    \quad
    =\sum_{t=0}^{\min(n_1-k,n_4-s)}
    D_t^{n_1-k,n_4-s} q_V^t \times 
    \begin{tikzpicture}[scale=0.35, baseline={([yshift=0cm]current bounding box.center)}]
    \definecolor{AGHblue}{rgb}{0.118, 0.294, 0.812}
    \definecolor{AGHred}{rgb}{0.843, 0.267, 0.153}
       \draw[black] (0,0) circle (1.5cm);
       \draw[very thick] (5,0) arc (0:45:5);
       \draw[very thick] (5,0) arc (0:-45:5);
       \draw[very thick] (0,5) arc (90:135:5);
       \draw[very thick] (0,5) arc (90:45:5);
       \draw[very thick] (-5,0) arc (180:135:5);
       \draw[very thick] (-5,0) arc (180:225:5);
       \draw[very thick] (0,-5) arc (270:225:5);
       \draw[very thick] (0,-5) arc (270:315:5);
       \draw (-6.5,0)  node[text width=.1 cm,align=center] { \small $\tau_1$};
       \draw (0,-5.8)  node[text width=.1 cm,align=center] { \small $\tau_2$};
       \draw (5.8,0)  node[text width=.1 cm,align=center] { \small $\tau_3$};
       \draw (0,5.8)  node[text width=.1 cm,align=center] { \small $\tau_4$};
       \draw[fill=black, color=AGHblue] (-3.5, -3.5) circle (0.2);
       \draw (-4.5,-4.) node[text width=1 cm,align=center,AGHblue] { \small $V$};
       \draw[fill=black, color=AGHred] (3.5, 3.5) circle (0.2);
       \draw (4.5,4.) node[text width=1 cm,align=center,AGHred] { \small $W$};
       \draw[fill=black, color=AGHred] (3.5, -3.5) circle (0.2);
       \draw (-4.5,4.) node[text width=1 cm,align=center,AGHblue] { \small $V$};
       \draw[fill=black, color=AGHblue] (-3.5, 3.5) circle (0.2);
       \draw (4.5,-4.) node[text width=1 cm,align=center,AGHred] { \small $W$};
        \draw[thick,AGHblue] (-3.5, -3.5) .. controls (0.8,-1) and (0.8,1) .. (-3.5, 3.5);
       \draw[thick,AGHred] (3.5, -3.5) .. controls (0.1,-0.5) and (0.1,0.5) .. (3.5, 3.5);
       \draw (-3.2,-0.1) node[text width=1.cm,align=center,rotate = 0] {  \fontsize{5pt}{6pt}\selectfont $n_1-k-t$};
       \draw (0.2,2.5) node[text width=1 cm,align=center,rotate = 0] {  \fontsize{5pt}{6pt}\selectfont $n_4-s-t$};
       \draw (3.3,-0.1) node[text width=1.1 cm,align=center,rotate = 0] { \fontsize{5pt}{6pt}\selectfont  $n_3-m-s$};
       \draw (0.2,-2.5) node[text width=1.2 cm,align=center,rotate=0] { \fontsize{5pt}{6pt}\selectfont $n_2-k-m$};
       \draw (-2.,-1.5) node[text width=1 cm,align=center,rotate = 0,AGHblue] { \scriptsize $k$};
       \draw (-2.,1.8) node[text width=1 cm,align=center,rotate = 0,AGHblue] { \scriptsize $t$};
       \draw (2.8,-2.) node[text width=1 cm,align=center,rotate = 0,AGHred] { \scriptsize $m$};
       \draw (2.8,2.2) node[text width=1 cm,align=center,rotate = 0,AGHred] { \scriptsize $s$};
    \end{tikzpicture}
\end{align}
The last step to get the graph in (\ref{eq:TOC-C16}) is to connect the remaining open chords from $\tau_1$ to $\tau_3$ and from $\tau_2$ to $\tau_4$.
This is achieved by multiplying the following factors:
\begin{itemize}
    \item $\delta_{n_1-k-t,n_3-m-s} \delta_{n_2-k-m,n_4-s-t}$ by the conservation of chord numbers,
    \item $(n_1-k-t)! (n_2-k-m)!$ from the combinatorics of connecting the remaining open chords,
    \item $q_V^{n_1-k-t} q_W^{n_3-m-s}$ for the crossings of matter cords and Hamiltonian chords.
\end{itemize}
Collecting all these elements and using (\ref{eq:product-of-4-segment}) for the product state $|n(\tau_1)\ran \otimes|n(\tau_2)\ran \otimes|n(\tau_3)\ran \otimes|n(\tau_4)\ran$, we obtain
\begin{align}
    &\begin{tikzpicture}[scale=0.25, baseline={([yshift=0cm]current bounding box.center)}]
    \definecolor{AGHblue}{rgb}{0.118, 0.294, 0.812}
    \definecolor{AGHred}{rgb}{0.843, 0.267, 0.153}
       \draw[very thick] (0,0) circle (5cm);
       \draw (-6.5,0)  node[text width=.1 cm,align=center] { \small $\tau_1$};
       \draw (0,-5.8)  node[text width=.1 cm,align=center] { \small $\tau_2$};
       \draw (5.8,0)  node[text width=.1 cm,align=center] { \small $\tau_3$};
       \draw (0,5.8)  node[text width=.1 cm,align=center] { \small $\tau_4$};
       \draw[fill=black, color=AGHblue] (-3.5, -3.5) circle (0.2);
       \draw (-4.5,-4.) node[text width=1 cm,align=center,AGHblue] { \small $V$};
       \draw[fill=black, color=AGHred] (3.5, 3.5) circle (0.2);
       \draw (4.5,4.) node[text width=1 cm,align=center,AGHred] { \small $W$};
       \draw[fill=black, color=AGHred] (3.5, -3.5) circle (0.2);
       \draw (-4.5,4.) node[text width=1 cm,align=center,AGHblue] { \small $V$};
       \draw[fill=black, color=AGHblue] (-3.5, 3.5) circle (0.2);
       \draw (4.5,-4.) node[text width=1 cm,align=center,AGHred] { \small $W$};
       \draw[thick,AGHblue] (-3.5, -3.5) .. controls (-2,-0.5) and (-2,0.5) .. (-3.5, 3.5);
       \draw[thick,AGHred] (3.5, -3.5) .. controls (2,-0.5) and (2,0.5) .. (3.5, 3.5);
    \end{tikzpicture}
    = \left(\prod_{i=1}^{4}\sum_{n_i=0}^{\infty} \int {dE_i \over \sqrt{\pi}} e^{-E_i^2-E_i \tau_i} \frac{H_{n_i}(E_i)}{2^{n_i \over 2}n_i!}\right)\times
    \delta_{n_1-k-t,n_3-m-s} \delta_{n_2-k-m,n_4-s-t} \notag \\
    &\times
    \sideset{}{'}\sum_{k,m,s,t}
    \frac{n_1! n_2! n_3 ! n_4!}{k! m! s! t! (n_3-m-s)! (n_4-s-t)!}\times 
    q_V^{(n_1-k-t)+k+t} q_W^{(n_3-m-s)+m+s} \notag \\
    =&e^{\sum_{i=1}^4 \frac{\tau_i^2}{4}} \sum_{n_1,n_2,n_3,n_4=0}^{\infty} \left(-\frac{\tau_1}{\sqrt{2}}\right)^{n_1}
    \left(-\frac{\tau_2}{\sqrt{2}}\right)^{n_2}
    \left(-\frac{\tau_3}{\sqrt{2}}\right)^{n_3}
    \left(-\frac{\tau_4}{\sqrt{2}}\right)^{n_4} \delta_{n_1-k-t,n_3-m-s} \delta_{n_2-k-m,n_4-s-t} \notag \\
    &\times \sideset{}{'}\sum_{k,m,s,t} \frac{1}{k! m! s! t! (n_3-m-s)! (n_4-s-t)!}\times 
    q_V^{n_1} q_V^{n_3}
    \label{eq:TOC-C21}
\end{align}
Here $\sideset{}{'}\sum_{k,m,s,t}$ is a shorthand for $\sum_{k=0}^{\min(n_1,n_2)} \sum_{m=0}^{\min(n_2-k,n_3)} \sum_{s=0}^{\min(n_4,n_3-m)} \sum_{t=0}^{\min(n_1-k,n_4-s)}$.
In the second step, we performed the integrals using (\ref{eq:integral_C12}).
As in the OTOC calculation, we make use of the Kronecker delta constraints to express $n_1$ and $n_2$ in terms of $n_3$ and $n_4$, which leads again to the same replacement of sums (\ref{eq:sum-transform}).
By further changing the dummy variables to $\tilde n_3 \equiv n_3 - m -s$ and $\tilde n_4 \equiv n_4-s-t$, we get
\begin{align}
    \text{eq.(\ref{eq:TOC-C21})} =&\ e^{\sum_{i=1}^4 \frac{\tau_i^2}{4}}\sum_{k,m,s,t,\tilde n_3 \tilde n_4=0}^{\infty}\left[\frac{1}{k!}\left(\frac{\tau_1 \tau_2}{2}q_V\right)^k
    \right]
    \left[
    \frac{1}{m!}\left(\frac{\tau_2 \tau_3}{2}q_W\right)^m\right]
    \left[\frac{1}{s!}\left(\frac{\tau_3 \tau_4}{2}q_W\right)^s\right] \notag \\
    &\times
    \left[\frac{1}{t!}\left(\frac{\tau_1 \tau_4}{2}q_V\right)^t\right]
    \left[\frac{1}{\tilde n_3!} \left(\frac{\tau_1 \tau_3}{2} q_V q_W\right)^{\tilde n_3}\right]
    \left[\frac{1}{\tilde n_4!} \left(\frac{\tau_2 \tau_4}{2}\right)^{\tilde n_4}\right]~.
\end{align}
Note that each sum gives an exponential, we finally obtain
\begin{align}
    \begin{tikzpicture}[scale=0.25, baseline={([yshift=0cm]current bounding box.center)}]
    \definecolor{AGHblue}{rgb}{0.118, 0.294, 0.812}
    \definecolor{AGHred}{rgb}{0.843, 0.267, 0.153}
       \draw[very thick] (0,0) circle (5cm);
       \draw (-6.5,0)  node[text width=.1 cm,align=center] { \small $\tau_1$};
       \draw (0,-5.8)  node[text width=.1 cm,align=center] { \small $\tau_2$};
       \draw (5.8,0)  node[text width=.1 cm,align=center] { \small $\tau_3$};
       \draw (0,5.8)  node[text width=.1 cm,align=center] { \small $\tau_4$};
       \draw[fill=black, color=AGHblue] (-3.5, -3.5) circle (0.2);
       \draw (-4.5,-4.) node[text width=1 cm,align=center,AGHblue] { \small $V$};
       \draw[fill=black, color=AGHred] (3.5, 3.5) circle (0.2);
       \draw (4.5,4.) node[text width=1 cm,align=center,AGHred] { \small $W$};
       \draw[fill=black, color=AGHred] (3.5, -3.5) circle (0.2);
       \draw (-4.5,4.) node[text width=1 cm,align=center,AGHblue] { \small $V$};
       \draw[fill=black, color=AGHblue] (-3.5, 3.5) circle (0.2);
       \draw (4.5,-4.) node[text width=1 cm,align=center,AGHred] { \small $W$};
       \draw[thick,AGHblue] (-3.5, -3.5) .. controls (-2,-0.5) and (-2,0.5) .. (-3.5, 3.5);
       \draw[thick,AGHred] (3.5, -3.5) .. controls (2,-0.5) and (2,0.5) .. (3.5, 3.5);
    \end{tikzpicture}
    =&\ \exp\Bigg[\sum_{i=1}^4 \frac{\tau_i^2}{4} + 
    \frac{e^{-\Delta_V}}{2}(\tau_1 \tau_2 + \tau_1 \tau_4)+\frac{e^{-\Delta_W}}{2}(\tau_2 \tau_3 + \tau_3 \tau_4) \notag \\
    &+\frac{e^{-\Delta_V-\Delta_W}}{2} \tau_1 \tau_3 + \frac{\tau_2 \tau_4}{2}
    \Bigg]~.
\end{align}
Including the normalization factor $Z^{-1} = e^{-\frac{\beta^2}{4}}$ for $\sum_{i=1}^4 \tau_i = \beta$, we get (\ref{eq:TOC}).

\subsection{All point correlation functions}

In the previous subsections, we worked out the two- and four-point functions. Here we give the derivation for the general all-point correlation function building on the analysis in appendix \ref{alllengthgen}.

In the probe limit approximation, general matter correlation functions are evaluated via Wick contractions. Suppose we have an even number $k$ of operators inserted around the thermal circle. This divides the boundary into $k$ intervals. In appendix \ref{alllengthgen}, we saw that the distribution of the number of chords connecting any two such intervals of lengths $\tau_i,\tau_j$ is 
\begin{align}
    {1 \over m_{ij}!} \left( \tau_i \tau_j \over 2\right)^{m_{ij}}
\end{align}
Therefore, if a matter cord separates regions $i$ and $j$, then we have to multiply this distribution by the penalty factor $e^{-\Delta}$ where $\Delta$ is the dimension of the matter operator. This slightly distorts the $m_{ij}$ distribution to 
\begin{align}
    {1 \over m_{ij}!} \left( e^{-\Delta}\tau_i \tau_j \over 2\right)^{m_{ij}}
\end{align}
Accounting for all the matter insertions and which boundary regions they separate and summing over $m_{ij}$ we land on
\begin{align}
    \mathrm{Exp}\left[ -{\beta^2 \over 4} + \sum_{i=1}^{4} {\tau^2_i \over 4} + \sum_{i<j=1}^n {e^{-\sum_k \Delta_k} \over 2}\tau_i \tau_j   \right]e^{\sum \prod_i \Delta_i}.
\end{align}
The first two terms in the exponent are for normalizing the correlator; the correlator goes to one if all $\Delta$'s are set to zero. The $\sum_k \Delta_k$ in the exponential refers to all the matter cords encountered by the Hamiltonian chords connecting $i$ and $j$. Matter cords intersections generate the last factor on the right.

\section{\texorpdfstring{$f(r)$}{f(r)} }
\label{appendix:f(r)}
In this appendix, we compute $f(r)$ in the metric ansatz (\ref{eq:rotationally invariant metric ansatz}) that gives rise to the length formula (\ref{ltheta}).\footnote{We thank Douglas Stanford, Xiaoliang Qi, and Alexey Milekhin for the discussion on this derivation.}

The central angle $\Theta$, swept by a geodesic of length $l$, is given by (\ref{thetaint}). Our goal is to derive $f(r)$ from it. To proceed, we need to express $\Theta$ as a function of the conserved quantity $M= r^2 \dot \theta$. For this purpose, we select the geodesic length itself as the affine parameter $\lambda$ such that the geodesic starts off from the boundary at $\lambda =0$ and reaches the turning point $r_{\text{min}}$ at $\lambda = l/2$.
For such a half of a geodesic, we have
\begin{align}
  \frac{\Theta}{2} 
  =\int_0^{l/2} d\lambda \ \dot \theta 
  = \int_0^{l/2} d\lambda\ \frac{M}{r(\lambda)^2} 
  = \frac{1}{M} \int_0^{l/2} d\lambda\ (1-f^2 \dot r^2)~.
\end{align}
In getting the last expression, we used the geodesic equation.
Taking the derivative with respect to $l$ and using the fact that at the turning point, $\dot r|_{\lambda = l/2} = 0$, we immediately get $M = \frac{dl}{d\Theta}$ and then
\begin{align}
  \Theta 
  = \pi\left(1 - \frac{r_{\text{min}}}{r_{\partial}}\right)
  =\pi\left(1-\frac{4\pi M}{\beta^2}\right) ~.
  \label{eq:theta-M}
\end{align}
Substituting (\ref{eq:theta-M}) for $\Theta$, we can rewrite (\ref{ltheta}) as
\begin{align}
  \pi\left(\frac{1}{M} - \frac{4\pi}{\beta^2}\right) = \int_{M^2}^{r_{\partial}^2} dr^2 \frac{\tilde f(r^2)/r^2}{ \sqrt{r^2 - M^2}}
\end{align}
where we introduced $\tilde f(r^2)\equiv f(r)$.

This equation takes the form of a generalized Abel equation \cite{Srivastav_1963}
\begin{align}
  \int_x^b dt \left( F(t) - F(x)\right)^{-\gamma} H(t) = G(x)~, 
  \quad\text{where } 
  x<b~, \quad 0<\gamma<1~,
  \label{eq:GAE}
\end{align}
upon the identification of variables and functions:
\begin{align}
  x = M^2~, \quad 
  \gamma&=\frac{1}{2}~, \quad
  b = r_{\partial}^2~, \quad 
  t=r^2~, 
  \label{eq:GAE-identification1}
  \\ 
  F: x \to x~, \quad 
  H&: t \to \frac{\tilde f(t)}{t}~, \quad 
  G: x \to \pi \left( \frac{1}{\sqrt{x}} - \frac{4\pi}{\beta^2}\right)~.
  \label{eq:GAE-identification2}
\end{align}
The solution to the generalized Abel equation (\ref{eq:GAE}) is given by \cite{Srivastav_1963}
\begin{align}
  H(t) = -\frac{\sin(\gamma \pi)}{\pi} \frac{d}{dt}\int_{t}^{b} du\ F'(u) \left( F(u) - F(t)\right)^{\gamma-1} G(u)~.
\end{align}
By applying the mappings (\ref{eq:GAE-identification1}) and (\ref{eq:GAE-identification2}), we obtain
\begin{align}
  f(r) = \sqrt{1-\frac{r^2}{r_{\partial}^2}} = \sqrt{1-\frac{16\pi^2 r^2}{\beta^4}}~.
\end{align}

\bibliographystyle{ssg}
\bibliography{Biblio}

\begingroup\raggedright\begin{thebibliography}{10}

\bibitem{Sachdev:1992fk}
S.~Sachdev and J.~Ye, ``{Gapless spin fluid ground state in a random, quantum
  Heisenberg magnet},'' {\em Phys. Rev. Lett.} {\bf 70} (1993) 3339,
  \href{http://xxx.lanl.gov/abs/cond-mat/9212030}{{\tt cond-mat/9212030}}.

\bibitem{kitaevTalks}
A.~Kitaev, ``{Talks given at the Fundamental Physics Prize Symposium and KITP
  seminars},''.
\newblock \url{https://www.youtube.com/watch?v=OQ9qN8j7EZI},
  \url{http://online.kitp.ucsb.edu/online/joint98/kitaev/},
  \url{http://online.kitp.ucsb.edu/online/entangled15/kitaev}.

\bibitem{Maldacena:2016hyu}
J.~Maldacena and D.~Stanford, ``{Remarks on the Sachdev-Ye-Kitaev model},''
  {\em Phys. Rev. D} {\bf 94} (2016), no.~10 106002,
  \href{http://xxx.lanl.gov/abs/1604.07818}{{\tt 1604.07818}}.

\bibitem{Cotler:2016fpe}
J.~S. Cotler, G.~Gur-Ari, M.~Hanada, J.~Polchinski, P.~Saad, S.~H. Shenker,
  D.~Stanford, A.~Streicher, and M.~Tezuka, ``{Black Holes and Random
  Matrices},'' {\em JHEP} {\bf 05} (2017) 118,
  \href{http://xxx.lanl.gov/abs/1611.04650}{{\tt 1611.04650}}. [Erratum: JHEP
  09, 002 (2018)].

\bibitem{Berkooz:2018jqr}
M.~Berkooz, M.~Isachenkov, V.~Narovlansky, and G.~Torrents, ``{Towards a full
  solution of the large N double-scaled SYK model},'' {\em JHEP} {\bf 03}
  (2019) 079, \href{http://xxx.lanl.gov/abs/1811.02584}{{\tt 1811.02584}}.

\bibitem{Berkooz:2018qkz}
M.~Berkooz, P.~Narayan, and J.~Simon, ``{Chord diagrams, exact correlators in
  spin glasses and black hole bulk reconstruction},'' {\em JHEP} {\bf 08}
  (2018) 192, \href{http://xxx.lanl.gov/abs/1806.04380}{{\tt 1806.04380}}.

\bibitem{Blommaert:2023opb}
A.~Blommaert, T.~G. Mertens, and S.~Yao, ``{Dynamical actions and
  q-representation theory for double-scaled SYK},'' {\em JHEP} {\bf 02} (2024)
  067, \href{http://xxx.lanl.gov/abs/2306.00941}{{\tt 2306.00941}}.

\bibitem{Blommaert:2023wad}
A.~Blommaert, T.~G. Mertens, and S.~Yao, ``{The q-Schwarzian and Liouville
  gravity},'' \href{http://xxx.lanl.gov/abs/2312.00871}{{\tt 2312.00871}}.

\bibitem{Okuyama:2022szh}
K.~Okuyama, ``{Hartle-Hawking wavefunction in double scaled SYK},'' {\em JHEP}
  {\bf 03} (2023) 152, \href{http://xxx.lanl.gov/abs/2212.09213}{{\tt
  2212.09213}}.

\bibitem{Okuyama:2023iwu}
K.~Okuyama, ``{High temperature expansion of double scaled SYK},''
  \href{http://xxx.lanl.gov/abs/2304.01522}{{\tt 2304.01522}}.

\bibitem{Okuyama:2023byh}
K.~Okuyama, ``{End of the world brane in double scaled SYK},'' {\em JHEP} {\bf
  08} (2023) 053, \href{http://xxx.lanl.gov/abs/2305.12674}{{\tt 2305.12674}}.

\bibitem{Okuyama:2023kdo}
K.~Okuyama, ``{Discrete analogue of the Weil-Petersson volume in double scaled
  SYK},'' {\em JHEP} {\bf 09} (2023) 133,
  \href{http://xxx.lanl.gov/abs/2306.15981}{{\tt 2306.15981}}.

\bibitem{Okuyama:2023aup}
K.~Okuyama and T.~Suyama, ``{Solvable limit of ETH matrix model for
  double-scaled SYK},'' \href{http://xxx.lanl.gov/abs/2311.02846}{{\tt
  2311.02846}}.

\bibitem{Okuyama:2023yat}
K.~Okuyama, ``{Matter correlators through a wormhole in double-scaled SYK},''
  {\em JHEP} {\bf 02} (2024) 147,
  \href{http://xxx.lanl.gov/abs/2312.00880}{{\tt 2312.00880}}.

\bibitem{Okuyama:2024yya}
K.~Okuyama, ``{Doubled Hilbert space in double-scaled SYK},''
  \href{http://xxx.lanl.gov/abs/2401.07403}{{\tt 2401.07403}}.

\bibitem{Boruch:2023bte}
J.~Boruch, H.~W. Lin, and C.~Yan, ``{Exploring supersymmetric wormholes in $
  \mathcal{N} $ = 2 SYK with chords},'' {\em JHEP} {\bf 12} (2023) 151,
  \href{http://xxx.lanl.gov/abs/2308.16283}{{\tt 2308.16283}}.

\bibitem{Berkooz:2022mfk}
M.~Berkooz, M.~Isachenkov, M.~Isachenkov, P.~Narayan, and V.~Narovlansky,
  ``{Quantum groups, non-commutative AdS$_{2}$, and chords in the double-scaled
  SYK model},'' {\em JHEP} {\bf 08} (2023) 076,
  \href{http://xxx.lanl.gov/abs/2212.13668}{{\tt 2212.13668}}.

\bibitem{Berkooz:2020xne}
M.~Berkooz, N.~Brukner, V.~Narovlansky, and A.~Raz, ``{The double scaled limit
  of Super--Symmetric SYK models},'' {\em JHEP} {\bf 12} (2020) 110,
  \href{http://xxx.lanl.gov/abs/2003.04405}{{\tt 2003.04405}}.

\bibitem{Xu:2024hoc}
J.~Xu, ``{Von Neumann Algebras in Double-Scaled SYK},''
  \href{http://xxx.lanl.gov/abs/2403.09021}{{\tt 2403.09021}}.

\bibitem{Goel:2023svz}
A.~Goel, V.~Narovlansky, and H.~Verlinde, ``{Semiclassical geometry in
  double-scaled SYK},'' \href{http://xxx.lanl.gov/abs/2301.05732}{{\tt
  2301.05732}}.

\bibitem{Lin_2022}
H.~W. Lin, ``The bulk Hilbert space of double scaled SYK,'' {\em Journal of
  High Energy Physics} {\bf 2022} (Nov., 2022).

\bibitem{berkooz:2024ofm}
M.~Berkooz, N.~Brukner, Y.~Jia, and O.~Mamroud, ``{A Path Integral for Chord
  Diagrams and Chaotic-Integrable Transitions in Double Scaled SYK},''
  \href{http://xxx.lanl.gov/abs/2403.05980}{{\tt 2403.05980}}.

\bibitem{Berkooz:2024evs}
M.~Berkooz, N.~Brukner, Y.~Jia, and O.~Mamroud, ``{From Chaos to Integrability
  in Double Scaled SYK},'' \href{http://xxx.lanl.gov/abs/2403.01950}{{\tt
  2403.01950}}.

\bibitem{Gao:2023gta}
P.~Gao, ``{Commuting SYK: a pseudo-holographic model},'' {\em JHEP} {\bf 01}
  (2024) 149, \href{http://xxx.lanl.gov/abs/2306.14988}{{\tt 2306.14988}}.

\bibitem{Gao:2016bin}
P.~Gao, D.~L. Jafferis, and A.~C. Wall, ``{Traversable Wormholes via a Double
  Trace Deformation},'' {\em JHEP} {\bf 12} (2017) 151,
  \href{http://xxx.lanl.gov/abs/1608.05687}{{\tt 1608.05687}}.

\bibitem{Maldacena:2017axo}
J.~Maldacena, D.~Stanford, and Z.~Yang, ``{Diving into traversable
  wormholes},'' {\em Fortsch. Phys.} {\bf 65} (2017), no.~5 1700034,
  \href{http://xxx.lanl.gov/abs/1704.05333}{{\tt 1704.05333}}.

\bibitem{Aalsma:2020aib}
L.~Aalsma and G.~Shiu, ``{Chaos and complementarity in de Sitter space},'' {\em
  JHEP} {\bf 05} (2020) 152, \href{http://xxx.lanl.gov/abs/2002.01326}{{\tt
  2002.01326}}.

\bibitem{Aalsma:2019rpt}
L.~Aalsma, M.~Parikh, and J.~P. Van Der~Schaar, ``{Back(reaction) to the Future
  in the Unruh-de Sitter State},'' {\em JHEP} {\bf 11} (2019) 136,
  \href{http://xxx.lanl.gov/abs/1905.02714}{{\tt 1905.02714}}.

\bibitem{Lin:2023trc}
H.~W. Lin and D.~Stanford, ``{A symmetry algebra in double-scaled SYK},'' {\em
  SciPost Phys.} {\bf 15} (2023) 234,
  \href{http://xxx.lanl.gov/abs/2307.15725}{{\tt 2307.15725}}.

\bibitem{Haehl:2021tft}
F.~M. Haehl, A.~Streicher, and Y.~Zhao, ``{Six-point functions and collisions
  in the black hole interior},'' {\em JHEP} {\bf 08} (2021) 134,
  \href{http://xxx.lanl.gov/abs/2105.12755}{{\tt 2105.12755}}.

\bibitem{Hartman:2013qma}
T.~Hartman and J.~Maldacena, ``{Time Evolution of Entanglement Entropy from
  Black Hole Interiors},'' {\em JHEP} {\bf 05} (2013) 014,
  \href{http://xxx.lanl.gov/abs/1303.1080}{{\tt 1303.1080}}.

\bibitem{Hashimoto:2020mrx}
K.~Hashimoto, ``{Building bulk from Wilson loops},'' {\em PTEP} {\bf 2021}
  (2021), no.~2 023B04, \href{http://xxx.lanl.gov/abs/2008.10883}{{\tt
  2008.10883}}.

\bibitem{Milekhin:2023bjv}
A.~Milekhin and J.~Xu, ``{Revisiting Brownian SYK and its possible relations to
  de Sitter},'' \href{http://xxx.lanl.gov/abs/2312.03623}{{\tt 2312.03623}}.

\bibitem{Susskind:2022dfz}
L.~Susskind, ``{Scrambling in Double-Scaled SYK and De Sitter Space},''
  \href{http://xxx.lanl.gov/abs/2205.00315}{{\tt 2205.00315}}.

\bibitem{Susskind:2022bia}
L.~Susskind, ``{De Sitter Space, Double-Scaled SYK, and the Separation of
  Scales in the Semiclassical Limit},''
  \href{http://xxx.lanl.gov/abs/2209.09999}{{\tt 2209.09999}}.

\bibitem{Susskind:2023hnj}
L.~Susskind, ``{De Sitter Space has no Chords. Almost Everything is
  Confined.},'' {\em JHAP} {\bf 3} (2023), no.~1 1--30,
  \href{http://xxx.lanl.gov/abs/2303.00792}{{\tt 2303.00792}}.

\bibitem{Rahman:2023pgt}
A.~A. Rahman and L.~Susskind, ``{Comments on a Paper by Narovlansky and
  Verlinde},'' \href{http://xxx.lanl.gov/abs/2312.04097}{{\tt 2312.04097}}.

\bibitem{Lin:2022nss}
H.~Lin and L.~Susskind, ``{Infinite Temperature's Not So Hot},''
  \href{http://xxx.lanl.gov/abs/2206.01083}{{\tt 2206.01083}}.

\bibitem{Rahman:2024vyg}
A.~A. Rahman and L.~Susskind, ``{Infinite Temperature is Not So Infinite: The
  Many Temperatures of de Sitter Space},''
  \href{http://xxx.lanl.gov/abs/2401.08555}{{\tt 2401.08555}}.

\bibitem{Rahman:2022jsf}
A.~A. Rahman, ``{dS JT Gravity and Double-Scaled SYK},''
  \href{http://xxx.lanl.gov/abs/2209.09997}{{\tt 2209.09997}}.

\bibitem{Narovlansky:2023lfz}
V.~Narovlansky and H.~Verlinde, ``{Double-scaled SYK and de Sitter
  Holography},'' \href{http://xxx.lanl.gov/abs/2310.16994}{{\tt 2310.16994}}.

\bibitem{Verlinde:2024znh}
H.~Verlinde, ``{Double-scaled SYK, Chords and de Sitter Gravity},''
  \href{http://xxx.lanl.gov/abs/2402.00635}{{\tt 2402.00635}}.

\bibitem{Verlinde:2024zrh}
H.~Verlinde and M.~Zhang, ``{SYK Correlators from 2D Liouville-de Sitter
  Gravity},'' \href{http://xxx.lanl.gov/abs/2402.02584}{{\tt 2402.02584}}.

\bibitem{Aguilar-Gutierrez:2024nau}
S.~E. Aguilar-Gutierrez, ``{Towards complexity in de Sitter space from the
  double-scaled Sachdev-Ye-Kitaev model},''
  \href{http://xxx.lanl.gov/abs/2403.13186}{{\tt 2403.13186}}.

\bibitem{Maldacena:2018lmt}
J.~Maldacena and X.-L. Qi, ``{Eternal traversable wormhole},''
  \href{http://xxx.lanl.gov/abs/1804.00491}{{\tt 1804.00491}}.

\bibitem{Jafferis:2022crx}
D.~Jafferis, A.~Zlokapa, J.~D. Lykken, D.~K. Kolchmeyer, S.~I. Davis, N.~Lauk,
  H.~Neven, and M.~Spiropulu, ``{Traversable wormhole dynamics on a quantum
  processor},'' {\em Nature} {\bf 612} (2022), no.~7938 51--55.

\bibitem{Kobrin:2023rzr}
B.~Kobrin, T.~Schuster, and N.~Y. Yao, ``{Comment on ''Traversable wormhole
  dynamics on a quantum processor''},''
  \href{http://xxx.lanl.gov/abs/2302.07897}{{\tt 2302.07897}}.

\bibitem{Jafferis:2023moh}
D.~Jafferis, A.~Zlokapa, J.~D. Lykken, D.~K. Kolchmeyer, S.~I. Davis, N.~Lauk,
  H.~Neven, and M.~Spiropulu, ``{Comment on ''Comment on ''Traversable wormhole
  dynamics on a quantum processor'' ''},''
  \href{http://xxx.lanl.gov/abs/2303.15423}{{\tt 2303.15423}}.

\bibitem{Srivastav_1963}
R.~P. Srivastav, ``A Note on Certain Integral Equations of Abel-Type,'' {\em
  Proceedings of the Edinburgh Mathematical Society} {\bf 13} (1963), no.~3
  271–272.

\end{thebibliography}\endgroup

\end{document}